\documentclass[aps,prx,showpacs,twocolumn,superscriptaddress,longbibliography]{revtex4-1}
\usepackage{graphicx}
\usepackage{dcolumn}
\usepackage{bm}
\usepackage{slashed}
\usepackage{amssymb}
\usepackage{amsmath}
\usepackage{amsthm}
\usepackage{xcolor}
\usepackage{tikz}
\usetikzlibrary{arrows,chains,matrix,positioning,scopes}
\makeatother
\usetikzlibrary{matrix}

\AtBeginDocument{%
    \newwrite\bibnotes
    \def\bibnotesext{Notes.bib}
    \immediate\openout\bibnotes=\jobname\bibnotesext
    \immediate\write\bibnotes{@CONTROL{REVTEX41Control}}
    \immediate\write\bibnotes{@CONTROL{%
    apsrev41Control,author="08",editor="1",pages="1",title="0",year="1"}}
     \if@filesw
     \immediate\write\@auxout{\string\citation{apsrev41Control}}%
    \fi
}%

\begin{document}

\title{Fractional topological insulator precursors in spin-orbit fermion ladders}

\author{Raul A. Santos}
\affiliation{School of Physics \& Astronomy, University of Birmingham, Edgbaston, Birmingham, B15 2TT, United Kingdom}
\affiliation{T.C.M. Group, Cavendish Laboratory, University of Cambridge, J.J. Thomson Avenue, Cambridge, CB3 0HE, United Kingdom}
\author{Benjamin B\'eri}
\affiliation{School of Physics \& Astronomy, University of Birmingham, Edgbaston, Birmingham, B15 2TT, United Kingdom}
\affiliation{T.C.M. Group, Cavendish Laboratory, University of Cambridge, J.J. Thomson Avenue, Cambridge, CB3 0HE, United Kingdom}
\affiliation{DAMTP, University of Cambridge, Wilberforce Road, Cambridge, CB3 0WA, United Kingdom}

\begin{abstract}
We study precursor states of fractional topological insulators (FTIs) 
in interacting fermionic ladders with spin-orbit coupling. 
Within a microscopically motivated bosonization approach, we investigate different competing phases depending on same-spin and interspin interactions at fractional effective 
filling $\nu=1/3$ per spin. In the spin-decoupled limit, we find that strong repulsive interactions of already moderate range may lead to a partially gapped state with two 
time-reversed copies of a quasi-one dimensional Laughlin phase. This FTI precursor competes with an interleg partially gapped phase displaying quasi long-range density wave 
order, however it may be stabilized if interactions are SU($2$) symmetric, or have suitable anisotropy, in leg space.
When the FTI phase is present, it is moderately robust to small interspin interactions; these introduce competing partially gapped phases of orbital antiferromagnetic and 
bond density wave character. 
Performing a strong coupling analysis of the FTI precursor regime, we find that the main effect of interspin interactions is to induce
correlated quasiparticle backscattering between the precursor FTI edge modes. Although this process competes with the topological phase, we show,
by considering an array of ladders, that its influence may disappear upon approaching the two dimensional case.
Considering time-reversal symmetry breaking perturbations, we also describe a protocol that adiabatically pumps $1/6$ charge per half-cycle, thus providing a quantized FTI 
signature arising already in the single ladder regime. 
\end{abstract}
\date{June 2018}
\preprint{}

\maketitle

\section{Introduction}
Topological phases of matter are characterized by topological invariants rather than conventional local order parameters.
Their discovery has widely extended the landscape of possible states of matter, beyond the standard classification based on symmetry breaking \cite{Haldane1988,Hasan2010,Qi2011}.
Nonetheless, symmetry can play a crucial role in the description of topological systems, as highlighted by the discovery of topological
insulators whose defining topological invariants exist only in the presence of time-reversal (TR) symmetry \cite{Kane2005,Kane2005b,Bernevig2006,Konig2007}.
The recognition of this interplay between symmetry and topology has led to a complete symmetry-based classification of topological phases in non-interacting 
fermion systems \cite{Schnyder2008,Kitaev2009,Morimoto2013,Chiu2013,Chiu2016}, providing a guideline for the search of new materials.

The presence of interparticle interactions greatly modifies the landscape of possible phases. In particular, interactions can give rise to fractionalised 
excitations \cite{Leinaas1977,Wilczek1982}, potentially useful for fault tolerant topological quantum computation \cite{Nayak2008}. Arguably the most prominent example that shows this fractionalisation is the fractional quantum Hall (FQH) state \cite{Tsui1982,Laughlin1983}.
This two-dimensional state supports anyonic quasiparticles in the bulk, i.e. quasiparticles with exchange statistics different
from bosons or fermions \cite{Goldman1995,Piccioto1997,Martin2004}. FQH 
states are also characterized by one-dimensional (1D) fractionalised states at their edge \cite{Lee1991}, which in the simplest case take the form of one-way chiral Luttinger 
liquids \cite{Wen1990}. These modes are robust against disorder as long as they cannot backscatter between different edges \cite{Chang2003}. This inter-edge backscattering is 
usually prevented by the width of the sample, as the bulk of the system forms a gapped incompressible quantum state. 

Given this picture, it is interesting that precursors of FQH states already emerge in quasi-1D systems such as two-leg ladders, which are the minimal departure from the strict 
1D limit \cite{Petrescu2015,Cornfeld2015,Calvanese2017,Petrescu2017,Haller2018}. In addition to being of interest on their own right, such ladder-based precursors may also be 
viewed as elementary building blocks of two-dimensional (2D) lattice FQH systems formed from an array of ladders, in a spirit similar to the coupled wire approach put forward 
in Refs. \onlinecite{Kane2002,Teo2014}.   

While the existence of robust FQH states is well established both theoretically and experimentally, much less is known about the strongly correlated, fractional analogues 
of TR invariant topological insulators, i.e., fractional topological insulators (FTIs). 
The simplest picture for such an FTI state is that of two copies of FQH states, one for each spin direction, with opposite magnetic field for opposite spins. This gives a 
TR-symmetric state because TR flips both the magnetic field and spin.
The direction in which the edge modes propagate is set by the magnetic field. An FTI therefore supports ``helical" edge modes: a counterpropagating pair of opposite-spin 
edge modes. 
The spin-dependent magnetic field in this picture is a particular case of spin-orbit (SO) coupling. The possibility of generating FTI states beyond this simple picture 
(e.g., with the two copies being coupled by more general SO couplings and interspin interactions) are among the key questions we shall address. 

Several approaches exist to uncover the edge and bulk characteristics of FTIs. 
The stability analysis of the edge modes starting from the picture above has been given in Refs.~\onlinecite{Levin2009,Neupert2011,SantosL2011,Stern2016}. 
Results on the bulk stability of different systems that may develop FTI states are so far based on numerical exact diagonalization 
\cite{Repellin2014,Neupert2011,Neupert2011b,Regnault2011,Sheng2011,Wu2012,Repellin2014,Fialko2014}, being thus limited to small system sizes.

In this paper, we develop a complementary approach to FTIs: motivated by the existence of FQH precursors in ladder systems, we investigate under what conditions FTI 
precursors may emerge in two-leg spinful fermion ladders, and what signatures such precursors may have in experiments. 
We also examine the phases that compete with the FTI precursor state, characterize their properties, and identify the regions in the phase diagram where the FTI precursor 
dominates.

The quasi-1D nature of our approach allows us to make progress in connecting phenomenological and microscopic considerations without being restricted to small system sizes.
The price paid for this advantage is the challenge of extrapolating our results reliably towards the genuinely 2D scenario. Nevertheless, in an approach 
similar to that of Ref.~\onlinecite{Kane2002,Teo2014,Neupert2014,Klinovaja2014,Sagi2014,Santos2015},
we are able to extend the discussion to two spatial dimensions by coupling a series of ladders and thus comment on possible FTI physics arising in a 2D lattice. 

The spinful ladder physics we describe may 
be particularly relevant to ultracold-atom-based realizations. In addition to the great degree of control over interactions these systems provide, they also offer a number of ingredients needed for prospective FTI states, such as SO fluxes via artificial gauge fields
\cite{Liu2009,Aidelsburger2011,Lin2011,Dalibard2011,Beugeling2012,Mei2012,Juzeliunas2010,Aidelsburger2013,Miyake2013,Beeler2013,Galitski2013,Atala2014,Mancini2015,Garcia2015,Stuhl2015,Zhai2015,Li2016,Song2016,Grusdt2017}
and topological bandstructures \cite{Goldman2010,Cooper2011a,Cooper2011b,Beri2011,Zhu2011,Jiang2011,Mei2012,Wei2012,Liu2012,Cocks2012,Nascimbene2013,Orth2013,Gopa2013,Jotzu2014,Goldman2014,Goldman2014b,Wall2016,Goldman2016,Goldman2017,Cooper2018}.
Our results may also be useful for solid state considerations, in particular, if one takes the view advocated above, that the ladders serve as elementary building blocks towards 2D FTI states.

In this work, we focus on the possible emergence of (precursors of) FTI states that form the TR invariant analogue of the simplest, most robust, FQH state: the Laughlin state
Although such a Laughlin FTI is just the simplest possibility, understanding the conditions for its emergence depending on same-spin
and interspin interactions will already provide a useful guide on how fermionic FTIs may be created. In addition, studying this possibility
illustrates how quasi-one dimensional ladder systems may be utilised as a playground for studying 2D topological states protected by symmetries. A key parameter in FQH as 
well as FTI systems is the filling fraction $\nu$: the ratio of the number of fermions per plaquette $n_{\rm plaq}$ (per spin) and the flux $\Phi$ per plaquette of the 
effective magnetic field. In terms of these, $\nu=2\pi n_{\rm plaq}/\Phi$, where $\Phi$ is measured in units of $h/e$.
(In what follows we use $\hbar=e=1$.) The 
Laughlin state corresponds to $\nu=1/3$.

Before we would turn to the detailed problem setup and analysis, we first provide a road-map of our approach and findings. We start with a two-leg fermion ladder with 
nearest and next-nearest neighbor interactions, in the presence of the simplest (opposite flux for opposite spin) SO coupling compatible with an FTI phase. (For a review on 
imprinting such synthetic gauge fields for ultracold atoms in quasi-1D geometries, see e.g., Ref.~\onlinecite{Dalibard2011}.)

Our first step is to develop a low-energy field theory of our microscopic lattice model, including an explicit link between microscopic interaction parameters and field 
theory couplings.  The correspondence thus established will be accurate for weak interactions. However, the field theory itself is valid even for strong interactions; here our correspondence gives a qualitative picture for the interpretation of the now phenomenological field theory couplings in terms of microscopics, a picture we support by strong interaction analysis and comparision to numerical results near a high-symmetry line. The FTI precursor will be found in this strongly interacting regime, including the proximity of the high-symmetry line; it develops due to interleg fermion tunnelling processes dressed by density fluctuations. 

To disentangle the effect of the different interaction terms, we first focus on vanishing interspin interactions, as in the simplest, decoupled-spin, FTI picture.
In this regime, the system will be shown to 
support fractionalised gapless modes that we will be able to identify as the precursors of helical FTI edge modes. We shall then investigate the role of interspin 
interactions. Provided these are not too strong, the helical FTI precursor edge modes will be shown to retain their integrity.

The FTI precursor competes with several density-fluctuation and backscattering processes. For the regimes of interactions where these dominate, they induce various forms of 
quasi-long-range-orders (QLROs) which we shall characterize in detail in Sec.~\ref{subsec:Vd0}. The phase diagrams including the FTI and these competing phases are shown in 
Figs.~\ref{fig:phases_spindecoupled} and \ref{fig:interspin_ops}. 

The results highlighted thus far build on taking the density-fluctuation-dressed tunneling term responsible for the FTI precursor as a weak perturbation and investigating 
(through a perturbative renormalization group procedure) the regimes where this term may dominate the low-energy physics. To assess the consistency of the thus predicted FTI 
precursor, we also investigate the corresponding low-energy picture: here the FTI term is at strong coupling, i.e., we have a fully developed FTI precursor phase. Its 
consistency amounts to it being robust to all symmetry-allowed perturbations, including the strong-coupling remnants of the density-fluctuation and backscattering 
processes mentioned above. 
We find this to be the case for perturbations related to same-spin interactions. The lack of separation between bulk and the edge in our ladder system, however, leads to some limitations when considering interspin processes: the most important of these
generates correlated backscattering of FTI quasiparticles between the (precursors of) opposite edges which can gap the corresponding helical edge modes. 
This observation allows us identify the energy window in which the FTI-precursor physics dominates: the energy scales of interest should be larger than the backscattering induced gap $m_{\rm low}$ but smaller than the (precursor of) the bulk gap $m_{\rm FTI}$. We provide a detailed discussion on this in Sec.\ref{sec:strong_coupling}.

\begin{figure}[ht!]
	\centering
		\includegraphics[width=1\linewidth]{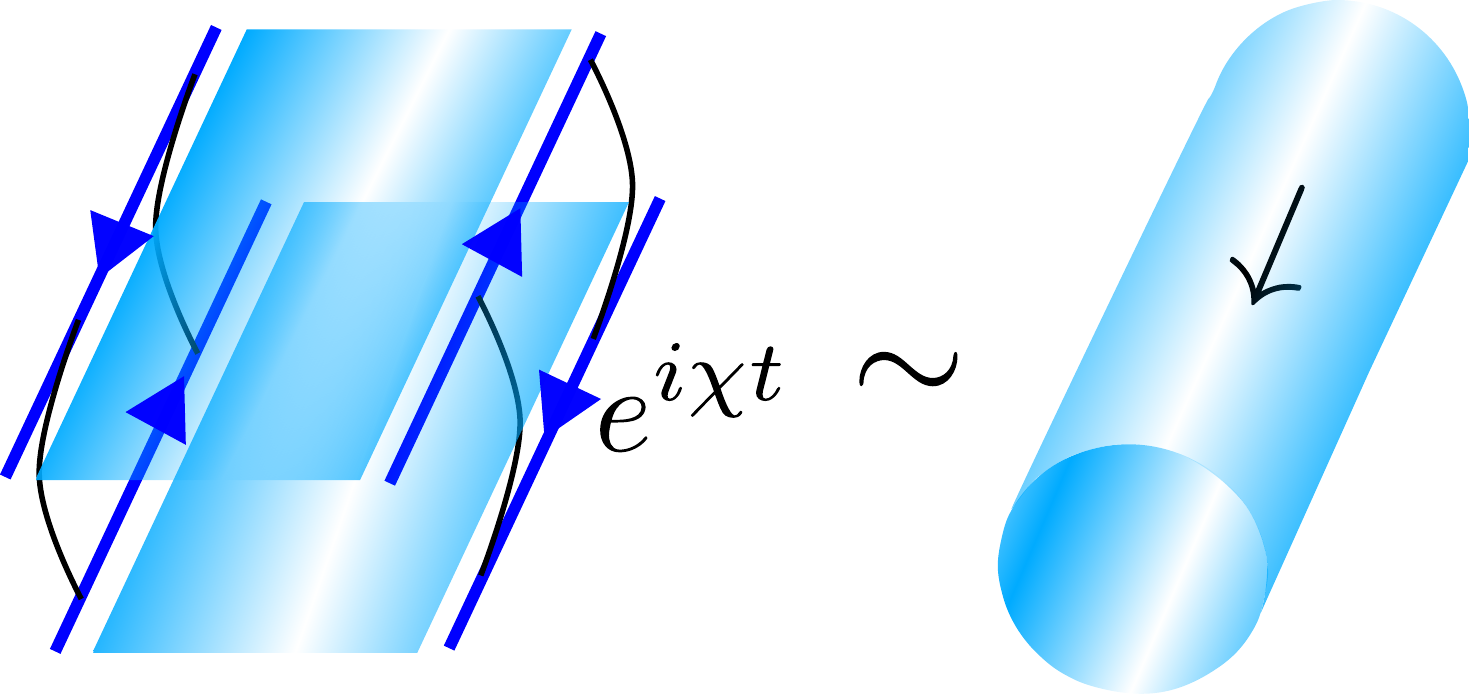}
	\caption{(Color online) Left: The bulk of the system (represented in light-blue) and the helical edge modes (blue arrows) in the FTI precursor phase. 
	Adding TR breaking processes (black lines) that connect TR partners, it is possible to gap the edge modes.
	In this case the system is similar to a FQH state on a cylinder. 
	Right: A suitable rotation of the TR breaking term along one edge effectively pierces flux through the cylinder (in the direction of the black arrow); this 
	pumps charge along the system. A protocol that rotates from a starting configuration to its TR conjugate pumps one sixth of a particle.}
	\label{fig:cylinder}
\end{figure}

One of our most striking finding is that this phase, despite being only a quasi-1D precursor of an FTI state, can give rise to fractionally quantized signatures typically expected only in the 2D regime. The signature we predict is based on the response of the FTI precursor to explicit TR symmetry breaking, in spirit similar to the idea in Ref.~\cite{Beri2012} for 2D systems. The protocol we describe involves gapping TR partner precursor edge-states by a suitable TR-breaking perturbation effectively implementing a Zeeman field. (The characteristic energy scale $m_{\text{TR}}$ of this is taken to be $m_\text{low}<m_\text{TR}<m_\text{FTI}$.) This effectively renders the system into an FQH state on a cylinder (Fig. \ref{fig:cylinder}). Then, a 
suitable local rotation of the Zeeman field has the same effect as piercing flux through the cylinder. According to Laughlin's argument \cite{Laughlin1981},
this flux insertion pumps charge along the system.
We find that the rotation of the Zeeman term between two TR-conjugate values pumps one-sixth of a particle (or of the electron charge in an electronic system), providing a 
fractionally quantised signature of the precursor state. We emphasize that this TR-conjugate protocol corresponds \emph{one-half} of a conventional (i.e., with identical initial and final Hamiltonian) adiabatic pumping cycle; this differentiates the signature we find from fractional charge pumping over a \emph{full} cycle in other systems, including those without (precursor) topological order \cite{Taddia2017}.   

Our findings indicate that a spinful fermionic ladder with SO coupling represents the minimal setup where the physics of the strongly interacting FTI 
phase appears. Nevertheless, due to the quasi-1D nature of this system, we find that it displays some characteristics which are not expected in a 
truly 2D system. These characteristics appear due to the absence of a separation of length scales between the edge and the bulk.
This is underscored by the existence of QLRO in the FTI phase of bond-density wave type, together with the presence of local 
perturbations that induce quasiparticle backscattering between the edges. In contrast, a 2D FTI 
phase does not have QLRO, and quasiparticle backscattering between its edges is suppressed by the presence of the bulk.
The FTI precursors in our ladder systems however can be viewed as the basic building blocks from which to construct such genuinely 2D phases. To illustrate this, we investigate the transition between the quasi-1D ladder and a 2D system by coupling many ladders side by side. In this way we show that the processes that compete with the development of true topological order are exponentially suppressed in the number of ladders (i.e., separation between the edges), suggestive of the FTI phase being stabilized in the 2D limit. 

Our results on FTI precursors, their quantized signatures, and their relation to 2D lattice systems exemplify a potentially much broader paradigm in which to investigate the interplay of topological phases and symmetries. We expect our approach centred on ladder-based precursors to generalize to a variety of novel 2D topological phases where symmetry plays an important role.

The paper is organised as follows. We start by introducing the microscopic model and its symmetries in Sec.~\ref{sec:SOladder}, 
followed by the formulation of its low energy description in Sec.~\ref{sec:lowEmodel}.  
In Sec.~\ref{sec:Weak_coupling}, we discuss the phase diagram of the model starting from microscopics in the weak interaction regime and expanding to stronger interactions 
using bosonization phenomenology. In Sec.~\ref{sec:strong_coupling}, we characterize the FTI precursor from a strong coupling perspective. To simplify our discussion, we 
mostly assume the presence of an additional inversion symmetry. Departures from this inversion symmetric point are considered in Sec.~\ref{sec:alpha}. In 
Sec.~\ref{sec:TRbreaking}, we discuss TR symmetry breaking perturbations, and propose a protocol that pumps a fractionally quantized charge per half-cycle in the FTI precursor phase.
A discussion of extending our findings towards 2D systems is given in Sec.~\ref{sec:2Dsyst}. In the last section we present our conclusions.

\section{The SO ladder}
\label{sec:SOladder}
We consider a TR symmetric ladder consisting of two one-dimensional legs
(legs labeled by $\beta=\{I,II\}=\{0,1\}$)
of spinful fermions with spin components \mbox{$\sigma=\{\uparrow,\downarrow\}=\{+1,-1\}$}. 
We assume that each leg contains $N_\text{leg}$ particles per spin
and has length $L$; there is a distance $d$ between the
legs and lattice spacing $a$ in the direction of the legs. 
Furthermore, we consider that the fermions are subject to SO coupling that 
generates flux $\pm\Phi$
per plaquette for opposite spins, and thus $\Phi_{\text{leg}}=\frac{L}{a}\frac{\Phi}{2\pi}$ flux quanta per leg and spin. 
The filling fraction $\nu$ (per spin) is given by the ratio of particle number to flux quanta, $\nu=N_\text{leg}/\Phi_\text{leg}=2\pi\frac{N_\text{leg}a}{\Phi L}$. 
Consequently the 
density of fermions per leg per spin is $\frac{N_\text{leg}}{L}=\frac{\nu\Phi}{2\pi a}$. In what follows we will be focusing on $\nu=\frac{1}{3}$. 

\begin{figure}[ht!]
	\centering
		\includegraphics[width=1\linewidth]{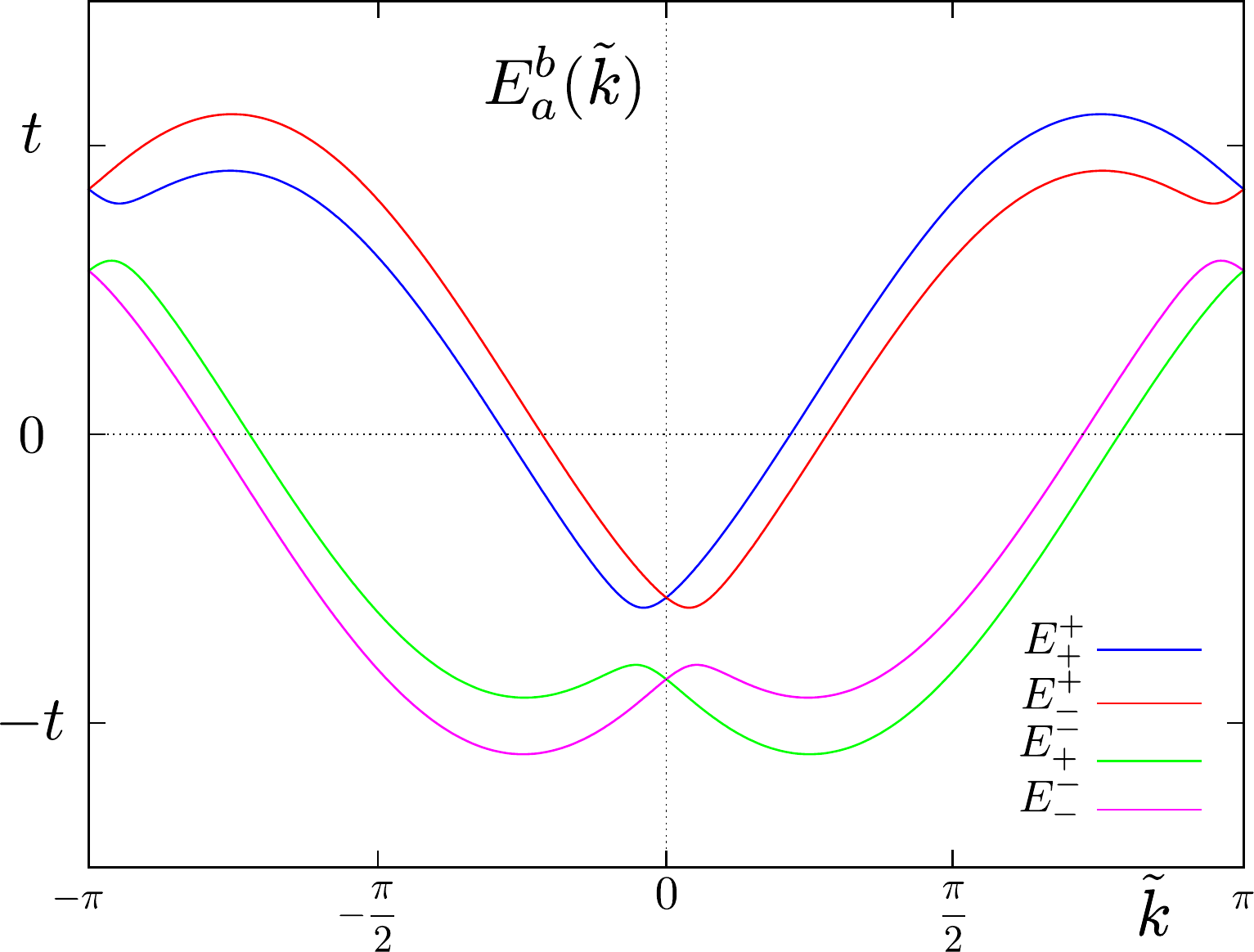}
	\caption{(Color online) Single particle spectrum for $\frac{t_{\perp}}{t}=0.1$, $\frac{\alpha_{\rm so}}{t}=0.1$, $\frac{\Phi}{2\pi}=\frac{1}{4}$,
  as a function of momentum $\tilde{k}$.}\label{fig:single_part}
\end{figure}

The single particle Hamiltonian for the system is
\begin{equation}\label{micro_quad}
 H_0=H_0^\parallel+H_0^\perp+H_{\rm so}^\perp,
\end{equation}
with (using $\bar{I}=II$ and $\bar{II}=I$)
\begin{eqnarray}
H_0^\parallel&=&-\frac{t}{2}\sum_{j,\sigma,\beta}\left[ e^{i\sigma\Phi \left(\beta-\frac{1}{2}\right)}(c^\beta_{j,\sigma})^{\dagger}c^\beta_{j+1,\sigma}+\text{h.c.}\right],\label{eq:H0par}\\
H_0^\perp&=&-t_\perp\sum_{j,\sigma,\beta} (c_{j,\sigma}^\beta)^{\dagger}c^{\bar{\beta}}_{j,\sigma},\\
H_{\rm so}^\perp&=&\alpha_{\rm so}\sum_{j,\sigma,\sigma'}\left[(i\sigma_2)_{\sigma\sigma'}(c^I_{j,\sigma})^{\dagger}c^{II}_{j,\sigma'}+\text{h.c}\right].
\end{eqnarray}
Here, $c_{j,\sigma}^\beta$ ($c_{j,\sigma}^{\beta\dagger}$) destroys (creates) fermions of spin $\sigma$ at site $j$ and leg $\beta$. 
The tunnelling amplitudes in the longitudinal (along the 1D legs) and perpendicular (across the legs) directions are $t$ and $t_{\perp}$ respectively. 
The Hamiltonian $H_0^\parallel + H_0^\perp$ is thus the fermion ladder with SO flux. These terms conserve the spin component along the quantization axis. 
With $H_{\rm so}^\perp$ we also include a SO coupling term (of strength $\alpha_{\rm so}$) that does not conserve this spin component, 
i.e. $[H_{\rm so}^\perp,S_z]\neq 0$. (The matrix $\sigma_2$ is the second Pauli matrix.)
The single particle Hamiltonian $H_0$ can be readily diagonalized in momentum space. The single particle spectrum is 
\begin{equation}\label{Sing_part_dispersion}
E_{\pm}^r(\tilde{k})=-t\cos(\tilde{k})\cos\frac{\Phi}{2}\pm\sqrt{\left(t\sin(\tilde{k})\sin\frac{\Phi}{2}+r\alpha_{\rm so}\right)^{2}+t_{\perp}^{2}},
\end{equation}
where $r=(+,-)$, $\tilde{k}=ka$ and $k$ is the momentum along the ladder. The single particle spectrum is shown in Fig.~\ref{fig:single_part}.

\subsection{Symmetries}

In the single particle Hamiltonian Eq.~\eqref{micro_quad}, the SO induced flux $\Phi$ is introduced via a (spin-dependent) vector potential aligned parallel to the legs of 
the system. This choice is convenient because allows us to 
diagonalize the noninteracting Hamiltonian easily. The physics of the system is invariant under changing $\Phi$ by a flux quantum: upon the combined change 
$\Phi\rightarrow 2\pi$ and the gauge transformation $c_{j,\sigma}^\beta\rightarrow(-1)^j c_{j,\sigma}^\beta$ the single particle Hamiltonian is unchanged. This gauge 
transformation also leaves invariant the electron densities and hence the interactions we will consider in Eq.~\eqref{interactions} below. 
Given this symmetry, together with the invariance of the physics under $\Phi\rightarrow-\Phi$, it is enough to consider flux values $\Phi\in[0,\pi]$.
The point symmetries of the model are crucial for establishing the various phases of the system. We consider two symmetries in detail: TR and inversion. 
Starting with TR, and introducing the vector 
$\mathbf{c}_{i}^{\dagger}=((c_{i,\uparrow}^{I})^{\dagger},(c_{i,\uparrow}^{II})^{\dagger},(c_{i,\downarrow}^{I})^{\dagger},(c_{i,\downarrow}^{II})^{\dagger})$
and its Fourier transform
$\mathbf{c}_{\tilde{k}}^{\dagger}=\sqrt{\frac{a}{L}}\sum_j e^{i\tilde{k}j}\mathbf{c}_{j}^{\dagger}$,  the TR transformation $\mathcal{T}$ 
acts on the operators as
\begin{equation}
 \mathcal{T}\mathbf{c}_{\tilde{k}}\mathcal{T}^{-1}=\begin{pmatrix}c_{-\tilde{k},\downarrow}^{I}\\
c_{-\tilde{k},\downarrow}^{II}\\
-c_{-\tilde{k},\uparrow}^{I}\\
-c_{-\tilde{k},\uparrow}^{II}
\end{pmatrix}=(i\sigma_{2}\otimes\openone_2)\mathbf{c}_{-k},
\end{equation}
with $\openone_2$ the two dimensional identity matrix and $A\otimes B$ the Kronecker tensor product between $A$ and $B$. In a system with time reversal symmetry, 
backscattering between Kramers pairs is forbidden. We will also consider inversion, i.e., the unitary operation that changes the momentum \mbox{$k\rightarrow -k$}.  
In the momentum basis inversion $\mathcal{I}$ acts as
\begin{equation}
 \mathcal{I}\mathbf{c}_{\tilde{k}}\mathcal{I}^{-1}=\mathbf{c}_{-\tilde{k}}.
\end{equation}

The microscopic model Eq.~\eqref{micro_quad} is invariant under
TR symmetry: keeping in mind that TR is antiunitary, one readily verifies that  $\mathcal{T}H_0\mathcal{T}^{-1}=H_0$. For vanishing $\alpha_{\rm so}$, inversion is also a 
symmetry of the system. The interactions that will be introduced below are
also assumed to keep both TR and inversion symmetry. Although TR symmetry will be considered as an exact symmetry throughout the discussion, inversion symmetry is just a 
symmetry 
of the $\alpha_{\rm so}=0$ point, and it will be explicitly broken after the Hamiltonian $H_{\rm so}^\perp$ is considered.
As a starting point, we consider $\alpha_{\rm so}=0$, and analyze the system in this limit, including interactions. The effect of $\alpha_{\rm so}\neq 0$ will be 
considered in Sec.~\ref{sec:alpha}. 

\subsection{Interactions}
\begin{figure}[ht!]
	\centering
		\includegraphics[width=0.8\linewidth]{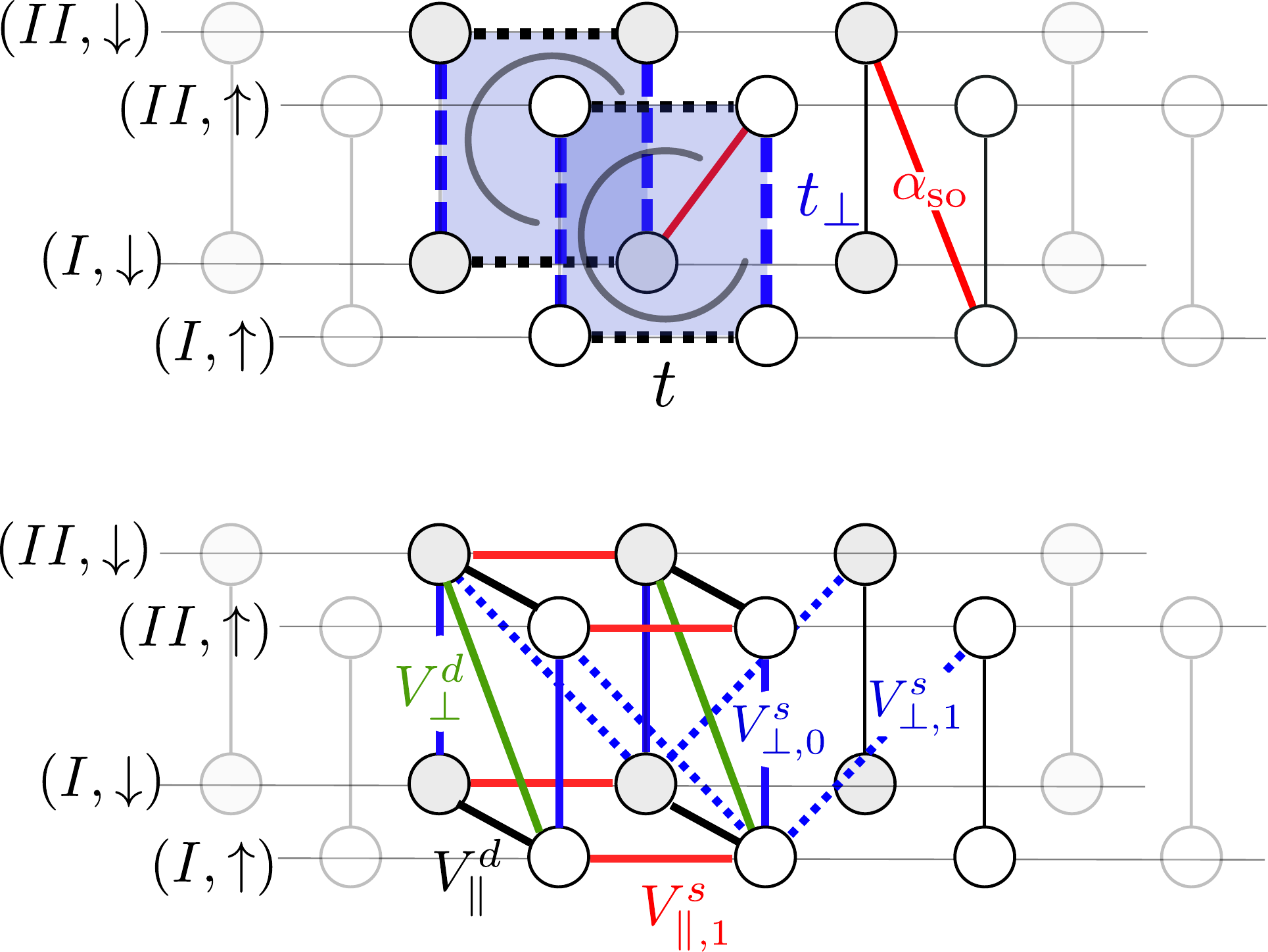}
	\caption{(Color online) Top diagram: Tunnelling parameters and SO coupling. The SO coupling that conserves the spin
	projections generates a spin dependent magnetic field. The effective magnetic flux through the plaquette is depicted by the 
	(counter)-clockwise arrow in the (down) up spin sectors. Bottom diagram: Different interaction parameters between nearest and next to nearest
	neighbors used in this work. }\label{fig:Ladder}
\end{figure}

FTIs are strongly correlated phases not adiabatically connected to a non-interacting system: their very existence hinges on the presence of interactions. To make the 
emergence of FTI phases possible in our system, we include interactions of the form
\begin{eqnarray}\label{interactions}
H_{\rm int}&=&\sum_{m,i,\sigma,\beta}V_{\parallel,m}^{s} n_{i,\sigma}^{\beta}n_{i+m,\sigma}^\beta+V_{\parallel,m}^{d} n_{i,\sigma}^{\beta}n_{i+m,\bar{\sigma}}^{\beta}\nonumber\\
&+& \sum_{m,i,\sigma,\beta} V_{\perp,m}^{s} n_{i,\sigma}^{\beta}n_{i+m,\sigma}^{\bar{\beta}}+V_{\perp,m}^{d} n_{i,\sigma}^{\beta}n_{i+m,\bar{\sigma}}^{\bar{\beta}}.
\end{eqnarray}
Here $\bar{\uparrow}=\downarrow$ and $\bar{\downarrow}=\uparrow$.
The electron densities at lattice site $i$ per leg $\beta$ and spin $\sigma$ are $n_{i,\sigma}^{\beta}=(c_{i,\sigma}^{\beta})^{\dagger}c_{i,\sigma}^{\beta}$. The letters
$s$ and $d$ in the interaction parameters refer to interaction between same or distinct spins, in the same leg ($\parallel$)
or between different legs ($\perp$). A diagram with the different interactions is presented in Fig. \ref{fig:Ladder}.
Note that for generic values of the interaction parameters $V_{\perp,m}^{s(d)}\neq V_{\parallel,m}^{s(d)}$, so the interactions are not ${\rm SU(2)}$ invariant in leg space.
This is the generic situation as the SU(2) symmetry of unitary transformations between the legs is broken already at the single particle level by SO and 
interleg tunnelling. 

For vanishing interspin interactions
$V^d_{\parallel},V^d_{\perp}$ and for  $\alpha_{\rm so}=0$, the system forms two time-reversed copies of its spinless counterpart. For a certain range of the $V^s_\parallel$, $V^s_\perp$ same-spin interactions, the system is expected to display FTI analogues of spinless quasi-one dimensional Laughlin state precursors. In 
this case, our study thus provides the FTI counterpart of earlier work on spinless states of bosons
\cite{Wei2014,Petrescu2015,Di2015,Natu2015,Piraud2015,Orignac2016,Shun2016,Greschner2016}, numerics on fermions
\cite{Sun2016,Calvanese2017}, spinful bosons \cite{Fialko2014}, and analytical approaches in strictly 1D spinful fermionic systems with a Zeeman field \cite{Oreg2014} or
systems using (leg) ${\rm SU(2)}$ invariant interactions \cite{Cornfeld2015}. We will see that, even in this spin-decoupled case, our microscopically motivated approach will allow us to reveal novel aspects, including the presence of a competing interleg gapped, density-wave type, phase. 

The inclusion of the interspin interaction terms $V^d_\parallel,V^d_\perp$ lets the two time-reversed copies interact, resulting in a genuinely spinful physics. In many 
systems, the interaction between different spin components $V^d$ is naturally present, for example due to the overall density being the main channel for interactions. The 
robustness of the FTI state to interspin interactions is therefore a key question to address. 

\section{Low energy description}
\label{sec:lowEmodel}

We are interested in exploring the SO ladder phase diagram, and in particular to study the regime where the putative fractional 
topological insulator (FTI) appears. To prepare for this analysis, we first describe our model at low energies, which will serve as a starting point for 
developing a bosonization approach.

As mentioned above, we start with  $\alpha_{\rm so}=0$. 
In this case,
TR and inversion symmetries together ensure that the bands are twofold degenerate. We focus on small fillings, so that
the Fermi level is below the avoided crossing at $k=0$, i.e. the chemical potential $\mu$ satisfies $\mu<-t\cos\frac{\Phi}{2}-t_{\perp}$ (see also Fig. \ref{fig:Dispersion_alphazero}).
The single particle Hamiltonian matrix is diagonalized by the unitary transformation
\begin{equation}\label{unitary}
\mathbb{U}=\begin{pmatrix}U_{k}\\
 & U_{-{k}}
\end{pmatrix}\text{ with}\quad U_{k}=\begin{pmatrix}\cos\alpha_{k} & -\sin\alpha_{k}\\
\sin\alpha_{k} & \cos\alpha_{k}
\end{pmatrix},
\end{equation}
where the rotation angle $\alpha_{k}$ is given by 
\begin{equation}\label{alpha_k}
\alpha_k= \begin{cases} \alpha^0_k &\mbox{if } k\in [0, \pi/a] \\
\alpha^0_k-\frac{\pi}{2} & \mbox{if } k\in [-\pi/a,0] \end{cases}, 
\end{equation}
and $\alpha^0_k=\frac{1}{2}\arctan\left(\frac{-t_\perp}{t\sin ka \sin \frac{\Phi}{2}}\right)$. The branches for the angle $\alpha_k$ in (\ref{alpha_k}) are chosen such that
the limit of zero tunnelling is recovered.

The fields corresponding to the diagonalizing basis are  $\bm{\psi}_k=\mathbb{U}^\dagger \mathbf{c}_k$, i.e.
\begin{eqnarray}\label{Diagonal_basis}
\psi_{{k},\sigma}^{+} & = & \cos\alpha_{k}c_{{k},\sigma}^{I}+\sin\alpha_{k}c_{{k},\sigma}^{II},\\
\psi_{{k},\sigma}^{-} & = & -\sin\alpha_{k}c_{{k},\sigma}^{I}+\cos\alpha_{k}c_{{k},\sigma}^{II},\\
c_{{k},\sigma}^{I} & = & -\sin\alpha_{k}\psi_{{k},\sigma}^{-}+\cos\alpha_{k}\psi_{{k},\sigma}^{+},\\
c_{{k},\sigma}^{II} & = & \cos\alpha_{k}\psi_{{k},\sigma}^{-}+\sin\alpha_{k}\psi_{{k},\sigma}^{+}.
\end{eqnarray}
As we will be working at small fillings, we will eventually project to the low energy band, corresponding to the fields $\psi_{k,\sigma}^-$ (see also Appendix 
\ref{app:Integrating_out}). We will be interested 
in the FTI phase at $1/3$ effective filling per spin. For this filling, four Fermi momenta $k_{F,\beta}^R, k_{F,\beta}^L$ exist. As 
a consequence of TR symmetry they satisfy \mbox{$k_{F,1}^{L}=-k_{F,2}^{R}$}
and \mbox{$k_{F,1}^{R}=-k_{F,2}^{L}$} (see also Fig. \ref{fig:Dispersion_alphazero}). 
For
\begin{equation}\label{small_t_perp}
 \frac{t_\perp}{t}\ll \left|\sin\left(\frac{\Phi}{2}\right)\sin\left(\frac{\Phi}{6}\right)\right|,
\end{equation}
the Fermi points are given by $k_{F,1}^L=-k_{F,2}^R=\frac{\Phi}{3a}$ and $k_{F,1}^R=-k_{F,2}^L=\frac{2\Phi}{3a}$.

\subsection{Low energy fermion branches in presence of tunnelling}\label{subsec:lowEfermion}

The presence of fermion tunnelling between the legs opens a gap at $k=0$ and $k=\pi/a$. This tunnelling also mixes the leg states into
the combinations $\psi^+_{k,\sigma}$ and $\psi^-_{k,\sigma}$. After projection, the lower band state $\psi^-_{k,\sigma}$ is the only remaining degree of freedom. 

\begin{figure}[ht!]
	\centering
		\includegraphics[width=1\linewidth]{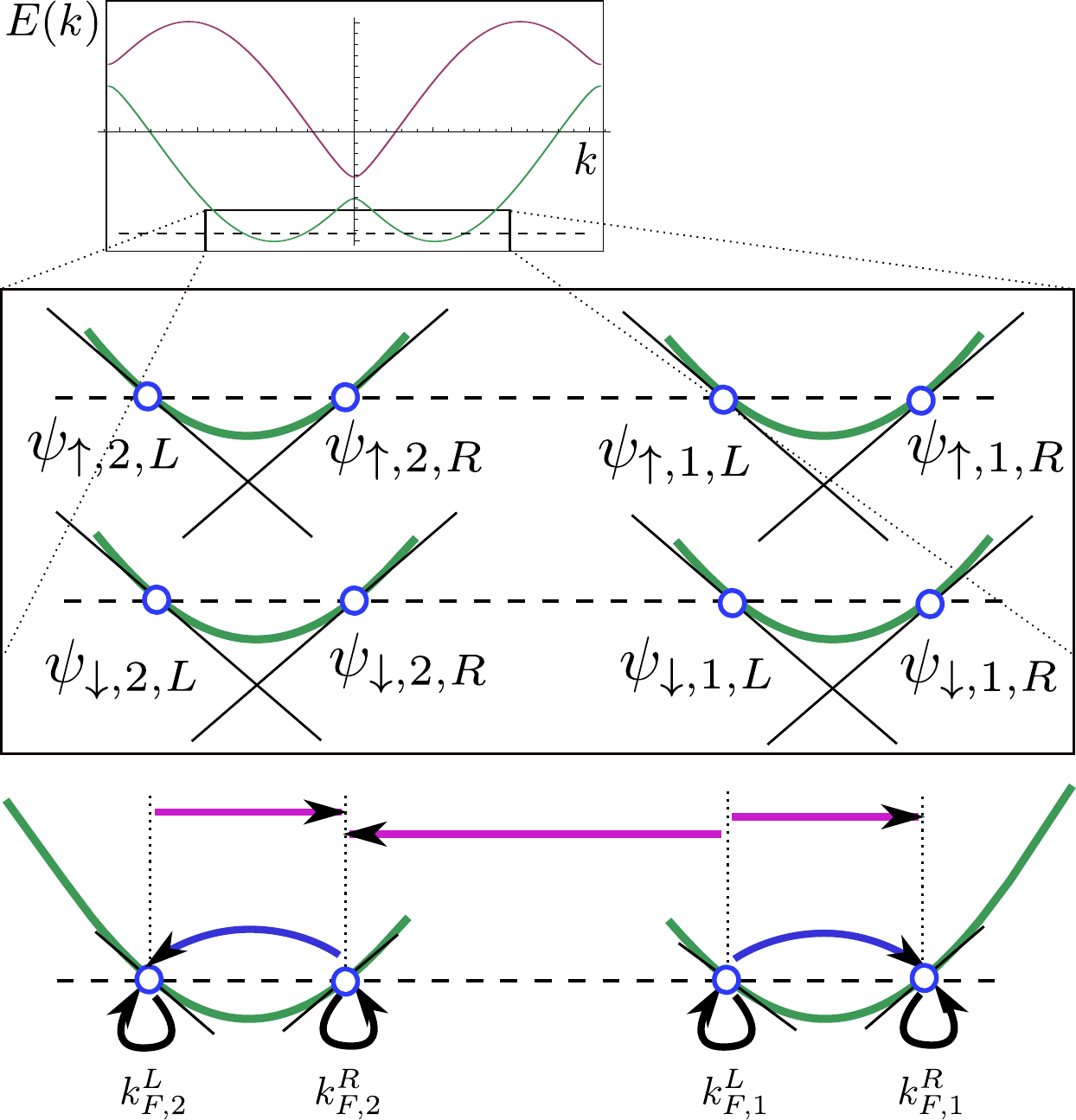}
	\caption{(Color online) Top: Band structure of the SO ladder for $\alpha_{\rm so}=0$. 
	Middle: Labelling of the different low energy fermion branches, in terms of their spin $\sigma=(\uparrow,\downarrow)$,
	their valley $\beta=1,2$ and their chirality $\eta=L,R$. We have separated the spin branches for clarity. The UV extension of the model corresponds to replacing the original spectrum
	with a linearised spectrum around the Fermi points.
	Bottom: Dispersion relation of fermions around the Fermi energy (dashed horizontal). For small filling fraction, around the 
	Fermi energy the lower band has four Fermi points per spin depicted by empty blue circles in the diagram above, together with the corresponding valleys $\beta$. An 
	arrow from $k_{F,\alpha}^{\eta}$ to $k_{F,\beta}^{\eta'}$ represents the process $\psi_{\sigma,\beta,\eta'}^{\dagger}\psi_{\sigma,\alpha,\eta}$.
	Momentum conservation requires that interaction generated processes are composed of arrows of zero sum.
	The most relevant processes in the sense of a weak coupling renormalization group are correlated backscattering between each minima (blue), 
	and Laughlin backscattering (purple). The black arrow-loops indicate forward scattering processes. }\label{fig:Dispersion_alphazero}
\end{figure}

In addition to the band projection, in our low energy description we will be focusing on the physics in a small energy window near the Fermi energy.  Taking this window  such that the corresponding window of momenta is much smaller than the scale on which $\alpha_k^0$ changes allows us to bring the rotation matrix $\mathbb{U}$ of Eq. (\ref{unitary}) out of the Fourier transform, which simplifies going to the real space representation. We find, 
taking into account the contribution of the fields $\psi_{k,\sigma}^-$ 
just near the Fermi points, that the original operators $c^\beta_{i,\sigma}$ can 
be expressed as
 \begin{eqnarray}\label{eq:fermion-branchdecomp}
 \begin{pmatrix}c^I_{i,\uparrow} \\
 c^{II}_{i,\uparrow}
\end{pmatrix}&=&\sum_\eta\begin{pmatrix} \cos\alpha^\eta_2 & - \sin\alpha^\eta_1\\
 \sin\alpha^\eta_2 &  \cos\alpha^\eta_1
\end{pmatrix}\begin{pmatrix} \psi_{i,\uparrow,2,\eta}\\
\psi_{i,\uparrow,1,\eta}
\end{pmatrix},\\
 \begin{pmatrix}c^I_{i,\downarrow} \\
 c^{II}_{i,\downarrow}
\end{pmatrix}&=&\sum_\eta\begin{pmatrix} \cos\alpha^\eta_1 & -\sin\alpha^\eta_2\\
 \sin\alpha^\eta_1 &  \cos\alpha^\eta_2
\end{pmatrix}\begin{pmatrix} \psi_{i,\downarrow,1,\eta}\\
\psi_{i,\downarrow,2,\eta}
\end{pmatrix},
\end{eqnarray}
with the angle $\alpha^\eta_\beta\equiv\alpha^0_{k^\eta_{F,\beta}}$. Here $\eta=(L,R)=(-,+)$ denotes the left and right branches around $k_F$ while 
$\beta=1,2$ the valley
around which the different branches appear (see also Fig. \ref{fig:Dispersion_alphazero}). 
Note that for weak interleg tunnelling there is an approximate correspondence between the  valley and leg indices; this is however the opposite for opposite spins due to these experiencing the opposite $\Phi$. Note also that as far as matrix elements between states involving excitations near the Fermi points are concerned, we may promote $\psi_{\sigma,\beta,\eta}$ 
to describing separate branches of excitations, e.g., with linear dispersions tangent to the low energy band at the Fermi points, as will be convenient for our subsequent 
bosonization \cite{vonDelft1998,Haldane1981}.

In terms of the branch decomposition \eqref{eq:fermion-branchdecomp}, the fermion densities become 
\begin{eqnarray}\label{densities_lat}
n_{i,\sigma}^I   & = & \sum_{\beta\beta',\eta\eta'}u^{\eta}_{\sigma\beta} u^{\eta'}_{\sigma\beta'}\psi_{i,\sigma,\beta,\eta}^{\dagger}\psi_{i,\sigma,\beta',\eta'},\\
n_{i,\sigma}^{II} & = & \sum_{\beta\beta',\eta\eta'}v^{\eta}_{\sigma\beta} v^{\eta'}_{\sigma\beta'}\psi_{i,\sigma,\beta,\eta}^{\dagger}\psi_{i,\sigma,\beta',\eta'},
\end{eqnarray}
where the tensors $u,v$ are
\begin{eqnarray}\nonumber
\begin{pmatrix}u^\eta_{\uparrow 2}\\u^\eta_{\uparrow 1}
                                \end{pmatrix}\equiv\begin{pmatrix}
                                 \cos\alpha_2^\eta\\
                                 -\sin\alpha^\eta_1
                                \end{pmatrix},\quad
                                \begin{pmatrix}v^\eta_{\uparrow2}\\v^\eta_{\uparrow1}
                                \end{pmatrix}\equiv\begin{pmatrix}
                                 \sin\alpha_2^\eta\\
                                 \cos\alpha^\eta_1
                                \end{pmatrix},\\
\begin{pmatrix}u^\eta_{\downarrow 1}\\u^\eta_{\downarrow 2}
                                \end{pmatrix}\equiv\begin{pmatrix}
                                 \cos\alpha_1^\eta\\
                                 -\sin\alpha^\eta_2
                                \end{pmatrix},\quad
                                \begin{pmatrix}v^\eta_{\downarrow 1}\\v^\eta_{\downarrow 2}
                                \end{pmatrix}\equiv\begin{pmatrix}
                                 \sin\alpha_1^\eta\\
                                 \cos\alpha^\eta_2
                                \end{pmatrix}.
\label{eq:uvtensor}\end{eqnarray}

Focusing on a small energy window around the Fermi points, also allows us to describe such branches using a continuum formulation of the lattice operators. 
We use the replacements (with $x=ja$)
\begin{equation}
 \sum_j\rightarrow \frac{1}{a}\int dx,\quad \frac{\psi_{j,\sigma,\alpha,\eta}}{\sqrt{a}}\rightarrow\psi_{\sigma,\alpha,\eta}(x).
\end{equation}

\subsection{Low energy Hamiltonian}
\label{subsec:lowEHam} 
 
In terms of the low energy continuum description, the kinetic energy of the fermions is given by
\begin{eqnarray}\label{kinetic_fermions}
H_{0} & = & i\sum_{\sigma,\alpha,\eta}\int dx\eta v_{F}\psi_{\sigma,\alpha,\eta}^{\dagger}(\partial_{x}-ik_{F,\alpha}^{\eta})\psi_{\sigma,\alpha,\eta},
\end{eqnarray}
where the Fermi velocity $v_F=ta|\sin(\Phi/6)|$ is assumed to be the same around all the Fermi points, which is valid for $(t_\perp/t)\ll 1$.
We use this assumption for convenience (e.g., near $t_\perp=0$ connections to standard Luttinger parameters of bosonization will become available), but it is not crucial as long as the interleg tunnelling satisfies $t_\perp/t<\frac{\sin^2(\Phi/2)}{\cos(\Phi/2)}$. For larger
values of $t_\perp$ the states around zero lattice momentum become minimum in energy, invalidating the analysis as two Fermi points disappear.

Next we turn to describing the low energy form of the interactions. Our explicit mapping between the microscopic interactions Eq. (\ref{interactions}) and our low energy model will be in terms of second order perturbation theory in the interaction strength to bandwidth ratio. This implies that this mapping is accurate only where the microscopic interactions are much smaller than the bandwidth. 
The physics beyond this regime will be accessible to us via phenomenological (but symmetry restricted) parameters of bosonization (see Sec.~\ref{sec:forwardscatt_Lutt}). Away from weak interactions, we will relate these parameters to microscopics for weak $t_\perp$, using a combination of symmetry based considerations, comparisons to numerics \cite{Sano94,Nakamura99,Nakamura00,Tsuchiizu04,Tsuchiizu04,Sandvik04,Ejima07}, and studying the interaction dominated limit (App.~\ref{app:KrhoKbeta}). 

First order perturbation theory involves matrix elements of the microscopic interactions between low energy states. These matrix elements are well 
captured using the branch decomposition Eq.~\eqref{eq:fermion-branchdecomp}. The interaction Hamiltonian (\ref{interactions}) in the continuum limit becomes
\begin{eqnarray}\nonumber
H_{\rm int}&= &\int dx dr\sum_{\rm all\, labels}A_{\beta\beta'\gamma\gamma'}^{\eta\eta'\tilde{\eta}\tilde{\eta}',\sigma\sigma'}(r)\psi_{\sigma,\beta,\eta}^{\dagger}(x)\psi_{\sigma,\beta',\eta'}(x)\\
&\times&\psi_{\sigma',\gamma,\tilde{\eta}}^{\dagger}(x+r)\psi_{\sigma',\gamma',\tilde{\eta}'}(x+r),
\end{eqnarray}
where the coefficients are 
\begin{multline}\label{eq:lowEintcoeffs}
A_{{\beta}{\beta'}{\gamma}{\gamma'}}^{\eta\eta'\tilde{\eta}\tilde{\eta}',\sigma\sigma'}  = 
V_{\parallel}^{\sigma\sigma'}(u_{\sigma\beta}^{\eta} u_{\sigma\beta'}^{\eta'} u_{\sigma'\gamma}^{\tilde{\eta}}
u_{\sigma'\gamma'}^{\tilde{\eta}'}+v_{\sigma\beta}^{\eta} v_{\sigma\beta'}^{\eta'} v_{\sigma'\gamma}^{\tilde{\eta}} v_{\sigma'\gamma'}^{\tilde{\eta}'})\\
+   V_{\perp}^{\sigma\sigma'}(u_{\sigma\beta}^{\eta} u_{\sigma\beta'}^{\eta'} v_{\sigma'\gamma}^{\tilde{\eta}} v_{\sigma'\gamma'}^{\tilde{\eta}'}+
v_{\sigma\beta}^{\eta} v_{\sigma\beta'}^{\eta'} u_{\sigma'\gamma}^{\tilde{\eta}} u_{\sigma'\gamma'}^{\tilde{\eta}'}),
\end{multline}
with the couplings $V^{\uparrow\uparrow}=V^{\downarrow\downarrow}=V^s$ and $V^{\uparrow\downarrow}=V^{\downarrow\uparrow}=V^d$ being those of the microscopic interactions 
Eq.~\eqref{interactions}. In terms of Eq.~\eqref{interactions}, for the purposes of explicit expressions,
we will consider same-spin interactions up to next nearest neighbor range
[$V_{\parallel,m>1}^s=V_{\perp,m>1}^s=0$] and interspin interactions to be on-site and on-rung \mbox{[$V_{\parallel,m\neq0}^d=V_{\perp,m\neq0}^d=0$].} 
We use the notation $V_{\parallel,1}^s\equiv V_\parallel^s$ and $V^d_{\parallel(\perp),0}=V^d_{\parallel(\perp)}$.
We will mostly focus on the following
concrete case for same spin interactions: $V_{\perp,0}^s=V_{\perp,1}^s\equiv V_\perp^s$ providing access to a tunable breaking of the SU($2$) invariance of same spin 
interactions in leg space. Generally, such moderate range interactions result in effectively local 
interactions in the continuum limit with the couplings in Eq.~\eqref{eq:lowEintcoeffs} essentially being replaced by a delta functions 
$V(r)\rightarrow \tilde{V}(r)=(aV)\delta(r)$ with strengths set by the Fourier components of $V(r)$ at zero momentum and the various Fermi momenta differences. 
The results of this continuum procedure for the concrete interactions described above, and expressed in terms of the subsequent bosonization parameters, are given in Appendix~\ref{app:Defs}.

In second order perturbation theory, we find corrections to all the previously discussed terms. These corrections have their origin in the
transitions to intermediate virtual states outside of the low energy window.
Apart from modifying the interaction coefficients $A_{{\beta}{\beta'}{\gamma}{\gamma'}}^{\eta\eta'\tilde{\eta}\tilde{\eta}',\sigma\sigma'}$ by second order terms, three particle (and higher) processes are generated in the low energy Hamiltonian.
Of these \mbox{$H_{\rm proj}=\int dx g_{\rm FTI} \mathcal{B}^\sigma_{\rm FTI}+\text{h.c}$}, with
\begin{equation}\label{FTI_psi}
 \mathcal{B}_{\rm FTI}^{\sigma}=\psi_{\sigma,1,R}^{\dagger}\psi_{\sigma,2,L}\psi_{\sigma,2,R}^{\dagger}\psi_{\sigma,2,L}\psi_{\sigma,1,R}^{\dagger}\psi_{\sigma,1,L},
\end{equation}
will be seen to be responsible for the FTI (precursor) physics, see also Appendix \ref{app:Integrating_out}. 
This term corresponds to tunnelling of fermions between the legs of the ladder, dressed by modulations of the density in each leg. Under a change of
basis, this term becomes a normal tunnelling process of dressed fermions. The condensation of these dressed fermions, for each spin projection, 
leaves behind fractionalised edge modes and a gapped bulk. The natural quasiparticles in this system satisfy anyonic statistics, signalling that the
system develops FTI order. This transition is explored in detail in Sec.\ref{sec:strong_coupling}.
The second order terms (\ref{FTI_psi}) involve an energy denominator of the order 
of the bandwidth $~v_F/a$, hence generically they come with a coefficient of order $aV^2/v_F$, which in turn translates to couplings of order $a^3V^2/v_F$ in the local, 
long-wavelength formulation.  In particular
the coupling constant of the FTI operator $\mathcal{B}_{\rm FTI}^{\sigma}$ is given by
\begin{equation}
 g_{\rm FTI}=\frac{a}{v_F}\left[c_1aV^s_\parallel+c_2aV_{\perp,0}^s+c_3 aV^s_{\perp,1}\right]^2.
\end{equation}
Here $c_i$ are flux dependent parameters of order one (see Appendix \ref{app:pref_signs} for more details). Furthermore, for weak interleg tunnelling
$t_\perp\ll t$, we have $g_{\rm FTI}\propto t_\perp$.    

\subsection{Bosonization}
\label{sec:bosonization}
We now turn to expressing
the fields in terms of bosonization. In a long-wavelength description near the Fermi energy, 
we have eight fields (labelled by spin $\sigma=(\uparrow,\downarrow)$, valley $\alpha=(1,2)$, and chirality $\eta=(L,R)=(-,+)$). In bosonization these are given 
by \cite{GogolinBook2004}

\begin{eqnarray}\label{bosonization}
\psi_{\sigma,\alpha,\eta}(x)&=&\frac{\kappa_{\sigma,\alpha}^\eta}{\sqrt{2\pi a}}e^{i\eta\sqrt{4\pi}\phi_{\sigma,\alpha,\eta}(x)}e^{ik_{F,\alpha}^{\eta}x},
\label{eq:bosoexp}\end{eqnarray}
where $\kappa_{\sigma,\alpha}^\eta$ is a Klein factor
ensuring the anticommutation of different fermions (see Appendix \ref{app:Klein}). The commutation relations of the bosonic fields
are 
\begin{equation}
 [\phi_{\sigma,\alpha,\eta}(x),\phi_{\tilde{\sigma},\beta,\tilde{\eta}}(x')]=\frac{i}{4}(\sigma_{3})_{\eta\tilde{\eta}}\delta_{\sigma\tilde{\sigma}}\delta_{\alpha\beta}\text{sgn}(x-x'),
\end{equation}
with $\sigma_3$ the diagonal Pauli matrix. In our notation, the bosonized form of the small wavevector component of the particle (``charge") density operator reads 
\mbox{$ \rho_{\sigma,\alpha,\eta}(x)=-\frac{1}{\sqrt{\pi}}\partial_x\phi_{\sigma,\alpha,\eta}.$ }

Note that in contrast to existing descriptions of Laughlin ladders \cite{Petrescu2015,Cornfeld2015,Petrescu2017}, we do not include higher harmonics in the bosonization 
formula Eq.~\eqref{bosonization}. Instead, as mentioned above, we assume that focusing on the physics sufficiently close to the Fermi momenta, one may replace the 
dispersion with unbounded branches extending the low energy window, in which case Eq.~\eqref{eq:bosoexp} becomes exact \cite{Haldane1981,vonDelft1998}. 
The contributions that would be generated phenomenologically by harmonics are obtained using perturbation theory; this approach helps us keeping the microscopic origin of various terms transparent, and highlight, even beyond the weak interaction limit, the presence of interrelations between various bosonization parameters.

The kinetic energy \eqref{kinetic_fermions} in terms of our low energy bosonic description is given by
\begin{equation}
H_{0} =\sum_{\sigma,\alpha}\int dxv_{F}[(\partial_{x}\phi_{\sigma,\alpha,R})^{2}+(\partial_{x}\phi_{\sigma,\alpha,L})^{2}].
\end{equation}

Introducing the vector of bosonic fields $\bm{\phi}=(\bm{\phi}_\uparrow,\bm{\phi}_\downarrow)$ where
\begin{equation}\label{vector_basis}
\bm{\phi}^T_\sigma=(\phi_{\sigma,1,L},\phi_{\sigma,1,R},\phi_{\sigma,2,L},\phi_{\sigma,2,R}),
 \end{equation}
we find that TR symmetry in this basis acts as
\begin{eqnarray}
\mathcal{T}\bm{\phi}\mathcal{T}^{-1}=-[\sigma_{1}\otimes\mathbb{\sigma}_{1}\otimes\sigma_{1}]\bm{\phi}+\bm{t},
\end{eqnarray}
with $\sigma_1$ the first Pauli matrix and $\bm{t}$ the constant vector $\bm{t}^T=(0,0,0,0,\pi,\pi,\pi,\pi)$. 
Note that as TR flips momentum, it acts non-trivially on the valley index. 

Inversion symmetry, on the other hand, acts as
\begin{eqnarray}
\mathcal{I}\bm{\phi}\mathcal{I}^{-1}=-[\openone_{2}\otimes\mathbb{\sigma}_{1}\otimes\sigma_{1}]\bm{\phi}.
\end{eqnarray}

\subsection{Forward scattering terms and Luttinger parameters}
\label{sec:forwardscatt_Lutt}

The different processes induced by the interactions become either quadratic terms in the bosonic representation, 
or cosine nonlinear operators in the boson fields. 
The quadratic bosonic Hamiltonian, together with the quadratic terms from the kinetic energy, are encoded in the forward scattering matrix $\mathcal{M}$.
This matrix determines the Luttinger parameters of the system \cite{Haldane1981,GiamarchiBook2003,GogolinBook2004}
and the fate of the nonlinear operators under scale renormalization. We will be comparing the relevance of the different
operators under the renormalization group (RG) to determine the different phases of the system.

The quadratic sector of the Hamiltonian is
\begin{equation}\label{Gaussian}
 H_{\rm{fwd}}=\int dx \partial_x\bm\phi^T\mathcal{M}\partial_x\bm\phi,
\end{equation}
with the symmetric $\mathcal{M} =\openone_{2}\otimes\left(v_{F}\openone_{4}+\frac{1}{4\pi}\mathbb{V}\right)+\sigma_{1}\otimes\frac{1}{4\pi}\mathbb{W}.$
The $4\times 4$ symmetric matrices $\mathbb{V},\mathbb{W}$ have the form
\begin{equation}\label{int_matrix_same_spin}
\mathbb{V}=\begin{pmatrix}
f_{22} & f_{12} & g_{12} & g_{22} \\
f_{12} & f_{11} & g_{11} & g_{12}  \\
g_{12} & g_{11} & f_{11} & f_{12}  \\
g_{22} & g_{12} & f_{12} & f_{22}
\end{pmatrix},\quad
\mathbb{W} =\begin{pmatrix}
h_{22} & h_{12} & \tilde{h}_{12} & \tilde{h}_{22}\\
h_{12} & h_{11} & \tilde{h}_{11} & \tilde{h}_{12}\\
\tilde{h}_{12} & \tilde{h}_{11} & h_{11} & h_{12}\\
\tilde{h}_{22} & \tilde{h}_{12} & h_{12} & h_{22}
\end{pmatrix}.
\end{equation}
The forward interaction matrices  $\mathbb{V},\mathbb{W}$  have the the most general structure allowed by time reversal and inversion symmetry. Note that a $4\times 4$ real 
symmetric matrix is specified in general by 10 parameters. The symmetries impose relations between them, leaving just six independent parameters. By the unitary 
transformation $S=\frac{\sigma_3\otimes\openone_2+\sigma_1\otimes\sigma_1}{\sqrt{2}}$, the matrices $\mathbb{V}$ and  $\mathbb{W}$ can be put in a block diagonal form, 
composed of two $2\times 2$ symmetric matrices. These sub-matrices are independent, corresponding to the 6 different parameters. 
The relationship between the microscopic parameters and $f_{mn},g_{mn},h_{mn},\tilde{h}_{mn}$ for weak interactions is given in Appendix \ref{app:Defs}.

We will also make use
of a number of simplifications that arise for $t_\perp\ll t$. As established explicitly in App.~\ref{app:Defs} for weak interactions, the first small $t_\perp$ correction 
to the forward scattering parameters is of order $(t_\perp/t)^2$. This, however, should be a generic feature valid also for strong interactions, because forward scattering 
conserves the number of particles in a given leg while $O(t_\perp)$ processes involve a single interleg tunnelling event. Working to linear order in $t_\perp$ (where the FTI 
term is already operative), the forward scattering part can thus be taken at $t_\perp=0$. In this case, the theory has an additional reflection symmetry in each leg 
separately, which ensures that $f_{11}=f_{22}\equiv f$, $g_{11}=g_{22}\equiv g$, $h_{11}=h_{22}\equiv h$ and $\tilde{h}_{11}=\tilde{h}_{22}\equiv \tilde{h}$. 
This allows one to discuss the physics, including the qualitative behavior away from weak interactions, in terms of simple Luttinger parameters summarized in 
App.~\ref{app:Defs} and Eqs. (\ref{Krho}-\ref{Kbeta}), below.
In particular, for $h=\tilde{h}=0$ we recover the familiar charge Luttinger parameter $K_\rho$ and the leg analogue, $K_\beta$, of the spin Luttinger parameter, given by
\begin{equation}
K_\rho=\sqrt{\frac{v_F+\frac{f-g-f_{12}+g_{12}}{4\pi}}{v_F+\frac{f+g+f_{12}+g_{12}}{4\pi}}},\,\,
K_\beta=\sqrt{\frac{v_F+\frac{f+g-f_{12}-g_{12}}{4\pi}}{v_F+\frac{f-g+f_{12}-g_{12}}{4\pi}}}.
\end{equation}

Note that the quadratic Hamiltonian defines a quadratic action, which is invariant under scale transformations, i.e. an action that does not change if we change 
$(x,t)$ to $(x',t')=\lambda (x,t)$. Under a scale transformation this is a fixed point. In the next section we will see how the inclusion of non-quadratic
terms changes this picture.

\subsection{Interaction operators and scaling dimensions}
\label{sec:IO_Delta}

The presence of interactions also generates cosine terms in the bosonic description. The only terms that may affect the low energy description
are the ones allowed by momentum conservation. The momentum non-conserving terms acquire an oscillation with wavelength 
$1/k_F$, which averages out the operators at large distances. The remaining terms appear in the Hamiltonian as
\begin{equation}
 H_{\rm int}=\sum_i\int dx  \bar{g}_i\mathcal{O}_i(x),
\end{equation}
where the operators $\mathcal{O}_i$ correspond either to same-spin interactions or opposite spin interactions. The strength of interactions is $\bar{g}_i$.
The different $\mathcal{O}_i$ terms are generally combinations of  exponentials in the bosonic fields. The Klein factors that appear from bosonization are not dynamical and
can be dealt appropriately, as shown in the Appendix \ref{app:Klein}.
These exponential terms, viewed as perturbations to the Gaussian Hamiltonian (\ref{Gaussian}), induce an RG flow of the parameters after integrating out short distance degrees of freedom, as the system is not scale invariant anymore. 
As we are interested in a low energy, long-wavelength theory, we will analyze how the system changes as we approach the physics of longer and longer lengthscales. 
To first order in the couplings $\bar{g}_i$, the RG flow can be determined by the behavior of the action under a scale transformation
$(x,t)\rightarrow (x',t')=\lambda (x,t)$, ($\lambda$ is usually parameterized as $\lambda=e^\ell$) \cite{CardyBook,DiFrancescoBook}.
The action transforms as
\begin{equation}
 S=S_{\rm quad}-\sum_i\int dt dx  \bar{g}_i\lambda^{2-\Delta_i}\mathcal{O}_i,
\end{equation}
where we have characterized the change of the operators $\mathcal{O}_i$ by their scaling dimensions $\Delta_i$. In particular, using the parameterization $\lambda=e^\ell$, we find 
that the coupling constants $\bar{g}_i$ satisfy the RG equation
\begin{equation}\label{RG_eq}
 \frac{d\bar{g}_i}{d\ell}=(2-\Delta_i)\bar{g}_i.
\end{equation}
We see that if
$\Delta_i<2$, the coupling constant of the operator $\mathcal{O}_i$ 
grows larger under scale  transformation, which renders it relevant to the physics at low energies, large wavelengths. Operators of this type are dubbed relevant in the RG
sense. On the other hand, operators whose scaling dimension is larger than $2$, are dubbed irrelevant in RG sense. For a given operator $\mathcal{O}_i$ the value of the 
scaling dimension $\Delta_i$ is set by the quadratic term $S_\text{quad}$.  To first order in $\bar{g}_i$, this term does not change under the RG transformation, and hence 
neither do the scaling dimensions \cite{GogolinBook2004,GiamarchiBook2003,CardyBook}. 

Such first order RG equations
are sufficient where $\bar{g}_i$ are sufficiently weak and $\Delta_i$ is sufficiently away from $2$. It may, however, happen that these two conditions are not independent, 
e.g., for weak interactions if $\bar{g}_i$, similarly to the forward scattering parameters, is first order in interactions and cannot be suppressed, e.g., by small 
$t_\perp$. In this case, the RG has to be taken at second order, 
\begin{equation}\label{RG_eq_2ndorder}
 \frac{d\bar{g}_i}{d\ell}=(2-\Delta_i)\bar{g}_i-\sum_{jk} C_{ijk}\bar{g}_j\bar{g}_k, 
\end{equation}
where $g_i$ now include not only exponentials
but also corrections to the forward scattering terms, which thus also flow. The coefficient matrix $C_{ijk}$ is set by
the behaviour of products of operators under a short distance expansion, i.e., the operator product expansion. 

We now focus on the scaling dimensions of the different $\mathcal{O}_i$ operators, which are determined by the quadratic part of the action
\begin{equation}\label{action}
 S_{\rm quad}=\int dtdx \left(\partial_t\bm{\phi}^T\mathcal{K}\partial_x\bm{\phi}-\partial_x\bm{\phi}^T\mathcal{M}\partial_x\bm{\phi}\right),
\end{equation}
where the symmetric matrix $\mathcal{K}$ encodes the commutation relations of the fields \cite{Haldane1995}. In the basis of chiral fields (\ref{vector_basis}),
it corresponds to $\mathcal{K}=-\openone_N\otimes\sigma_3$ for $N$ right and $N$ left movers, with $\openone_N$ the $N\times N$ identity matrix 
and $\sigma_3$ is a left-right mover grading.
Given the action (\ref{action}), the operator $\mathcal{O}_{\bm{\eta}}=e^{i\bm{\eta}^T\bm{\phi}}$ has scaling dimension $\Delta_{\bm{\eta}}=\bm{\eta}^T \Lambda \bm{\eta}$, where  $\Lambda$ is
the matrix \cite{Haldane1995}
\begin{equation}\label{Lambda}
\Lambda=\frac{1}{8\pi}\mathcal{M}^{-\frac{1}{2}}|\mathcal{M}^{\frac{1}{2}}\mathcal{K}^{-1}\mathcal{M}^{\frac{1}{2}}|\mathcal{M}^{-\frac{1}{2}},
\end{equation}
with $|B|\equiv\sqrt{B^\dagger B}$ the absolute value of the matrix $B$ (see Appendix \ref{app:Scaling} for a derivation of this result).
Parameterizing the forward scattering matrix as
$\mathcal{M}=v_F\openone+\frac{1}{4\pi}\mathcal{V}$ we can expand the scaling dimension matrix to first order in $\mathcal{V}/v_F$. We obtain for $N$ legs 
\begin{equation}\label{scaling_pert}
 \Lambda=\frac{1}{8\pi}\left(\openone_{2N}+\frac{1}{8\pi v_F}[\mathcal{K},\mathcal V]\mathcal{K}\right)+{O}(\mathcal{V}^2).
\end{equation}
Below we summarize the four-fermion and six-fermion processes arising in our problem and calculate their scaling dimensions. (Though eight-fermion processes also 
arise from second order perturbation theory, we do not detail these as they are expected to be less relevant under RG than four- and six-fermion terms.)
While explicit links between these and microscopics we can obtain only for weak interactions, the expressions we provide for the scaling dimensions will be in terms of the 
general forward scattering parameters $f$, $g$, $h$, $\tilde{h}$ and in terms of Eq.~\eqref{Lambda}, thus being valid beyond the weakly interacting regime.

\subsubsection{Four fermion processes}\label{fourfermionscaling}

It is useful to define the slow modes around the Fermi points
as $R_{\sigma\alpha}=e^{i\sqrt{4\pi}\phi_{\sigma,\alpha,R}}$ and $L_{\sigma,\alpha}=e^{-i\sqrt{4\pi}\phi_{\sigma,\alpha,L}}$.
To first order in the interaction couplings, the interaction terms proportional to $V^s_{\parallel, \perp}$ between alike spins result in one cosine term, 
\begin{equation}
 \mathcal{O}^{s}_{1\sigma}= (R^\dagger_{\sigma,1}L_{\sigma,1}L^\dagger_{\sigma,2}R_{\sigma,2}+\text{h.c.}). 
\end{equation}
Including also the corrections to the same process from second order perturbation theory, $\mathcal{O}^s_{1\sigma}$ has coupling constant 
with $\bar{g}_{1\sigma}^s=a\frac{c^s_{1}V^s_\parallel+c^s_{2}V^s_{\perp,0}+c^s_3 V^s_{\perp,1}}{(2\pi a)^2}+O(V^2)$.
The coefficients $c^s_{i}$ are functions of the flux and of order one (see also Appendix~\ref{app:pref_signs} for a detailed discussion of this prefactor).

\begin{figure}[t]
	\centering
		\includegraphics[width=1\linewidth]{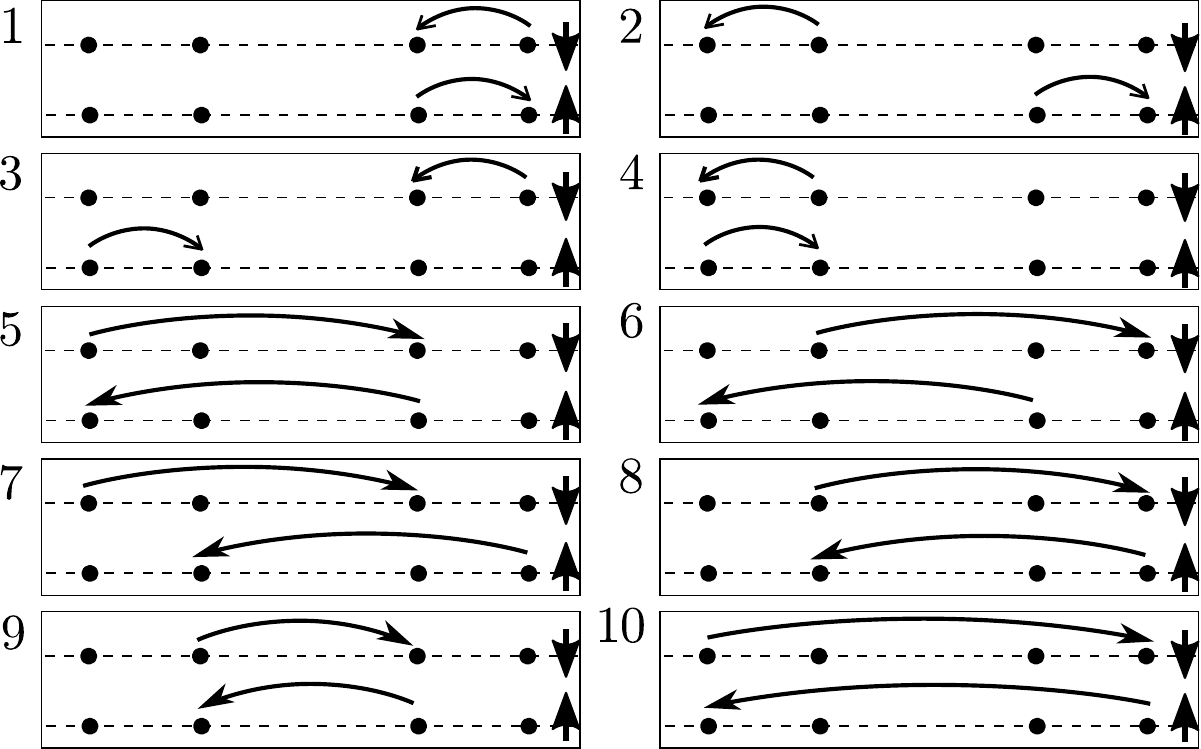}
	\caption{Four fermion processes in the presence of interspin interaction. The four Fermi points are denoted by black dots around the Fermi energy (dashed line).
	Scattering processes between Fermi points are denoted by long black arrows. Short arrows indicate the spin of the band. The numbering of the processes is the same 
	as in Eqs. 
	(\ref{int_interspin}-\ref{int_interspin_final}).}\label{fig:process_spin}
\end{figure}

The interaction terms that appear in the case of nonvanishing interactions between different spin components are
\begin{eqnarray}\label{int_interspin}
\mathcal{O}^{d}_1=\mathcal{O}_{1,\uparrow\downarrow} &=& (R_{\uparrow 1}^{\dagger}L_{\uparrow 1}L_{\downarrow 1}^{\dagger}R_{\downarrow 1}+\text{h.c.}),\\
\mathcal{O}^{d}_2=\mathcal{O}_{2,\uparrow\downarrow} &=& (R_{\uparrow 1}^{\dagger}L_{\uparrow 1}L_{\downarrow 2}^{\dagger}R_{\downarrow 2}+\text{h.c.}),\\
\mathcal{O}_3^{d}=\mathcal{O}_{5,\uparrow\downarrow} &=& (L_{\uparrow 2}^{\dagger}L_{\uparrow 1}L_{\downarrow 1}^{\dagger}L_{\downarrow 2}+\text{h.c.}),\\
\mathcal{O}_4^{d}=\mathcal{O}_{6,\uparrow\downarrow} &=& (L_{\uparrow 2}^{\dagger}L_{\uparrow 1}R_{\downarrow 1}^{\dagger}R_{\downarrow 2}+\text{h.c.}),\\
\mathcal{O}_5^{d}=\mathcal{O}_{9,\uparrow\downarrow} &=& (R_{\uparrow 2}^{\dagger}L_{\uparrow 1}L_{\downarrow 1}^{\dagger}R_{\downarrow 2}+\text{h.c.}),\\
\mathcal{O}_6^{d}=\mathcal{O}_{10,\uparrow\downarrow}&=& (L_{\uparrow 2}^{\dagger}R_{\uparrow 1}R_{\downarrow 1}^{\dagger}L_{\downarrow 2}+\text{h.c.}),
\end{eqnarray}
and
\begin{eqnarray}
\mathcal{O}_{3,\uparrow\downarrow} &=& (L_{\downarrow 1}^{\dagger}R_{\downarrow 1}R_{\uparrow 2}^{\dagger}L_{\uparrow 2}+\text{h.c.}),\\
\mathcal{O}_{4,\uparrow\downarrow} &=& (R_{\uparrow 2}^{\dagger}L_{\uparrow 2}L_{\downarrow 2}^{\dagger}R_{\downarrow 2}+\text{h.c.}),\\
\mathcal{O}_{7,\uparrow\downarrow} &=& (R_{\uparrow 2}^{\dagger}R_{\uparrow 1}L_{\downarrow 1}^{\dagger}L_{\downarrow 2}+\text{h.c.}),\\
\mathcal{O}_{8,\uparrow\downarrow} &=& (R_{\uparrow 2}^{\dagger}R_{\uparrow 1}R_{\downarrow 1}^{\dagger}R_{\downarrow 2}+\text{h.c.}).\label{int_interspin_final}
\end{eqnarray}
The coefficients of these interactions are
\begin{equation}\label{eq:1st_order}
\bar{g}^{\uparrow\downarrow}_i=a\frac{c^d_{1,i}V^d_\parallel+c^d_{2,i}V^d_\perp}{(2\pi a)^2}+O(V^2),
\end{equation}
where the coefficients $c_{1,i}^d,c_{2,i}^d$  are functions of the flux of order one $O(1)$ (see Appendix \ref{app:pref_signs} for more details). 
These processes are illustrated in Fig.\ref{fig:process_spin}. Note that the processes $\mathcal{O}_{i\geq5,\uparrow\downarrow}$ involve interleg tunnelling for each spin, hence for weak $t_\perp$ one has $\bar{g}^{\uparrow\downarrow}_{i\geq5}\sim (t_\perp/t) ^2$.
Under TR symmetry, the operators above satisfy the relations $\mathcal{O}_{1\sigma}^{s\mathcal{T}}\equiv\mathcal{T}\mathcal{O}_{1\sigma}^s\mathcal{T}^{-1}=\mathcal{O}_{1,\bar{\sigma}}$
together with $\mathcal{O}_{1,\uparrow\downarrow}^\mathcal{T}=\mathcal{O}_{4,\uparrow\downarrow}$ and $\mathcal{O}_{5,\uparrow\downarrow}^\mathcal{T}=\mathcal{O}_{8,\uparrow\downarrow}$. 
Under inversion, $\mathcal{O}_{2,\uparrow\downarrow}^\mathcal{I}=\mathcal{O}_{3,\uparrow\downarrow}$, $\mathcal{O}_{6,\uparrow\downarrow}^\mathcal{I}=\mathcal{O}_{7,\uparrow\downarrow}$.
Given these symmetries, in the analysis of scaling dimensions we just consider the subset of operators $\{\mathcal{O}^{s}_{1\sigma},\mathcal{O}^{d}_{i}\}$ with $i=1\dots 6$, as the remaining
operators have the same scaling dimension as the operators to which they are related by symmetry.
The scaling dimensions of $\{\mathcal{O}^{s}_{1\sigma},\mathcal{O}^{d}_{5},\mathcal{O}^{d}_{6}\}$ are
\begin{eqnarray}
\Delta_{1\sigma}^s&=&\frac{1}{2}\sum_{r=+,-}K_{34}^{r}+K_{43}^{r}+\sin(\zeta_2^r)(K_{43}^{r}-K_{34}^{r}),\\
 \Delta_5^d&=&K_{12}^{-}+K_{21}^{-}+\cos(\zeta_1^-)(K_{12}^{-}-K_{21}^{-}),\\
 \Delta_6^d&=&K_{12}^{-}+K_{21}^{-}+\cos(\zeta_1^-)(K_{21}^{-}-K_{12}^{-}).
\end{eqnarray}
The Luttinger parameters $K_{rs}^\pm$ and angles $\zeta_{1,2}^\pm$ are defined in Appendix \ref{app:Defs}.
Note that $\Delta_{1\uparrow}^s=\Delta_{1\downarrow}^s\equiv\Delta_{1}^s$ does not depend on the spin.
The scaling dimensions of $\{\mathcal{O}^d_i\}$, $i=1..4$ can be written compactly by defining
\begin{eqnarray}
  \Delta_{ab}&\equiv&\frac{K_{12}^{-}+K_{21}^{-}+K_{34}^{a}+K_{43}^{a}}{2}\\\nonumber
	   &+&b\sin(\zeta_1^-)\frac{K_{21}^{-}-K_{12}^{-}}{2} +\sin(\zeta_2^a)\frac{K_{43}^{a}-K_{34}^{a}}{2},
\end{eqnarray}
with $a,b=+,-$, such that $\Delta^d_1=\Delta_{-+}$, $\Delta^d_2=\Delta_{++}$, $\Delta^d_3=\Delta_{--}$ and $\Delta^d_4=\Delta_{+-}$.

The scaling dimensions of the different operators satisfy the following relations
\begin{eqnarray}\label{relation_deltas1}
 \Delta_1^s&=&\Delta_1^d+\Delta_4^d-\frac{1}{2}(\Delta_5^d+\Delta_6^d),\\
 \Delta_3^d&=&\Delta_1^d-\Delta_2^d+\Delta_4^d.\label{relation_deltas2}
\end{eqnarray}

We find that $\Delta_3^d\geq 2$ implying that the operator $\mathcal{O}_{3}^d=e^{i\bm{\eta}_{3}^d\bm{\phi}}$ 
(with $\bm{\eta}_3^d=\sqrt{4\pi}(-1,0,1,0,1,0,-1,0)$) is never relevant in the RG sense. This process does not contain backscattering terms. That $\Delta_3^d\geq 2$ can be seen considering
\begin{eqnarray}\nonumber
 \Delta_3^d&=&(\bm{\eta}_3^d)^T\Lambda\bm{\eta}_3^d=\frac{1}{4\pi}(\bm{\eta}_3^d)^T\mathcal{M}^{-\frac{1}{2}}|\mathcal{M}^{\frac{1}{2}}\mathcal{K}^{-1}\mathcal{M}^{\frac{1}{2}}|\mathcal{M}^{-\frac{1}{2}}\bm{\eta}_3^d\\
          &\geq& \frac{1}{4\pi}(\bm{\eta}_3^d)^T(-\mathcal{K}^{-1})\bm{\eta}_3^d=2, 
\end{eqnarray}
where we have used that $v^T|A|v\geq v^TAv$, as a simple consequence of the properties of the absolute value.

Given that $\Delta_3^d\geq 2$ for all interactions, it follows from (\ref{relation_deltas1},\ref{relation_deltas2}) that
\begin{equation}
 \Delta_1^s+\frac{1}{2}(\Delta_5^d+\Delta_6^d)\geq 2+\Delta_2^d,
\end{equation}
which implies that the operators $\mathcal{O}_{1\sigma}^s,\mathcal{O}_{5}^d$ and $\mathcal{O}_{6}^d$ cannot be all relevant in the same region 
of parameters, unless $\mathcal{O}_2^d$ is also relevant. An expanded discussion of the different relations between the scaling dimensions is given in Appendix \ref{app:Defs}.

\subsubsection{Six fermion processes}\label{sec2:six_fermion}

At second order perturbation theory in the interactions, in addition to the correction to the forward scattering parameters and the four fermion couplings mentioned 
above,  two sets of six fermion operators appear involving same or different spin components. We discuss first the cosine terms that appear from the interaction of alike spins. Among these, the terms
\begin{eqnarray}\nonumber
R_{\sigma,2}^{\dagger}L_{\sigma,2}(R_{\sigma,2}^{\dagger}R_{\sigma,2})L_{\sigma,1}^{\dagger}R_{\sigma,1}+\text{h.c.},\\\nonumber
(R_{\sigma,1}^{\dagger}R_{\sigma,1})L_{\sigma,1}^{\dagger}R_{\sigma,1}R_{\sigma,2}^{\dagger}L_{\sigma,2}+\text{h.c.},\\
L_{\sigma,2}^{\dagger}L_{\sigma,1}R_{\sigma,1}^{\dagger}(L_{\sigma,1}L_{\sigma,1}^{\dagger})R_{\sigma,2}+\text{h.c.},\\\nonumber
(L_{\sigma,2}^{\dagger}L_{\sigma,2})R_{\sigma,1}^{\dagger}L_{\sigma,1}R_{\sigma,2}^{\dagger}L_{\sigma,2}+\text{h.c.},
\end{eqnarray}
are all related to the first order operator $\mathcal{O}^s_{1\sigma}$ by the insertion of a density operator at some of the Fermi 
points. The scaling dimension of such operators is then $\Delta_{1}^s+1$. This means that these type of operators are never 
more relevant than $\mathcal{O}^s_{1\sigma}$, so we do not consider them further.

The operators (one per spin projection) that in principle can open a gap leading to a FTI (precursor) are
\begin{equation}
 \mathcal{O}_{\rm FTI}^{\sigma}=(R_{\sigma,1}^{\dagger}L_{\sigma,2}R_{\sigma,2}^{\dagger}L_{\sigma,2}R_{\sigma,1}^{\dagger}L_{\sigma,1}+\text{h.c.}),
\end{equation}
which we introduced earlier in (\ref{FTI_psi}) in terms of the fermionic degrees of freedom $\psi$. A detailed discussion of how the relevance of this 
operator induces the FTI precursor phase, with fractional edge modes and gapped bulk anyonic quasiparticles is presented in Sec.\ref{sec:strong_coupling}.
The scaling dimension 
of these operators is independent of the spin $\sigma$ and given by
\begin{eqnarray}\nonumber
 \Delta_{\rm FTI}&=&\frac{5}{4}\sum_{r=+,-}K_{12}^{r}+K_{21}^{r}+\cos(\zeta_1^r-\epsilon)(K_{21}^{r}-K_{12}^{r}),\\
&\mbox{with}&\quad \tan\epsilon=4/3.
 \end{eqnarray}
The bare coupling for $\mathcal{O}_{\rm FTI}^{\sigma}$ is
\begin{equation}
 \bar{g}_{\rm FTI}=\frac{g_{\rm FTI}}{(2\pi a)^3}=\frac{c_1(V^s_\parallel)^2+c_2V^s_\parallel V_\perp^s+c_3 (V^s_\perp)^2}{(2\pi)^3v_F},
\label{eq:gFTI}
 \end{equation}
 with $c_i\sim O(1)$ being functions of the flux (and $\propto t_\perp$ if  interleg tunnelling is weak).  The full expression can be seen in Appendix \ref{app:pref_signs}.

The last pair of operators of a single spin species are
\begin{eqnarray}
\mathcal{O}_2^s&=&(R_{\sigma,1}^{\dagger}L_{\sigma,2}R_{\sigma,2}^{\dagger}L_{\sigma,2}R_{\sigma,2}^{\dagger}L_{\sigma,2}+\text{h.c.}),\\
\mathcal{O}_3^s&=&(L_{\sigma,1}^{\dagger}R_{\sigma,1}L_{\sigma,1}^{\dagger}R_{\sigma,1}L_{\sigma,2}^{\dagger}R_{\sigma,1}+\text{h.c.}),
\end{eqnarray}
with scaling dimension $\Delta_2^s=\Delta_3^s=\Delta_1^s+\Delta_{\rm FTI}$. As these are both less relevant than $\mathcal{O}^\sigma_{\rm FTI}$, we do not consider them further. 

Similar to the case of interactions between alike spins considered above, some of the second order terms containing both species are just density insertions on top of the first order terms. 
All these terms have scaling dimension larger by one than the corresponding first order process to which they are related.  This implies that they are always less relevant 
in RG sense than they first order counterparts. For this reason we do not consider them.
The processes that are not just insertions of density operators on the lowest order terms are 
\begin{eqnarray}\label{second_ord_int}
\mathcal{O}^I_{11,\uparrow\downarrow}&=&(R_{\uparrow,1}^{\dagger}L_{\uparrow,1}R_{\uparrow,1}^{\dagger}L_{\uparrow,2}R_{\downarrow,1}^{\dagger}L_{\downarrow,1}+ \text{h.c.}),\\
\mathcal{O}^I_{12,\uparrow\downarrow}&=&(R_{\uparrow,1}^{\dagger}L_{\uparrow,2}R_{\uparrow,2}^{\dagger}L_{\uparrow,2}R_{\downarrow,2}^{\dagger}L_{\downarrow,2}+ \text{h.c.}),\\
\mathcal{O}^I_{13,\uparrow\downarrow}&=&(R_{\downarrow,1}^{\dagger}L_{\downarrow,1}R_{\downarrow,1}^{\dagger}L_{\downarrow,2}R_{\uparrow,1}^{\dagger}L_{\uparrow,1}+ \text{h.c.}),\\
\mathcal{O}^I_{14,\uparrow\downarrow}&=&(R_{\downarrow,1}^{\dagger}L_{\downarrow,2}R_{\downarrow,2}^{\dagger}L_{\downarrow,2}R_{\uparrow,2}^{\dagger}L_{\uparrow,2}+ \text{h.c.}),\\
\mathcal{O}^{II}_{15,\uparrow\downarrow}&=&(R_{\uparrow,1}^{\dagger}L_{\uparrow,1}R_{\uparrow,1}^{\dagger}L_{\uparrow,2}R_{\downarrow,2}^{\dagger}L_{\downarrow,2}+ \text{h.c.}),\\
\mathcal{O}^{II}_{16,\uparrow\downarrow}&=&(R_{\uparrow,1}^{\dagger}L_{\uparrow,2}R_{\uparrow,2}^{\dagger}L_{\uparrow,2}R_{\downarrow,1}^{\dagger}L_{\downarrow,1}+ \text{h.c.}),\\
\mathcal{O}^{II}_{17,\uparrow\downarrow}&=&(R_{\downarrow,1}^{\dagger}L_{\downarrow,2}R_{\downarrow,2}^{\dagger}L_{\downarrow,2}R_{\uparrow,1}^{\dagger}L_{\uparrow,1}+ \text{h.c.}),\\
\mathcal{O}^{II}_{18,\uparrow\downarrow}&=&(R_{\downarrow,1}^{\dagger}L_{\downarrow,1}R_{\downarrow,1}^{\dagger}L_{\downarrow,2}R_{\uparrow,2}^{\dagger}L_{\uparrow,2}+ \text{h.c.}).\label{second_ord_int_fin}
\end{eqnarray}
These operators are related by TR symmetry as 
\begin{eqnarray}\nonumber
(\mathcal{O}^I_{11,\uparrow\downarrow})^\mathcal{T}=\mathcal{O}^I_{14,\uparrow\downarrow},\quad
(\mathcal{O}^I_{12,\uparrow\downarrow})^\mathcal{T}=\mathcal{O}^I_{13,\uparrow\downarrow},\\
(\mathcal{O}^{II}_{15,\uparrow\downarrow})^\mathcal{T}=\mathcal{O}^{II}_{17,\uparrow\downarrow},\quad 
(\mathcal{O}^{II}_{16,\uparrow\downarrow})^\mathcal{T}=\mathcal{O}^{II}_{18,\uparrow\downarrow}. 
\end{eqnarray}
Inversion symmetry, on the other hand, relates them as
\begin{eqnarray}\nonumber
(\mathcal{O}_{11,\uparrow\downarrow}^I)^\mathcal{I}=\mathcal{O}^{I}_{12,\uparrow\downarrow},\quad 
(\mathcal{O}_{15,\uparrow\downarrow}^I)^\mathcal{I}=\mathcal{O}^{I}_{16,\uparrow\downarrow},\\
(\mathcal{O}_{13,\uparrow\downarrow}^{II})^\mathcal{I}=\mathcal{O}^{II}_{14,\uparrow\downarrow},\quad 
(\mathcal{O}_{17,\uparrow\downarrow}^{II})^\mathcal{I}=\mathcal{O}^{II}_{18,\uparrow\downarrow}. 
\end{eqnarray}
These symmetries split the operators above into two families $\{I,II\}$ in terms of scaling dimensions. The scaling dimension of the operators in each family are
\begin{eqnarray}\label{delta_sec_ord_spin}
 \Delta_I&=&\Delta_{\rm FTI}+\Delta_1^s+\frac{\Delta_3^d}{2}-\frac{1}{2}(3\Delta_1^d+\Delta_6^d),\\\label{delta_sec_ord_spin2}
\Delta_{II}&=&\Delta_I+\Delta_4^d-\Delta_3^d.
\end{eqnarray}

\subsubsection*{Scaling dimensions for $t_\perp=0$ forward scattering}
The above scaling dimensions dramatically simplify in the case when the forward scattering parameters can be taken at $t_\perp=0$. We find
\begin{eqnarray}\nonumber
\Delta_1^s&=&K_{\beta,-}+K_{\beta,+}, \quad \Delta_1^d=K_{\rho,-}+K_{\beta,-},\\\nonumber
\Delta_2^d&=&K_{\rho,-}+K_{\beta,+}, \quad \Delta_3^d=K_{\beta,-}+K_{\beta,-}^{-1},\\\nonumber
\Delta_4^d&=&K_{\beta,-}^{-1}+K_{\beta,+},\quad  \Delta_5^d=\Delta_6^d=K_{\rho,-}+K_{\beta,-}^{-1},\\ 
\Delta_{\rm FTI}&=&\frac{9}{4}(K_{\rho,+}+K_{\rho,-})+\frac{1}{4}(K_{\beta,+}^{-1}+K_{\beta,-}^{-1}),
\end{eqnarray}
where the Luttinger parameters are
\begin{align}\label{Krho}
 K_{\rho,\pm}= \sqrt{\frac{1+\frac{f-g-f_{12}+g_{12}\pm(h-\tilde{h}-h_{12}+\tilde{h}_{12})}{4\pi v_F}}{1+\frac{f+g+f_{12}+g_{12}\pm(h+\tilde{h}+h_{12}+\tilde{h}_{12})}{4\pi v_F}}},\\\label{Kbeta}
 K_{\beta,\pm}=\sqrt{\frac{1+\frac{f+g-f_{12}-g_{12}\pm(h+\tilde{h}-h_{12}-\tilde{h}_{12})}{4\pi v_F}}{1+\frac{f-g+f_{12}-g_{12}\pm(h-\tilde{h}+h_{12}-\tilde{h}_{12})}{4\pi v_F}}}.
\end{align}

\section{Weak coupling phase diagram}\label{sec:Weak_coupling}
In this section, we discuss the different phases of the SO ladder at 1/3 effective filling, as seen from a weak coupling perspective. We emphasize that by weak coupling we mean small $\bar{g}_i$, but not necessarily that the interactions are weak (e.g., our discussion of the FTI precursor phase will be for \emph{strong} interactions, but weak $t_\perp$). In this approach, the phases are 
determined by which operators are the most relevant in the RG sense. For most operators, we restrict ourselves to an analysis first order in $\bar{g}_i$  where RG 
(ir)relevancy is determined by the scaling dimensions [see Eq.~\eqref{RG_eq}]. As noted at Eq.~\eqref{RG_eq_2ndorder}, this works when $\bar{g}_i$ 
is small while $\Delta_i$ is sufficiently away from $2$. For weak $ \bar{g}_{1\sigma}^s$ and a nearly leg-SU($2$) invariant system, we have $\Delta_1^s\approx 2$ which necessitates a second order RG 
treatment for $\mathcal{O}_{1\sigma}^s$. [We also note that for certain six-fermion terms, given that their $\bar{g}_i$ is already second order in interactions when those are weak, one may worry that first order RG may not suffice if the $(2-\Delta_i)\bar{g}_i$ and $\bar{g}_j\bar{g}_k$ terms in Eq.~\eqref{RG_eq_2ndorder} give comparable contributions due to $\bar{g}_j$ and $\bar{g}_k$ being 
first order in interactions. However, we did not find that such scenario would occur.]

In what follows we will first start from weak interactions and combine first order RG for $\bar{g}_i\neq \bar{g}_{1\sigma}^s$ with a second order RG approach to $\mathcal{O}_{1\sigma}^s$.
This will provide a starting point from which the qualitative landscape and competition of various phases may be discussed. 

We will find that the appearance of the FTI phase requires going beyond the weakly interacting regime. This prompts us to complement our analysis with a $K_{\rho,\beta}$  Luttinger parameter based formulation suited also for strong interactions (and in this case work 
to linear order in $t_\perp$ as discussed in Sec.~\ref{sec:forwardscatt_Lutt}). To connect our findings in this regime to microscopics, we will exploit strong interaction analysis (Appendix~\ref{app:KrhoKbeta}) and numerical results \cite{Sano94,Nakamura99,Nakamura00,Tsuchiizu04,Tsuchiizu04,Sandvik04,Ejima07} around leg-SU($2$) invariance. 

We first discuss the case of vanishing interspin interactions. In order to characterize the different phases, we will begin by introducing order parameters, which differentiate between the different QLRO.  

\subsection{Vanishing interspin interaction}
\label{subsec:Vd0}

In the context of spin-decoupled (or spinless) ladders, the possible local,
fermion bilinear order parameters include \cite{GogolinBook2004,GiamarchiBook2003} the particle number conserving order parameters
\begin{equation}\label{order_params}
 O_{\mu,\sigma,x}=\sum_{\beta,\beta'}\left(c^{\dagger \beta}_{x,\sigma} (\tau_\mu)_{\beta\beta'}c^{\beta'}_{x,\sigma}\right),
 \end{equation}
 and the superconducting order parameters	
 \begin{equation}
 S_{\mu,\sigma,x}=\left(\sum_{\beta,\beta'}\left(c^{\dagger \beta}_{x,\sigma} (i\tau_\mu\tau_2)_{\beta\beta'}c^{\dagger \beta'}_{x+a,\sigma}\right)+\text{h.c.}\right).
\end{equation}
Here $\tau_\mu$, $\mu=0,1,2,3$ is the identity matrix $(\mu=0)$ and the three Pauli matrices in the space of leg degrees of freedom of the ladder.
In terms of the low energy theory, given by the four Fermi points in the system, these order parameters have the structure (see also Appendix \ref{app:Order_params})
\begin{equation}
O_{\mu,\sigma}(x)=O_{\mu,\sigma}^0(x)+\sum_\alpha \left(e^{i\Delta k_\alpha x}O_{\mu,\sigma}^\alpha(x)+\text{h.c.}\right),
\end{equation} 
where $\alpha\equiv b,b^\prime,\eta,\eta^\prime$ in the differences $\Delta k_\alpha=k_{F,b}^\eta-k_{F,b^\prime}^{\eta^\prime}$ and
$O_{\mu,\sigma}^\alpha(x)$ are slowly varying operators. A similar expansion holds for the superconducting order parameters. (The bosonized expressions of the order parameters are given in Appendix \ref{app:Order_params}.)

The order parameter $O_{0,\sigma}$ measures the total particle density per spin and thus a nonvanishing expectation value of $O_{0,\sigma}^{\alpha\neq0}$ indicates
the presence of a charge density wave for the spin projection $\sigma$.
We denote this type of order as CDW. 
The order parameter $O_{1,\sigma}$ measures bond densities: in terms of the single particle states $|I\rangle$ and $|II\rangle$ of a given rung, $O_{1,\sigma}$ measures the density of 
$\tau_1$ eigenstates $|I\rangle\pm |II\rangle$, i.e., of bonding and anti-bonding orbitals. Hence $O_{1,\sigma}^{\alpha\neq0}$ is a 
bond density wave (BDW) order parameter. The order parameter $O_{2,\sigma}\sim c^{\dagger 1}_{x,\sigma} c^{2}_{x,\sigma}-c^{\dagger 2}_{x,\sigma} c^{1}_{x,\sigma}$ measures the 
particle current between the ladder's legs. The order parameter $O_{2,\sigma}^{\alpha\neq0}$, in turn, measures a spatially alternating current pattern.
Due to the alternating orbital moments corresponding to this, it is an orbital antiferromagnet (OAF) order parameter (other names include staggered flux or $d$-density wave order parameter). 
Finally ${O}_{3,\sigma}$ measures polarisation along the $\tau_3$ eigenstates $|I\rangle$ and $|II\rangle$, hence  $O_{3,\sigma}^{\alpha\neq0}$ is a relative density wave (RDW) order 
parameter. 

The order parameter \mbox{$S_{0,\sigma}\sim c^{\dagger 1}_{x,\sigma} c^{\dagger 2}_{x+a,\sigma}-c^{\dagger 2}_{x,\sigma} c^{\dagger 1}_{x+a,\sigma}$}
indicates the presence of orbital singlet pairing order (i.e., singlet in leg space, for a given value of $\sigma$). 
Similarly, the order parameters $S_{\mu\neq 0,\sigma}$ describe the three orbital triplet order parameters \cite{GiamarchiBook2003,Fradkin_Book}. 

It is important to note that due to the incommensurability of lattice and the particle density, umklapp terms are absent, and hence the system is not completely gapped in 
any region. Instead, we start
with a gapless theory with central charge $c=4$ (a single fermionic chain has $c=1$, so $c=2$ per spin)
and, in the presence of cosine terms, end up with either a $c=4$ or a partially gapped $c=2$ system. Some of these $c=2$ systems can be characterised in terms of the $O_{\mu,\sigma}^{\alpha\neq0}$ 
or $S_{\mu,\sigma}$ order parameters, with the gap related to the order parameter amplitude fluctuation, and the gapless sector describing fluctuations of its 
phase, i.e.,  the Goldstone mode for, e.g., spontaneous breaking of translation symmetry (of the low energy continuum theory, due to working with fillings away from lattice commensurability). For our quasi-1D quantum system, these Goldstone modes preclude the 
appearance of true long-range order, and allow at most QLRO, where certain  $O_{\mu,\sigma}^{\alpha}$ or $S_{\mu,\sigma}$
correlators decay as power laws. 
When more than one 
order parameter is QLRO, the phases may be characterized by which of these has the dominant (i.e., slowest decaying) correlation function.  A complementary 
$c=2$ case, partially gapped by $\mathcal{O}^\sigma_{\rm FTI}$, will be identified with the FTI precursor. 
 In this case, the $O_{0,\sigma}^{\alpha}$ and $O_{3,\sigma}^\alpha$ correlators decay exponentially, while the correlators $O_{1,\sigma}^\alpha$ and 
 $O_{2,\sigma}^\alpha$ develop QLRO.
 In the Luttinger liquid phase where no gap develops ($c=4$) a characterization in terms of the order parameters $O_{\mu,\sigma}^{\alpha}$ and $S_{\mu,\sigma}$
 is possible, although here \emph{both} the amplitude and the phase are QLRO only.

For vanishing interspin interactions, all the operators $\mathcal{O}^d_i$ in the Hamiltonian have vanishing coupling constant, so we concentrate on the competition between 
$\mathcal{O}_{1\sigma}^s$ and $\mathcal{O}^\sigma_{\rm FTI}$. 
Using Eq. (\ref{scaling_pert}), which serves to find the first order correction in the forward scattering parameters to the scaling dimensions for any operator, we find 
the regions
\begin{eqnarray}\label{eq:linscale_bdry1}
 0<\frac{2f_{12}-g_{11}-g_{22}}{2\pi v_F},\quad&\mathcal{O}^s_{1\sigma}&\,\,\,\text{relevant},\\\label{eq:linscale_bdry2}
 \frac{3}{2}<\frac{g_{22}+4(f_{12}+g_{11})}{8\pi v_F},\quad&\mathcal{O}_{\rm FTI}^\sigma &\,\,\,\text{relevant},
\end{eqnarray}
where relevancy is understood in terms of first order RG. While the relation for $\mathcal{O}_{1\sigma}^s$ is consistent with the $\mathcal{V}\ll v_F$ regime of scaling dimension 
linearization, the relation for $\mathcal{O}^\sigma_{\rm FTI}$ is not, which indicates that system requires strong interactions for $\mathcal{O}^\sigma_{\rm FTI}$ to 
govern the physics. 

The different phases that arise are shown in Fig. \ref{fig:phases_spindecoupled}. In the microscopics 
for the weakly interacting regime (top panel), we use $V_{\parallel,1}^s\equiv V_\parallel^s$ and $V_{\perp,0}^s=V_{\perp,1}^s\equiv V_\perp^s$ as interaction 
variables. The diagram has been calculated for $t_\perp/t=0.1$, $\Phi=\pi/3$, using a second order RG procedure, discarding $O(t_\perp^4$) and/or 
$O[(V^s_{\parallel,\perp})^3]$ terms as well as RG irrelevant terms with $\Delta_i$ well away from $2$. The details of the calculations are {given in Appendix~\ref{app:RGeqs}}. 
Depending on the character of the interactions we encounter three possible phases:
\begin{figure}
	\centering
		\includegraphics[width=0.9\linewidth]{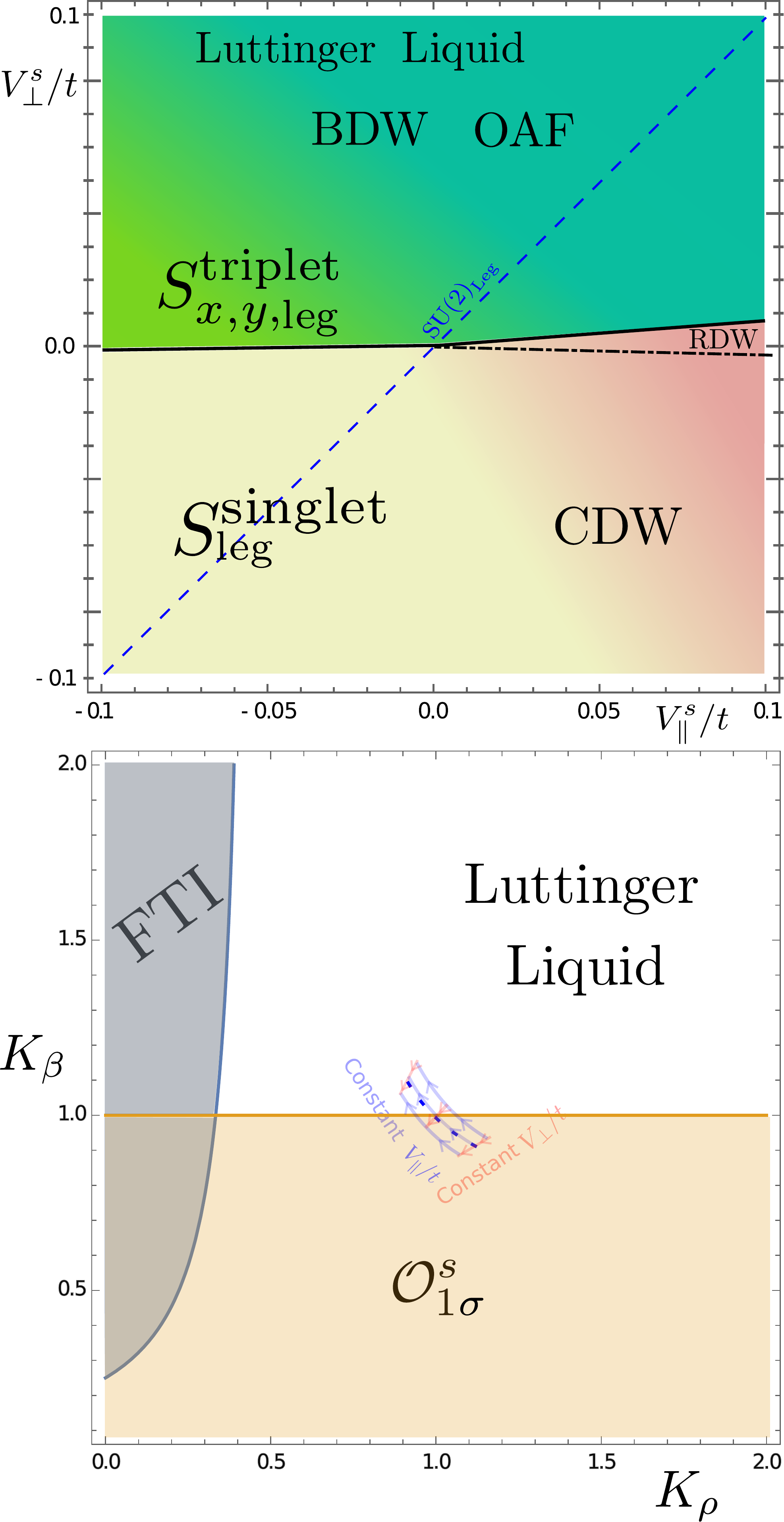} 
	\caption{(Color online) Phase diagram for vanishing interspin interactions.	
	Top panel: phase diagram for weak interactions for $t_\perp/t=0.1$ and $\Phi=\pi/3$. In this range 
	of interactions two possible phases appear (separated by solid black line), distinguished by the presence or absence of the partial $\mathcal{O}_{1\sigma}^s$ gap. 
	The dominant order parameters are also indicated, with coloring/shading illustrating the crossovers between the various cases. The blue dashed line shows $V^s_\parallel=V^s_\perp$ where interactions are SU($2$) symmetric in leg space. The dash-dotted line shows where $\bar{g}_{1\sigma}^s$ changes sign, changing the QLRO in the partially gapped regime between CDW and RDW.
	Bottom panel: Scaling dimension phase diagram in terms of Luttinger parameters $K_{\rho,\beta}$. We consider the most general quadratic term allowed by symmetries (Sec.~\ref{sec:forwardscatt_Lutt}) and work to first order in $t_\perp$.  In the FTI region (grey area) $\Delta_{\rm FTI}<2$; the $\mathcal{O}_{1\sigma}^s$ region (yellow) has $\Delta_1^s<2$. For the $c=4$ Luttinger liquid (white), $\Delta_{\rm FTI},\ \Delta_1^s > 2$. The FTI gap dominates for $K_\beta \geq 1$ in the FTI region. We also show a coordinate grid linking bare Luttinger parameters and microscopics for weak interactions $V_\parallel^s/t=-1,0,1$ and $V_\perp^s/t=-0.1,0,0.1$, as well as the leg-SU($2$) invariant line (blue dashed). Arrows are in the direction of increasing interactions.	
	}\label{fig:phases_spindecoupled}
\end{figure}

\subsubsection{$\mathcal{O}_{1\sigma}^s$ partial gap and Luttinger liquid phases}

In the weakly interacting regime, the physics is governed by the $\mathcal{O}_{1\sigma}^s$  term. The phases that arise are a $c=4$ Luttinger liquid phase (two Luttinger liquids per spin) and a $c=2$ phase characterized by a $\mathcal{O}_{1\sigma}^s$ partial gap (the leg analogue of the familiar spin gap of spinful one-dimensional fermion systems), with the phase boundary between them being a function of the flux and the interleg tunnelling. The Luttinger liquid phase exists for repulsive or moderately attractive interleg interactions, and for attractive to moderately repulsive intraleg interactions. When both interactions are repulsive ($V^s_\parallel,\ V^s_\perp \geq 0$) it is delimited by a line below the $V^s_\parallel=V^s_\perp$ line of leg-SU($2$) invariant interactions. For $t_\perp\rightarrow 0$, this boundary line approaches the SU($2$) invariant line.  In the complementary regions of weak interactions, the system displays the $\mathcal{O}_{1\sigma}^s$ partial gap. It is worthwhile to note that were one to use simple first order RG (i.e., describe the phase diagram solely using scaling dimensions), the Luttinger liquid phase would gain area at the expense of the $\mathcal{O}_{1\sigma}^s$ partial gap [the boundaries in this case would be implicitly determined by Eq.~\eqref{eq:linscale_bdry1}]. 
For repulsive interactions, if the system is precisely on the Luttinger liquid - $\mathcal{O}_{1\sigma}^s$ partial gap boundary, the parameters flow to $\bar{g}_{1\sigma}^s=0$ and $K_\beta=1$. For weak interactions, the resulting phase again is the $c=4$ Luttinger liquid. Conversely, for attractive interactions, the flow on the boundary is to the  $\mathcal{O}_{1\sigma}^s$ partial gap phase.  

In the regions with a $\mathcal{O}_{1\sigma}^s$ partial gap, the strong coupling behavior of  $\mathcal{O}_{1\sigma}^s=\cos(\sqrt{4\pi}(\phi_{\sigma,1,R}+\phi_{\sigma,1,L}-\phi_{\sigma,2,R}-\phi_{\sigma,2,L}))\equiv\cos(\sqrt{4\pi}\Theta_\sigma)$ means that the field $\Theta_\sigma$ becomes locked to the minimum of $\bar{g}_{1\sigma}^s\mathcal{O}_{1\sigma}^s$. For most of the $V_\perp^s<0$ side of the  $\mathcal{O}_{1\sigma}^s$ gap we have $\bar{g}_{1\sigma}^s<0$ and hence $\Theta_\sigma=n\sqrt{\pi}$. Conversely, above the dashed-dotted line in Fig.~\ref{fig:phases_spindecoupled}
we have $\bar{g}_{1\sigma}^s>0$ and hence $\Theta_\sigma=(n+1/2)\sqrt{\pi}$. (In both cases, $n\in \mathbb{Z}$.)  
Towards the $V_\parallel^s>0$ region of the $\mathcal{O}_{1\sigma}^s$ partial gap, the dominant order parameters are RDW (for $\bar{g}_{1\sigma}^s>0$) and CDW (for $\bar{g}_{1\sigma}^s<0$). The dominant component of the corresponding order parameters are  
$O_{\sigma,{\rm RDW}}\sim e^{i\sqrt{\pi}\Upsilon^c_{\sigma}}\sin(\sqrt{\pi}\Theta_\sigma)$
and 
$O_{\sigma,{\rm CDW}}\sim e^{i\sqrt{\pi}\Upsilon^c_{\sigma}}\cos(\sqrt{\pi}\Theta_\sigma)$. 
The Goldstone mode $\Upsilon^c_{\sigma}=\phi_{\sigma,1,R}+\phi_{\sigma,1,L}+\phi_{\sigma,2,R}+\phi_{\sigma,2,L}$ 
commutes with $\Theta_\sigma(x)$ and appears due to the spontaneous breaking of continuous translation symmetry (of the continuum theory) by the CDW or RDW. In fact, $\Upsilon^c_{\sigma}$ is the total charge mode (of each spin); this is consistent with the soft mode causing distortions of the density wave orders and thus local particle density accumulation. The wavenumber associated with these  QLRO is $\Delta k_\alpha=\frac{\Phi}{3a}$. 

For sufficiently attractive intraleg interactions \mbox{($V_\parallel^s<0$)} the dominant order parameter crosses over to orbital singlet pairing with dominant component $S^{\rm singlet}_{\sigma,\text{leg}} \sim e^{i\sqrt{\pi}\Theta^c_{\sigma}}\cos(\sqrt{\pi}\Theta_\sigma)$ 
where $\Theta^c_{\sigma}$ is conjugate to $\Upsilon^c_{\sigma}$, i.e., it is the  phase, as befits superconducting QLRO. The wavenumber for this QLRO is $\Delta k_\alpha =0$.

In the $c=4$ Luttinger liquid phase all order parameters have power law correlations. For sufficiently repulsive interleg interactions, the dominant ones are BDW and OAF with wavenumber $\Delta k_\alpha=\frac{2\Phi}{3a}$. 
For sufficiently strong ($|V_\parallel^s|\gg V_\perp^s$) intraleg attraction
there is a crossover towards orbital triplet pairing $S^{\rm triplet}_{\sigma;x,y,\text{leg}}$ with wavenumber $\Phi/a$.

\subsubsection{FTI precursor phase}
\label{sec:FTIprecursor1}

As noted above, reaching the FTI precursor phase requires going beyond weak interactions. To discuss this phase, we now turn to a formulation which is nonperturbative in interactions but works with nearly decoupled legs, i.e., to first order in $t_\perp$. 
To connect to our preceding discussion, in the bottom panel of Fig.~\ref{fig:phases_spindecoupled} we show the $c=4$ Luttinger liquid and $\mathcal{O}_{1\sigma}^s$ partial gap regions in terms of a (first order in $t_\perp$) Luttinger parameter diagram, together with the FTI region to be discussed below. As noted in Sec.~\ref{sec:forwardscatt_Lutt}, the parameters $K_\rho$ and $K_\beta$ remain valid descriptors of the system even away from weak interactions so long as the system can be viewed as a weakly perturbed Luttinger liquid at low energies. Therefore, the Luttinger parameters here are to be interpreted as those describing the low-energy physics. We delineate the borders of the different phases in terms of their scaling dimensions. Below we identify a key region of microscopic interactions where this is sufficient, and which includes the FTI phase, by linking our weak interaction phase diagram to strong interaction numerics, using a combination of symmetries and bosonization phenomenology.

Starting from a $t_\perp=0$ system, a key condition for the FTI precursor to develop upon turning on $\bar{g}_\text{FTI}\propto t_\perp$ is that (i) the $t_\perp=0$ system form a $c=4$ Luttinger liquid, and (ii) its Luttinger parameters $K_{\rho,\beta}$ fall into the  FTI gap dominated region ($\Delta_{\rm FTI}\!<\!2$, $K_\beta \geq 1$) of the bottom panel of Fig.~\ref{fig:phases_spindecoupled}. We now show that both of these conditions can be met in our model. 

We start with (i). For $t_\perp=0$, the system displays U($1$)$\times$U($1$) symmetry owing to the particle number being conserved in the two legs separately. (There are further symmetries in spin space but they play no role in the argument that follows.) Our nonperturbative (in interactions) analysis (Secs.~\ref{sec:forwardscatt_Lutt} and \ref{sec:IO_Delta}) shows that the most important U($1$)$\times$U($1$) symmetric, potentially gap opening, term to consider is $\mathcal{O}_{1\sigma}^s$. It is thus this term we now focus on. (We emphasize that the density is incommensurate with the lattice, hence umklapp terms are inoperative.) To perform an analysis in terms of the microscopic interactions, we first consider the repulsive part of the leg-SU($2$) invariant line. [For $t_\perp=0$ the full system, not only the interactions, is leg-SU($2$) symmetric here.] On this line, each spin sector of the ladder provides an instance of the extended Hubbard (or $U$-$V$) model \cite{GiamarchiBook2003} with a constant interaction parameter ratio.  
Phrased in terms of our ladders, a key feature of this model, numerically observed for several particle densities \cite{Sano94,Nakamura99,Nakamura00,Tsuchiizu04,Tsuchiizu04,Sandvik04,Ejima07}, is that if $\bar{g}_{1\sigma}^s>0$ and hence the $\mathcal{O}_{1\sigma}^s$ gap is absent for a given interaction ratio at weak interactions, then, for the same ratio it remains absent away from weak interactions. 
For weak interactions, we have $\bar{g}_{1\sigma}^s=V^s_\perp(1+\cos \Phi/3)/(2\pi^2 a)$ at $\nu=1/3$ (see App.~\ref{app:pref_signs}), 
therefore $\bar{g}_{1\sigma}^s>0$ 
on the repulsive part of leg-SU($2$) invariant line; the $\mathcal{O}_{1\sigma}^s$ gap is absent, as we found. This, then, means that the $\mathcal{O}_{1\sigma}^s$ gap is absent on the entire repulsive part of the line, even for strong interactions.
The $\mathcal{O}_{1\sigma}^s$ gap phase boundary thus does not cross the $V^s_\perp= V^s_\parallel>0$ line. 
The $c=4$ Luttinger liquid then prevails for $V^s_\perp\geq V^s_\parallel>0$ beyond weak interactions, provided that no disconnected $\mathcal{O}_{1\sigma}^s$ gap phases appear in that region (consistently with $V^s_\perp\gg V^s_\parallel>0$ suppressing interleg singlets, and hence the $\mathcal{O}_{1\sigma}^s$ gap).

We now turn to condition (ii). For the interleg sector, the preceding analysis coupled with the standard second order RG picture for $\bar{g}_{1\sigma}^s$ (with $\bar{g}_{1\sigma}^s$ \emph{weak} at low energies owing to the system being a $c=4$ Luttinger liquid, even if $V^s_{\parallel,\perp}$ are \emph{strong}) implies \cite{GiamarchiBook2003} that the $t_\perp=0$ low-energy physics has $\bar{g}_{1\sigma}^s\rightarrow 0$, $K_\beta\rightarrow1$ on the leg-SU($2$) invariant line and $\bar{g}_{1\sigma}^s\rightarrow 0$, $K_\beta>1$ for $V^s_\perp>V^s_\parallel>0$. [This also shows that the leg-SU($2$) invariant line marks the boundary where second order RG is required; for $V^s_\perp>V^s_\parallel>0$ a scaling dimension analysis suffices.] As seen from the bottom panel of Fig.~\ref{fig:phases_spindecoupled}, $\Delta_{\rm FTI}\!<\!2$ requires $K_\rho\! \lesssim\! 0.4$.   The precise value depends on $K_\beta$, e.g., we need $K_\rho < 1/3$ for $K_\beta=1$. As we show in Appendix~\ref{app:KrhoKbeta} (see also \cite{G-Santos1993,Cornfeld2015,Mila1993,GiamarchiBook2003,Hohenadler2012}), with strong, repulsive, leg-SU($2$) invariant interactions our model can reach $K_\rho \approx 0.15$
for densities compatible with the two partially filled lower pockets (Fig.~\ref{fig:Dispersion_alphazero}) required for FTI physics, thus the leg-SU($2$) invariant line itself reaches well into the FTI region of the parameter space. 

With the $t_\perp=0$ system in the FTI region, turning on weak interleg tunneling adds $\mathcal{O}^\sigma_{\rm FTI}$ as the most relevant perturbation, hence establishing the FTI precursor phase. 
In the FTI phase, $\mathcal{O}^\sigma_{\rm FTI}$ open a partial gap in the spectrum and leave behind two chiral modes corresponding to the edge modes of a FQH state at 
filling fraction 1/3 for each spin; the central charge is thus $c=2$. As a consequence of time 
reversal symmetry, different spin projections have different chiralities. These helical gapless modes are the precursors of FTI edge modes. The dominant order parameters, 
with power law correlations, are of BDW and OAF type.
The rest exhibit exponential decay.
The existence of these power law correlations for the local order parameters can be understood as a consequence of the quasi-one dimensional nature of the system. As 
shown in the Appendix \ref{app:Order_params}, the BDW and OAF order parameters contain contributions from counterpropagating gapless edge modes, which can be connected 
by a local operator in the quasi-one dimensional system.

\subsubsection{Intermediate discussion and comparison to FQH ladders}
\label{sec:spindecoupled_discuss}

Our findings in the case of vanishing interspin interactions may be contrasted to results on spinless fermion ladders under a magnetic flux. An exhaustive analysis of 
different phases was carried out in Ref. \onlinecite{Carr2006}, focusing on the case that both single particle bands are partially filled. This is a regime complementary 
to our analysis, where the upper band is empty and the bottom band is partially filled with the density tied to the flux, keeping a constant filling factor. Nevertheless, 
while a FQH precursor (the spinless counterpart of the FTI precursor) phase is absent in Ref. \onlinecite{Carr2006} due to the different filling, similar density 
wave and Luttinger liquid phases have been found for small interactions.

A study of a (spinless) fermion ladder system in a magnetic field at  $\nu=1/3$ has been performed in Ref.~\onlinecite{Cornfeld2015}, using phenomenological bosonization 
with higher harmonics in  Eq.~\eqref{bosonization}. Using first order RG (i.e., based on scaling dimensions), it was found that for sufficiently strong 
leg-space-${\rm SU(2)}$-invariant interactions a FQH precursor dominates. In this phenomenological approach, the various cosine coefficients $\bar{g}_{i}$  are 
undetermined, and therefore the interrelation of these with forward scattering parameters is not immediately obvious. It is, however, this interrelation that necessitates a 
second order RG at leg-SU($2$) symmetry, at least for $\mathcal{O}_{1\sigma}^s$. (The interrelation between $\bar{g}_{1\sigma}^s$ and $K_\beta$ holds even at low energies,
when these parameters may differ significantly from their bare values, as it is tied to the degree of leg-SU($2$) symmetry in the interactions.) While in this case, 
similarly to Ref.~\onlinecite{Cornfeld2015}, we find that $\mathcal{O}_{1\sigma}^s$ is irrelevant for repulsive interactions, we note that this holds only marginally. 
The corresponding, only logarithmic, suppression of $\mathcal{O}_{1\sigma}^s$ towards low energies may partly explain the numerical difficulties in earlier spinless ladder simulations \cite{Calvanese2017}. A more pronounced suppression of the $\mathcal{O}_{1\sigma}^s$ term can be achieved by going away from leg-SU($2$) 
symmetry.  Noting the presence of this $\mathcal{O}_{1\sigma}^s$ competition, and how interaction anisotropy may be used to suppress it, is a contribution of this 
work pertinent already for the spinless (i.e., FQH) ladder case.

Even though our considerations for the FTI precursor have been formulated working to first order in $t_\perp$, this does not imply that the FTI precursor (and the 
Laughlin precursor for spinless systems) may not arise away from weak $t_\perp$: indeed, the top panel of Fig.~\ref{fig:phases_spindecoupled} suggests that by reducing 
the region of repulsive interactions with a $\mathcal{O}_{1\sigma}^s$ partial gap, interleg tunnelling may help in suppressing the $\mathcal{O}_{1\sigma}^s$ competition. 
While examining this scenario away from weak interactions is outside the scope of our methods, it is an interesting direction to explore in the future, e.g., using 
numerical simulations.

Finally, it is worth mentioning that our analysis also applies to 1D spinful electron systems for which the ladder flux translates to spin-orbit coupling and $t_\perp$ to a 
Zeeman energy \cite{Atala2014}.
Such systems have been proposed to host fractional helical liquids \cite{Oreg2014}, 
(the phase corresponding to the Laughlin precursor), based on which parafermion modes with potential utility for quantum computation may be created. Taking into account 
the competition from $\mathcal{O}_{1\sigma}^s$ we observed may facilitate achieving the prerequisite fractional helical liquid state.

\subsection{Including interspin interactions}
\label{subsec:incl_spin_int}

Once interspin interactions are added, the previous phase diagram is modified. As our main interest is the exploration of possible FTI precursor phases, 
we focus on the $K_\beta\geq1$,  $K_\rho\lesssim 0.4$ region of the phase diagram where the FTI precursor may arise in the absence of interspin interactions, and study the 
competition of interspin 
interactions and the FTI term. A more complete exploration of the full phase diagram is left for a future study.

Upon including interspin interactions, all the operators $\mathcal{O}_j^d$, with $j=\{1,6\}$ have to be considered in the analysis of RG relevance. We will mostly focus on weak interspin interactions. In this case, first order RG suffices, because the same-spin interactions are strong and hence the zero-interspin Luttinger parameters $K_{\rho,\beta}$ (which remain useful characteristics of the system for small interspin-to-same-spin interaction ratios) largely set the interspin scaling dimensions, independently of the small interspin coupling constants $\bar{g}_j^d$.

For weak interspin interactions, irrespective of their sign and the details, we find that for most of the $K_\beta\geq1$ FTI regime, the operators $\mathcal{O}_{5,6}^d$ are more relevant than $\mathcal{O}_{\rm FTI}^\sigma$ in the RG; the exception is $K_\rho <1/7$, $K_\beta\gtrsim 1$ where $\mathcal{O}_{\rm FTI}^\sigma$ is the most relevant. (Reaching this regime requires going beyond the next-nearest neighbor interactions considered in our model.) 
As $\mathcal{O}_{5,6}^d$ are the most relevant competitors for $K_\beta >1$ and, apart from a patch near $K_\beta\gtrsim 1$, $K_\rho\approx 1/3$ (where $\mathcal{O}_{1,2}^d$ are also more relevant than $\mathcal{O}_{\rm FTI}^\sigma$), $\mathcal{O}_{5,6}^d$ are the only operators more relevant than $\mathcal{O}_{\rm FTI}^\sigma$, we focus on these in what follows, and show how, even in their presence, the FTI precursor phase may survive. [The discussion for $\mathcal{O}_{1,2}^d$ is analogous, in particular leading again to Eq.~\eqref{competition_interspin}.]

\begin{figure}[ht!]
	\centering
		\includegraphics[width=0.8\linewidth]{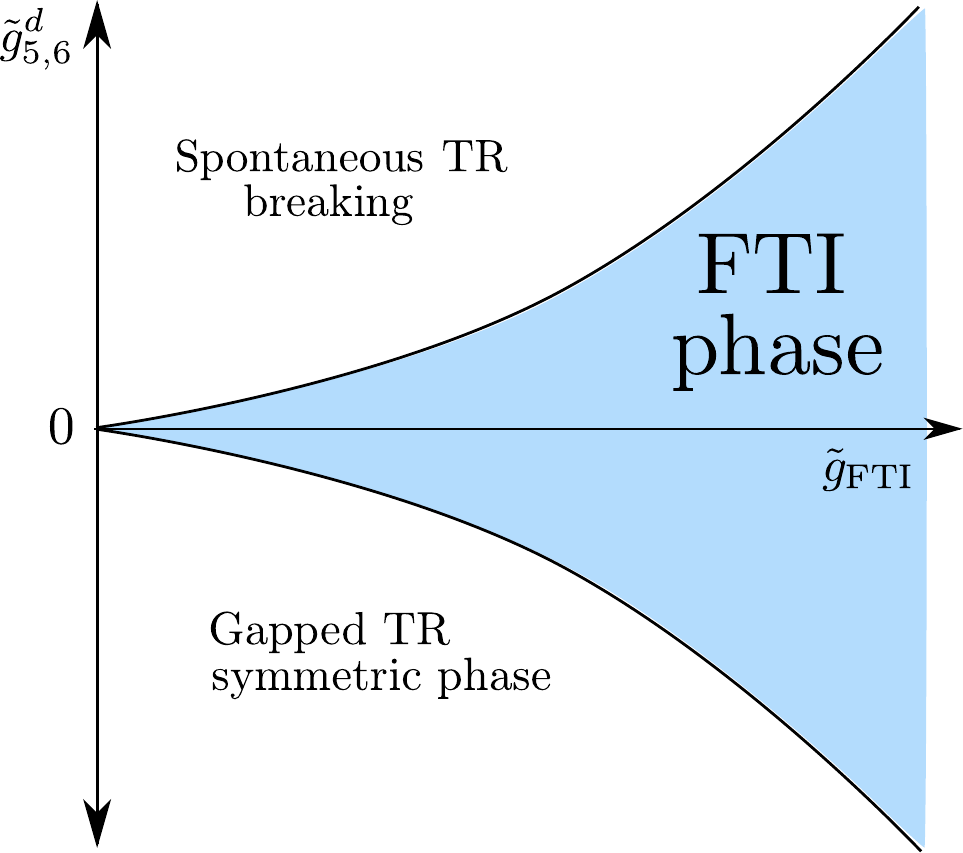} 
	\caption{(Color online) Sketch of the phase diagram
	for the SO ladder with interspin interactions, around the FTI phase. Interspin interactions generate competing operators that are always more relevant than the 
	operator leading to the FTI phase. The competition between these determines the ultimate fate of the system under RG [see Eq.~\eqref{competition_interspin}]. 
	Outside the FTI regime, the dominant order parameter parameter is odd (even) under time reversal symmetry for $\tilde{g}_{5,6}^d>0$ ($\tilde{g}_{5,6}^d<0$).
	}	
        \label{fig:interspin_ops}
\end{figure}

For small interspin interactions $V^d_{\parallel,\perp}$, the couplings $\bar{g}_{5,6}^d$ of the interspin operators 
can be much smaller than $\bar{g}_\text{FTI}$.
This implies that the FTI operator can still grow larger under RG and hit the high energy cutoff scale 
$v_F/a$ before the other operators' coupling would grow comparable. Physically this corresponds to the FTI operator having opened a gap; the interspin operators are 
perturbations for the low energy theory of the remaining gapless FTI edge modes. 
An estimate of the boundaries can be found by identifying the bare couplings corresponding to which the $\tilde{g}_{\rm FTI}$ and $\tilde{g}^d_i$ processes (with 
$\tilde{g}_i = a^2 \bar{g}_i/v_F$ the dimensionless 
couplings) reach the high-energy cutoff at the same scale under renormalization. From the first order renormalization equations we find that the cutoff is reached at the 
same scale when 
\begin{equation}\label{competition_interspin}
 |\tilde{g}_{\rm FTI}^*|\propto |\tilde{g}_i^{d*}|^{\frac{2-\Delta_{\rm FTI}}{2-\Delta^d_{i}}},
\end{equation}
where $\tilde{g}^*_i$ is the bare value of the coupling $i$. 
The conclusions above are not influenced significantly by the second order interspin terms Eqs.~(\ref{second_ord_int}-\ref{second_ord_int_fin}). These operators can be 
separated into two families, each with a single scaling dimension $\Delta_{I,II}$ [see Eqs.~(\ref{delta_sec_ord_spin}) and (\ref{delta_sec_ord_spin2})]. 
For most of the  $K_\beta\geq1$ and $K_\rho\lesssim 0.4$ region, we find that $\Delta_{I,II}>\Delta_{\rm FTI}$
and hence the corresponding terms are less RG relevant than $\mathcal{O}_{\rm FTI}^\sigma$.
Considering also the weakness of interspin interactions, the  second order interspin processes are also suppressed compared 
to $\mathcal{O}_{\rm FTI}^\sigma$ and the pair $\mathcal{O}^d_{5,6}$ 
in terms of their bare couplings. Thus, for small interspin interactions, the terms in Eqs. (\ref{second_ord_int}-\ref{second_ord_int_fin}) can be ignored, and the FTI precursor survives in a region according to Eq.~\eqref{competition_interspin}. A diagram of the different phases is shown in Fig.~\ref{fig:interspin_ops}.

For stronger interspin interactions, analogous weak $t_\perp$ considerations may be developed to the ones we presented in Sec.~\ref{sec:FTIprecursor1} for the spin 
decoupled case, again provided the $t_\perp=0$ system forms a $c=4$ Luttinger liquid.
Though our analysis of scaling dimensions considering stronger interspin interactions has not indicated cases where only $\mathcal{O}_{\rm FTI}^\sigma$ would be relevant, 
scenarios where $\mathcal{O}_{\rm FTI}^\sigma$ only competes with $\mathcal{O}_{5,6}^d$ do arise, e.g. for $K_{\beta,\pm}>2$ and $K_{\rho,\pm}\leq 0.4$.
In this case, for small $t_\perp$, these operators give small perturbations with $\bar{g}_\text{FTI}\propto t_\perp$ and $\bar{g}_{5,6}^d\propto t_\perp^2$ so that 
$\bar{g}_{5,6}^d\ll\bar{g}_\text{FTI}$, thus making Eq.~\eqref{competition_interspin} and Fig.~\ref{fig:interspin_ops} again applicable. Exploring for what microscopic interactions this scenario may 
occur is left as a subject for future investigations. 

The characterization of the phases involved in the competition described by Eq.~\eqref{competition_interspin} and Fig.~\ref{fig:interspin_ops} requires a family of order 
parameters with both leg and spin degrees of freedom. In the regions of our interest, the dominant order parameters conserve particle number. The possible such local 
fermion bilinears  now include \cite{GogolinBook2004,GiamarchiBook2003} 
\begin{equation}\label{order_params2}
 O_{\mu,\lambda,x}=\sum_{\beta,\beta',\sigma,\sigma'}\left(c^{\dagger \beta}_{x,\sigma} (\tau_\mu)_{\beta\beta'}(\sigma_\lambda)_{\sigma \sigma^\prime}c^{\beta'}_{x,\sigma}\right). 
\end{equation}
Here in addition to the $\tau_\mu$ matrices in leg space that appeared previously in (\ref{order_params}), we also use the matrices
$\sigma_\lambda$ that denote the identity matrix $(\lambda=0)$ and the three Pauli matrices ($\lambda=1,2,3$) in spin space. In terms of the low energy theory, we now have
\begin{equation}
O_{\mu,\lambda}(x)=O_{\mu,\lambda}^0(x)+\sum_\alpha \left(e^{i\Delta k_\alpha x}O_{\mu,\lambda}^\alpha(x)+\text{h.c.}\right),
\end{equation}
with slowly varying operators $O_{\mu,\eta}^\lambda(x)$.

In bosonized language $\mathcal{O}_{5}^d=\cos(\sqrt{4\pi}\vartheta^-_5)$ and
$\mathcal{O}_{6}^d=\cos(\sqrt{4\pi}\vartheta^-_6)$; here and for Eqs.~\eqref{eq:OmPhi-boso1} and \eqref{eq:OmPhi-boso2} below, we have introduced
$\vartheta^\pm_5=\phi_{\uparrow,1,L}+\phi_{\uparrow,2,R}\pm(\phi_{\downarrow,1,L}+\phi_{\downarrow,2,R})$ and 
$\vartheta^\pm_6=\phi_{\uparrow,1,R}+\phi_{\uparrow,2,L}\pm(\phi_{\downarrow,1,R}+\phi_{\downarrow,2,L})$ which satisfy $[\vartheta^s_j,\vartheta^{s'}_{j'}]=0$ for 
$(s,j)\neq(s',j')$. The variables $\vartheta^-_5$ and $\vartheta^-_6$ become locked to the minimum of the cosine potentials $\mathcal{O}_5^d$ and $\mathcal{O}_6^d$ 
respectively once these terms run to strong coupling. As $[\vartheta^-_5,\vartheta^-_6]=0$,  the locking of these two variables can occur simultaneously. 
The field values minimising these cosines are
\begin{equation}
\vartheta^-_i=\begin{cases} n\sqrt{\pi} &\mbox{if } \bar{g}^d_{i} < 0, \\
(n+\frac{1}{2})\sqrt{\pi} & \mbox{if }  \bar{g}^d_{i} > 0, \end{cases}
\end{equation}
for $i=5,6.$ For weak interspin interactions, if $\bar{g}^d_{5}$ and $\bar{g}^d_{6}$ change sign they do so simultaneously (Appendix \ref{app:pref_signs}).
Once the $\vartheta^-_i$ fields are pinned, the phase displays QLRO characterised by $O^{\alpha\neq0}_{\mu,\lambda}$, with $\mu=1,2$ (BDW and OAF) 
and $\lambda=0,3$; specifically its component at wavenumber $\Delta k_\alpha a = \frac{2\Phi}{3}$ and $\Delta k_\alpha a = \frac{4\Phi}{3}$  
\begin{eqnarray}\label{eq:munuop}
e^{i \frac{2\Phi}{3a}x}O^{2\Phi/3}_{\mu,\lambda}+\text{h.c.}=\sum_{\sigma\sigma'}\tilde{\Psi}^\dagger_\sigma \tau_\mu (\sigma_\lambda)_{\sigma\sigma'}\tilde{\Psi}_{\sigma'},\\\label{eq:munuop2}
e^{i \frac{4\Phi}{3a}x}O^{4\Phi/3}_{\mu,\lambda}+\text{h.c.}=\sum_{\sigma\sigma'}\Psi^\dagger_\sigma \tau_\mu (\sigma_\lambda)_{\sigma\sigma'} \Psi_{\sigma'},
\end{eqnarray}
with $\tilde{\Psi}_\sigma=(\psi_{\sigma,1,L},\psi_{\sigma,2,R})^T$ and $\Psi_\sigma=(\psi_{\sigma,1,R},\psi_{\sigma,2,L})^T$. Note that $\psi_{\uparrow,\beta,\eta}$ and 
$\psi_{\downarrow,\beta,\eta}$ are predominantly on different legs,
hence the OAF order parameter in Eq.~\eqref{eq:munuop} is defined such that it is antiphase between spins for $\lambda=0$ and in phase for $\lambda=3$. 

In bosonization terms, we have
\begin{eqnarray}\label{eq:OmPhi-boso2}
O^{2\Phi/3}_{\mu,\lambda}\sim e^{i\sqrt{\pi}\vartheta^+_5}\begin{cases} \cos(\sqrt{\pi}\vartheta_5^-),\quad \mu=1,2,\ \lambda=0,\\ 
\sin(\sqrt{\pi}\vartheta_5^-),\quad \mu=1,2,\ \lambda=3.\end{cases}\\\label{eq:OmPhi-boso1}
O^{4\Phi/3}_{\mu,\lambda}\sim e^{i\sqrt{\pi}\vartheta^+_6}\begin{cases} \cos(\sqrt{\pi}\vartheta^-_6),\quad \mu=1,2,\ \lambda=0,\\ 
\sin(\sqrt{\pi}\vartheta^-_6),\quad \mu=1,2,\ \lambda=3.\end{cases}
\end{eqnarray} 
The $O^{m\Phi/3}_{\mu,\lambda}$ ($m=2,4$) order parameter correlators have approximately the same power law decay (the exponents of the two, similarly to the  $\mathcal{O}_{5,6}^d$ scaling dimensions, approach each other for $t_\perp \rightarrow 0$) and we will thus consider them as the two dominant order parameters in the regions simultaneously gapped by $\mathcal{O}_{5,6}^d$.

Due to the left/right mover and valley structure of $\Psi$, the TR transformation of the fields is implemented by 
$\mathcal{T}\Psi_\sigma\mathcal{T}^{-1}=\sum_{\sigma'}(i\sigma_2)_{\sigma'\sigma}\tau_1\Psi_{\sigma'}$. Therefore, $\tau_{1,2}$ and $\sigma_0$ are TR even, while 
$\sigma_3$ is TR odd.
In the region $\bar{g}^d_{5,6}>0$, the QLRO is characterised by $O^{m\Phi/3}_{\mu,3}$
which thus implies the onset of spontaneous TR symmetry breaking (note that there is no true TR breaking in the sense that there is no local order parameter, just QLRO). In the region $\bar{g}^d_{5,6}<0$, the QLRO is that of $O^{m\Phi/3}_{\mu,0}$
 which is compatible with TR symmetry. 

The nature of the corresponding phases involved in the competition between $\mathcal{O}_{\rm FTI}^\sigma$ and $\mathcal{O}_{5,6}^d$ is shown in Fig.~\ref{fig:interspin_ops}. 
For both the $\bar{g}_{5,6}^d>0$ and $\bar{g}_{5,6}^d<0$ regions outside of the FTI regime, the operators $\mathcal{O}_{5,6}^d$ open a gap, leaving two gapless modes behind 
corresponding to the total charge mode $\vartheta^+_5+\vartheta^{+}_6=\Upsilon^c_{\uparrow}+\Upsilon^c_{\downarrow}$ and the mode
$\vartheta^{+}_5-\vartheta^{+}_6=-(\Upsilon_\uparrow+\Upsilon_\downarrow)$. Here $\Upsilon_\sigma$ is the conjugate mode to $\Theta_\sigma$.

\section{Strong coupling analysis}\label{sec:strong_coupling}

In the FTI (precursor) phase, the  $\mathcal{O}^\sigma_{\rm FTI}$ operator opens a gap in the excitation spectrum. In what follows we will be interested in working deep in this 
(partially) gapped phase, focusing on energies much below the FTI gap. Before turning to this strong coupling analysis, we first establish the typical 
energy scale of the FTI gap. 

Starting from weak coupling, $\tilde{g}_{\rm FTI}$ grows exponentially with the scaling parameter $\ell$ in the FTI phase, where $\mathcal{O}_{\rm FTI}^\sigma$ is relevant. In terms of first order RG (and ignoring the flow of $K_\beta$ due to $\mathcal{O}_{1\sigma}^s$), the scale $\ell_1$ [see Eq. (\ref{RG_eq})] where this coupling becomes of order one is 
\begin{equation}\label{scale_strong}
 e^{\ell_1} \approx (\tilde{g}^*_{\rm FTI})^\frac{1}{\Delta_{\rm FTI}-2},
\end{equation}
where, as before, $\tilde{g}^*_{\rm FTI}$ is the bare value of the FTI coupling. 
At this RG scale, the argument of $\mathcal{O}_{\rm FTI}^\sigma$ is largely pinned to one of the minima of the corresponding cosine, with fluctuations being costly in energy. This physics 
can be described by expanding the cosine around this minimum, truncating the expansion up to second order. This introduces a mass scale in the quadratic
part of the Hamiltonian, implementing the opening of the FTI gap. The gap at this scale is given by $ m_{\rm FTI}(\ell_1)\sim\frac{v'}{a}\sqrt{2\pi K_{\rm eff}}$ with $v'$ and 
$K_{\rm eff}$ being the 
renormalized velocity and the effective Luttinger parameter of the hard modes respectively, obtained by decoupling the interaction between the hard and the soft modes,
and $a$ is the short distance cutoff of the renormalized theory \cite{GiamarchiBook2003}.
In terms of the forward scattering parameters, they read (see Appendix \ref{app:strong_coupling})
\begin{equation}\label{K_eff}
 v'=\frac{5}{9}\sqrt{(v_F+g_4)^2-g_2^2},\quad K_{\rm eff}=\sqrt{\frac{v_F+g_4+g_2}{v_F+g_4-g_2}},
\end{equation}
where we have introduced the $g$-parameters
\begin{eqnarray}
&g_4= \frac{f_{22}+4(f_{11}-f_{12})}{20\pi}, \,  g_{2}=\frac{4(g_{12}-g_{11})-g_{22}}{20\pi}.
\end{eqnarray}
In the region where the $\mathcal{O}_{\rm FTI}^\sigma$ are the most relevant operators $K_{\rm eff}< 1$.
The gap generated by $\mathcal{O}_{\rm FTI}^\sigma$ has units of energy, so it scales with $\ell$ as $m_{\rm FTI}(\ell_1)\sim e^{\ell_1} m_{\rm FTI}(0)$ with 
$m_{\rm FTI}(0)$ the bare gap. This provides a crude, leading order RG based, estimate
\begin{eqnarray}\nonumber
 m_{\rm FTI}(0)&\sim& m_{\rm FTI}(\ell_1)(\tilde{g}^*_{\rm FTI})^\frac{1}{2-\Delta_{\rm FTI}},\\
 &\sim & \frac{v'}{a}\sqrt{2\pi K_{\rm eff}}(\tilde{g}^*_{\rm FTI})^\frac{1}{2-\Delta_{\rm FTI}}\label{eq:FTIgap},
\end{eqnarray}
where $a$ now is the short distance cutoff at scale $\ell=0$. Though based on taking the weak coupling RG out of its domain of validity, such estimates are known to capture certain essential qualitative features, e.g., 
that the gap depends on the bare coupling $\tilde{g}_{\rm FTI}^*$ through a power law \cite{GiamarchiBook2003}.

The opening of a gap induced by the operators $\mathcal{O}_{\rm FTI}^{\sigma}$ defines the FTI precursor phase of the ladder. This is the FTI analogue of the term introduced for (two-dimensional) FQH systems in Ref.~\onlinecite{Teo2014}. Here we summarize the main properties of this phase, in terms of its bulk excitations. The edge properties of
the FTI precursor phase are addressed in the next section, when we discuss the complete transformation between the original and the effective degrees of freedom describing
the low energy theory of the topological phase. 

In the strong coupling limit of the FTI phase, the fields 
\begin{equation}\label{eq:FTIfield}
 \theta_{g\sigma}\equiv \phi_{\sigma,1,L}+2\phi_{\sigma,1,R}+2\phi_{\sigma,2,L}+\phi_{\sigma,2,R},
\end{equation}
are pinned to the minimum of the cosine potential defined by $\mathcal{O}_{\rm FTI}^{\sigma}=\cos(\sqrt{4\pi}\theta_{g\sigma})$.
This implies that those modes
become massive, i.e. it costs energy $\sim m_{\rm FTI}(\ell_1)$ to excite them. 
The (bulk) charge density per spin is
$\rho_c=-\frac{1}{3\sqrt{\pi}}\partial_x\theta_{g\sigma}$. 
Hence, a kink in the $\theta_{g\sigma}$ field connecting neighboring cosine minima ($\theta_{g\sigma}\rightarrow\theta_{g\sigma}+\sqrt{\pi}$) carries charge 1/3. 
It is important that the different cosine minima are physically equivalent: the compactness of the microscopic fields 
$\phi_{\sigma,\alpha,\eta}\equiv\phi_{\sigma,\alpha,\eta}+\sqrt{\pi}n$ ($n\in\mathbb{Z}$) implied by Eq.~\eqref{bosonization}, translates to 
$\theta_{g\sigma}\equiv\theta_{g\sigma}+\sqrt{\pi}n$ according to Eq.~\eqref{eq:FTIfield}. 
Hence such a kink configuration is local; it can be moved by local operators. 
The conjugate mode to $\theta_{g\sigma}$ is $\varphi_{g\sigma}$, and together
they satisfy the commutation relation
\begin{equation}
 [\theta_{g\sigma}(x),\varphi_{g\sigma'}(y)]={3i}\delta_{\sigma\sigma'}\Theta(x-y).
\end{equation}
where $\Theta(x)$ is the Heaviside step function (see appendix \ref{app:Klein} for more details).
Hence, the operator $e^{i\frac{\sqrt{\pi}}{3}\varphi_{g\sigma}}$,
by creating the corresponding kink in the $\theta_{g\sigma}$ field, creates a quasiparticle. 

In the quasi-1D geometry that we consider, this bulk 
quasiparticle cannot be braided. Extending the system by coupling different quasi-one dimensional systems (ladders) labelled by $\ell$, it is possible to show that these
bulk quasiparticles possess fractional statistics due to the field $\theta_{g\sigma}$ being locked in the bulk. Specifically, moving a quasiparticle along a loop in the bulk can be described by the extended operator 
\begin{equation}
{\circlearrowright}^\sigma(x_1)=\Gamma^\sigma_{\ell}(x_1,x_2)\chi^\sigma_{\ell,\ell+1}(x_2)
\Gamma^\sigma_{\ell+1}(x_2,x_1)\chi^{\sigma\dagger}_{\ell+1,\ell}(x_1)
\end{equation}
where $\Gamma^\sigma_{\ell}(x_1,x_2)= e^{i\frac{\sqrt{\pi}}{3}(\varphi^\ell_{g\sigma}(x_2)-\varphi^\ell_{g\sigma}(x_1))}$ 
is an operator that displaces the bulk quasiparticle
along the quasi-one dimensional system from $x_1$ to $x_2$, while $\chi_{\ell,\ell+1}(x)=e^{i\frac{\sqrt{\pi}}{3}(\varphi^{\ell+1}_{g\sigma}(x)-\varphi^\ell_{g\sigma}(x)+\theta^{\ell+1}_{g\sigma}(x)+\theta^\ell_{g\sigma}(x))}$ 
moves a bulk quasiparticle between neighbouring ladders. 
The operators $\Gamma^\sigma_{\ell}(x,y)$ and $\chi^\sigma_{\ell,\ell+1}(x)$ can be constructed using products of the original 
fermion operators. Using their explicit form, the loop operator becomes
\begin{equation}
{\circlearrowright}^\sigma(x_1)=e^{i2\pi N^\sigma_{QP}/3}
\end{equation}
where we used that the number of quasiparticles (of a given spin) inside the loop is  $N^\sigma_{QP}=\sum_{m=\ell}^{\ell+1}\langle\frac{1}{\sqrt{\pi}}(\theta^{m}_{g\sigma}(x_1)-\theta^{m}_{g\sigma}(x_1))\rangle/4\pi$.
This signals the anyonic statistics of the bulk quasiparticles.

In what follows we will be focusing on the regime of
momenta and frequencies small compared to the gap,  
which corresponds to the high energy cutoff of the low energy theory.
This implies that we can project out the high energy processes, which create excitations of the order the FTI gap (or larger). 
After the projection, the resulting operators constitute  perturbations to the low energy FTI sector formed by the precursor FTI edge modes. Depending on the RG scaling 
dimensions,  the FTI edge modes may be robust against these perturbations or they may become gapped. We first concentrate in the case of vanishing interspin interaction.

\subsection{Decoupled spin limit}

In the decoupled spins regime, we focus on the region where the operators $\mathcal{O}^s_{1\sigma}$ are irrelevant while the FTI operator flows to strong coupling.
As we discussed previously in section \ref{sec2:six_fermion}, higher order terms are more irrelevant than $\mathcal{O}^s_{1\sigma}$ in terms of the weak coupling analysis. 
All these terms can be present in a strong coupling description, with arbitrarily small coupling strengths. In the analysis of this section we consider the
largest of those, which corresponds to $\mathcal{O}_{1\sigma}^s$. The Hamiltonian consists of
$H=\sum_\sigma (H_{0}^{\sigma}+H_\text{FTI}^{\sigma}+H_{1}^{\sigma})$ where
\begin{equation}\label{H_0_strong}
 H_{0}^{\sigma}=\frac{1}{2}\int dx\partial_{x}\bm{\Omega}_{\sigma}^{T}\tilde{\mathcal{M}}\partial_{x}\bm{\Omega}_{\sigma},
\end{equation}
together with $H_\text{FTI}^\sigma=\int dx\bar{g}_{\rm FTI}\cos(\sqrt{4\pi}\theta_{g\sigma}),$ and
\begin{equation}\label{H_2_strong}
H_{1}^\sigma= \int dx \bar{g}_1^s\cos\left(\frac{\sqrt{4\pi}}{3}(\varphi_{g\sigma}+\tilde{\phi}_{L\sigma}-\tilde{\phi}_{R\sigma})\right).
\end{equation}
The Klein factors of these operators are given explicitly in 
Appendix \ref{app:Klein}. They do not play a role in the
following discussion.
Here we introduced the fields $\bm{\Omega}_\sigma^{T}=(\varphi_{g\sigma},\theta_{g\sigma},\tilde{\phi}_{L\sigma},\tilde{\phi}_{R\sigma})$ which form a natural choice of 
basis in the FTI phase. Their relation to our original fields $\phi_{\sigma,\beta,\eta}$ is given by
\begin{eqnarray}\label{U_strong}
\begin{pmatrix}\varphi_{g\sigma}\\
\theta_{g\sigma}\\
\tilde{\phi}_{L\sigma}\\
\tilde{\phi}_{R\sigma}
\end{pmatrix}\equiv\begin{pmatrix}1 & 2 & -2 & -1\\
1 & 2 & 2 & 1\\
2 & 1 & 0 & 0\\
0 & 0 & 1 & 2
\end{pmatrix}\begin{pmatrix}\phi_{\sigma,1,L}\\
\phi_{\sigma,1,R}\\
\phi_{\sigma,2,L}\\
\phi_{\sigma,2,R}
\end{pmatrix}. 
\end{eqnarray}
The forward scattering matrix that determines the Gaussian part of the Hamiltonian is $\tilde{\mathcal{M}}=\mathcal{U}^{T}M\mathcal{U}$
with $M=v_F\openone_{4}+\frac{1}{2\pi}\mathbb{V}$ and $\mathbb{V}$ given in Eq. (\ref{int_matrix_same_spin}). The similarity transformation 
$\mathcal{U}$ is the inverse of the matrix in Eq. (\ref{U_strong})
\begin{equation}\mathcal{U}=\left(\begin{array}{cccc}
-\frac{1}{6} & -\frac{1}{6} & \frac{2}{3} & 0\\
\frac{1}{3} & \frac{1}{3} & -\frac{1}{3} & 0\\
-\frac{1}{3} & \frac{1}{3} & 0 & -\frac{1}{3}\\
\frac{1}{6} & -\frac{1}{6} & 0 & \frac{2}{3}
\end{array}\right).
\end{equation}

The commutator of these fields is given by 
\begin{equation}\label{commutators_FTI}
[\Omega_{i,\sigma}(x),\partial_{y}\Omega_{j,\sigma'}(y)]=i\tilde{\mathcal{K}}^{-1}_{ij}\delta_{\sigma\sigma'}\delta(x-y),
\end{equation}
with the  $\tilde{\mathcal{K}}$ matrix being explicitly
\begin{equation}
\tilde{\mathcal{K}}=\left(\begin{array}{cc}
 \tilde{\mathcal{K}}_h & 0\\
 0 & \tilde{\mathcal{K}}_s
\end{array}\right), \quad
\tilde{\mathcal{K}}_h=\frac{1}{3}\sigma_1,\quad
\tilde{\mathcal{K}}_s=-\frac{2}{3}\sigma_3.
\end{equation}
The charge density per spin is given by
\begin{equation}
 \rho_c=-\frac{1}{3\sqrt{\pi}}(\partial_x\theta_{g\sigma}+\partial_x\tilde{\phi}_{L\sigma}+\partial_x\tilde{\phi}_{R\sigma}).
\end{equation}

For FQH and topological insulator states the $\tilde{\mathcal{K}}$ matrix is known to encode topological data, which  directly determine the commutator structure of the edge 
modes. In our case, the modes $\tilde{\phi}_{L,R}$ are seen to obey the commutator relations corresponding to FTI edge modes at 1/3 effective filling per spin, provided that the edge 
quasiparticle operators are proportional to  $\exp[\pm i (\sqrt{4\pi}/3)\tilde{\phi}_{\eta\sigma}]$, as suggested by Eq.~\eqref{H_2_strong}, which is also consistent 
with the observation that $\exp[ -i\eta (\sqrt{4\pi}/3)\tilde{\phi}_{\eta\sigma}]$ creates charge $1/3$.

To obtain a low energy description in the strong coupling regime, we project out the massive sector. To perform the projection, we first consider the situation
of vanishing $H_1^\sigma$ (i.e $g^s_1=0$). 
In this case, the low energy theory is obtained upon a quadratic expansion of the cosine term in $H_1^\sigma$ around one of the minima, and 
integrating out the massive degrees of freedom. 

\begin{figure}[ht!]
	\centering
		\includegraphics[width=\linewidth]{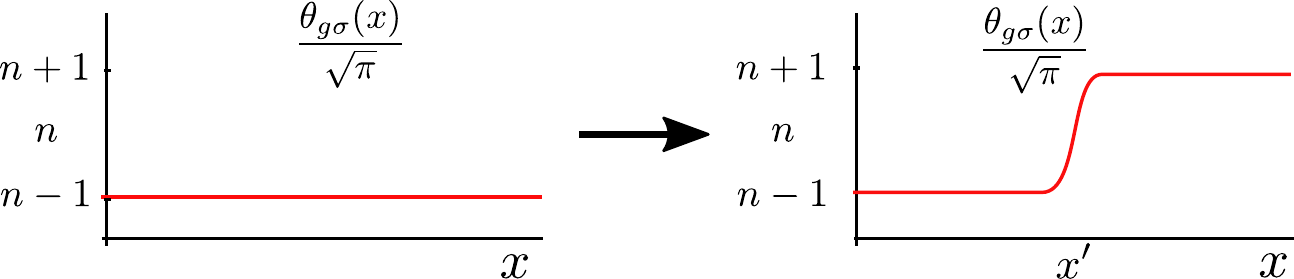} 
	\caption{(Color online) The operator $\mathcal{O}_1^s$ acting on a uniform configuration (represented by the black arrow) creates a kink in the field 
	$\theta_{g\sigma}$, which carries an extra charge of $2/3$. The operator also creates two quasiparticles in the gapless sector, an excitation of charge $-2/3$.}
        \label{fig:vacuum}
\end{figure}

Considering now $H_1^\sigma$, we observe that the operator $e^{-i\frac{\sqrt{4\pi}}{3}\varphi_{g\sigma}}$ creates a $\theta_{g\sigma}$ profile connecting second neighbor 
minima (Fig. \ref{fig:vacuum}), 
\begin{equation}
 e^{-i\frac{\sqrt{4\pi}}{3}\varphi_{g\sigma}(x')}\left|\frac{\theta_{g\sigma}}{\sqrt{\pi}}=n\right\rangle=\left|\frac{\theta_{g\sigma}}{\sqrt{\pi}}=n+2\Theta(x')\right\rangle.
\end{equation}
That is, it creates a (double) kink.
Due to the equivalence of the different minima, this is a local object \cite{Teo2014}. This object has charge 
$Q=\int dx \rho_{\sigma}=-\frac{1}{3\sqrt{\pi}}(\theta_{g\sigma}(\infty)-\theta_{g\sigma}(-\infty))=-2/3$ which is accumulated entirely in the gapped sector. This
process is identified with the creation of two quasiparticles in the ``FTI bulk". The term $H_1^\sigma$ thus transfers pairs of quasiparticles between the gapless 
(``edge modes") and the gapped (``FTI bulk") sector.

Physically, one expects that due to the ``bulk" quasiparticles that appear in the phase where the FTI term is dominant, $H_1^\sigma$ creates 
high energy excitations. This intuition is supported by analysing the effect of the perturbation $H_1^\sigma$ in the Hamiltonian 
$H_0^\sigma+H_\text{FTI}^\sigma$. This perturbation creates kink eigenstates of the unperturbed Hamiltonian $H_0^\sigma+H_\text{FTI}^\sigma$.
These states have an energy of the order of
 $\sim 2m_{\rm FTI}(\ell_1)/(\pi K_{\rm eff})$ \cite{GiamarchiBook2003}, which, as expected, is comparable to the FTI gap (measuring energies
on the RG scale of the low energy theory).
Due to $H_1^\sigma$ having no low energy to low energy  matrix elements, to first order in $\tilde{g}_1^s$, the projection to the low energy sector amounts to discarding 
$H_1^\sigma$ altogether.  
The first nonvanishing contribution comes from second order perturbation theory (in $\tilde{g}_1^s$), which allows for processes where ``bulk" quasiparticles appear only 
as intermediate states. 
These processes have a prefactor of the order of $(\bar{g}_1^s)^2/m_{\rm FTI}(\ell_1)$ and  their most RG relevant contribution
corresponds to density-density interactions between the right and left mover ``edge" modes of the FTI precursor phase in the low 
energy description. Although these can modify the effective Luttinger parameters of the low energy theory, they do not open a gap for the FTI ``edge" modes. This implies 
that (at least to second order in $\tilde{g}_1^s$) the $H_1^\sigma$ perturbation does not destroy the FTI precursor phase.

Defining the new basis of soft bosonic fields
\begin{equation}
\bar{\varphi}_{s\sigma}=\frac{\tilde{\phi}_{L\sigma}+\tilde{\phi}_{R\sigma}}{2},\quad\bar{\theta}_{s\sigma}=\frac{\tilde{\phi}_{L\sigma}-\tilde{\phi}_{R\sigma}}{2},
\end{equation}
the forward interaction matrix in the gapless sector becomes diagonal. In this new basis, the low energy theory splits into two uncoupled Luttinger liquids, described by the
Hamiltonian $H_{s\uparrow}$ and $H_{s\downarrow}$, with  
\begin{equation}
 H_{s\sigma}=\frac{1}{2}\int dx v_s\left((\partial_x\bar{\varphi}_{s\sigma})^2K+\frac{(\partial_x\bar{\theta}_{s\sigma})^2}{K}\right).
\end{equation}
In terms of the forward scattering parameters, the velocity $v_s$ and the Luttinger parameter $K$ are respectively $v_{s}=\frac{10}{9}\sqrt{(v_F+a_1)^2-b_1^2}$ and
\begin{equation}\label{Luttinger_nointer}
K=\sqrt{\frac{v_F+a_1+b_1}{v_F+a_1-b_1}},
\end{equation}
where $a_{1}=\frac{1}{20\pi}(f_{11}+4(f_{22}-f_{12}))$ and $b_{1}=\frac{1}{20\pi}(g_{11}+4(g_{22}-g_{12}))$. In the region of parameters considered, the Luttinger parameter is $K\gtrsim 3/2$, indicating a strong interedge interaction.
The fields $\bar{\varphi}_{s\sigma}$ and $\bar{\theta}_{s\sigma}$ are conjugates, and satisfy the commutation relations
\begin{equation}
 [\bar{\varphi}_\sigma(x),\partial_y\bar{\theta}_{\sigma'}(y)]=-i\frac{3}{4}\delta_{\sigma\sigma'}\delta(x-y).
\end{equation}

\subsection{Including interspin interactions}\label{sec:SCInterspin_int}

In the case with nonzero interspin interactions, there are
many operators that appear. To first order in the interspin interaction they are given by Eqs. (\ref{int_interspin}) to (\ref{int_interspin_final}). 
The high energy projection analysis is similar to the decoupled case, with the important difference that the forward scattering matrix couples all the fields. The quadratic part of the Hamiltonian
is now
\begin{equation}\label{H_inter_spin_strong}
 H_0=\frac{1}{2}\int dx \partial_{x}\bm{\Omega}^{T}{\mathcal{M}}\partial_{x}\bm{\Omega},\\
\end{equation}
while the nonlinear contributions are
 \begin{eqnarray}
 H_\text{FTI}&=&\sum_\sigma\int dx\bar{g}_{\rm FTI}\cos(\sqrt{4\pi}\theta_{g\sigma}),\\
 H_2&=&\sum_i\int dx \bar{g}_i\mathcal{O}_i(\bm{\Omega}). 
\end{eqnarray}
The vector of fields $\bm\Omega=(\bm\Omega_h,\bm\Omega_s)$ contains the hard ($\bm\Omega_h$) and soft modes ($\bm\Omega_s$). They are given by 
$ \bm\Omega_h=(\varphi_{g\uparrow},\theta_{g\uparrow},\varphi_{g\downarrow},\theta_{g\downarrow})$ and
$ \bm\Omega_s=(\tilde{\phi}_{L\uparrow},\tilde{\phi}_{R\uparrow},\tilde{\phi}_{L\downarrow},\tilde{\phi}_{R\downarrow}).$

The operators $\mathcal{O}_i$ correspond
to all the exponential operators considered in the previous discussion of scaling dimensions, apart from the FTI operators.
The forward scattering matrix ${\mathcal{M}}$ is given by  
\begin{equation}
\mathcal{M}=\begin{bmatrix}\mathcal{M}^{(4)}_{hh}&\mathcal{M}^{(4)}_{hs}\\
                                 (\mathcal{M}^{(4)}_{hs})^T&\mathcal{M}^{(4)}_{ss}
                                \end{bmatrix},
\end{equation}
where $\mathcal{M}^{(4)}_{ab}$ encodes the forward scattering interaction between $a$ and $b$ sectors. The $4\times 4$ matrices $\mathcal{M}^{(4)}_{ab}$ are given in Appendix 
\ref{app:strong_coupling}. The commutation relations of the bosonic operators are given in Eq.~(\ref{commutators_FTI}).
The exponential operators $\mathcal{O}_i$ can be written in compact form as 
\begin{equation}
 \mathcal{O}_{i,\uparrow\downarrow}=(e^{i\frac{\sqrt{\pi}}{3}\bm{\eta}_{i,\uparrow\downarrow}^T\bm{\Omega}}+\text{h.c.}),
\end{equation}
with the vectors $\bm{\eta}_{i,\uparrow\downarrow}$ given by
\begin{eqnarray}
 \bm{\eta}_{1,\uparrow\downarrow}^T&=&(-1,-1,1,1,-2,0,2,0),\\
 \bm{\eta}_{2,\uparrow\downarrow}^T&=&(-1,-1,-1,1,-2,0,0,2),\\
 \bm{\eta}_{3,\uparrow\downarrow}^T&=&(1,-1,1,1,0,-2,2,0),\\
 \bm{\eta}_{4,\uparrow\downarrow}^T&=&(1,-1,-1,1,0,-2,0,2),\\
 \bm{\eta}_{5,\uparrow\downarrow}^T&=&(-1,3,1,-3,-4,-2,4,2),\\
 \bm{\eta}_{6,\uparrow\downarrow}^T&=&(-1,3,-1,-3,-4,-2,2,4),\\
 \bm{\eta}_{7,\uparrow\downarrow}^T&=&(1,3,1,-3,-2,-4,4,2),\\
 \bm{\eta}_{8,\uparrow\downarrow}^T&=&(1,3,-1,-3,-2,-4,2,4),\\
 \bm{\eta}_{9,\uparrow\downarrow}^T&=&(0,2,0,-2,-4,-4,4,4),\\
 \bm{\eta}_{10,\uparrow\downarrow}^T&=&(0,-4,0,4,2,2,-2,-2).
\end{eqnarray}
 Due to the nonvanishing $\varphi_{g\sigma}$ components, 
the operators $\mathcal{O}_{i,\uparrow\downarrow}$ with ${i=1\dots 8}$ create kinks in a similar way as discussed previously for $\mathcal{O}_{1\sigma}^s$.
Projecting out the high energy states, these operators do not contribute in first order of $g_i$.
On the other hand, the operators $\mathcal{O}_{9,\uparrow\downarrow}\equiv \mathcal{O}^d_5$ and $\mathcal{O}_{10,\uparrow\downarrow}\equiv \mathcal{O}^d_6$ do not vanish after the projection.
To first order in the interaction parameter the projected operators become 
\begin{eqnarray}
&\mathcal{O}_{5,{\rm proj}}^d=\cos\left(\frac{2\sqrt{4\pi}}{3}(\tilde{\phi}_{L\uparrow}+\tilde{\phi}_{R\uparrow}-\tilde{\phi}_{L\downarrow}-\tilde{\phi}_{R\downarrow})\right),\\
&\mathcal{O}_{6,{\rm proj}}^d=\cos\left(\frac{\sqrt{4\pi}}{3}(\tilde{\phi}_{L\uparrow}+\tilde{\phi}_{R\uparrow}-\tilde{\phi}_{L\downarrow}-\tilde{\phi}_{R\downarrow})\right),
 \end{eqnarray}
with the Klein factors considered explicitly in Appendix \ref{app:Klein}.
In terms of the weak coupling analysis, the operators $\mathcal{O}^d_5,\mathcal{O}^d_6$ were the most relevant in RG sense.
After the projection performed above, both operators survive. Among them, the more relevant is $\mathcal{O}_{6,{\rm proj}}^d$ in terms of 
the low energy description of the FTI dominated phase.  
The second order operators (\ref{second_ord_int}) to (\ref{second_ord_int_fin}) are less relevant in RG sense according to the weak coupling analysis
of the previous section. Now we are concerned with the strong coupling regime, where these operators have flowed under RG as well.
Writing
them in the basis 
of fields $\bm{\Omega}$, we find that they also create high energy excitations as the processes considered above, and hence, to first order in their coupling constant, do 
not contribute after the low energy projection.
We thus concentrate on the effect of $\mathcal{O}_{6,{\rm proj}}^d$ as a perturbation in the FTI precursor phase. 

Introducing the quasiparticle operator $\psi^\eta_{{\rm qp},\sigma}=e^{i\eta\frac{\sqrt{4\pi}}{3}\tilde{\phi}_{\eta\sigma}}$, 
(using the same left-right notation as Eq.~\eqref{bosonization} and without Klein factors), the projected process $\mathcal{O}_{6,{\rm proj}}^d$ becomes 
\begin{equation}\label{proj_op_10}
\mathcal{O}_{6,{\rm proj}}^d=\left((\psi^R_{{\rm qp},\uparrow})^\dagger\psi^L_{{\rm qp},\uparrow}(\psi^L_{{\rm qp},\downarrow})^\dagger\psi^R_{{\rm qp},\downarrow}+\text{h.c.}\right).
\end{equation}
We see that this operator corresponds to correlated quasiparticle backscattering between opposite edges of the FTI precursor, with one backscattering factor for
each spin component.

Projecting out the massive modes $\theta_{g\sigma}$ of the FTI precursor following the steps discussed in the previous section,  the quadratic Hamiltonian
of the soft modes takes the form 	
\begin{eqnarray}\label{soft_action}
 H&=&\frac{1}{2}\int dx \partial_{x}\bm{\Omega}^{T}_s{\mathcal{M}_s}\partial_{x}\bm{\Omega}_s,
  \end{eqnarray}
 The soft-mode commutators are given by $[\tilde{\phi}_{\eta\sigma}(x),\partial_y\tilde{\phi}_{\eta'\sigma'}(y)]=i\frac{3}{2}\eta\delta_{\sigma\sigma'}\delta_{\eta\eta'}\delta(x-y)$ while the forward scattering matrix is
\begin{equation}
 \mathcal{M}_s=\frac{10}{9}\left(\openone_2\otimes\begin{bmatrix}  v_F+a_1&b_1  \\
 b_1 & v_F+a_1\end{bmatrix}+\sigma_1\otimes\begin{bmatrix} \beta_1&\beta_2  \\
 \beta_2 & \beta_1\end{bmatrix}\right),
\end{equation}
where $a_1$ and $b_1$ correspond to the same parameters defined in the decoupled spin limit. The parameters $\beta_{1,2}$ describe the interspin interactions
and are given by
$\beta_1=\frac{4(h_{22}-h_{12})+h_{11}}{20\pi}$ and $\beta_2=\frac{4(\tilde{h}_{22}-\tilde{h}_{12})+\tilde{h}_{11}}{20\pi}$
in terms of the forward scattering parameters.
This quadratic Hamiltonian
can be diagonalized defining the new bosonic fields
\begin{equation}
\begin{bmatrix}
\bar{\varphi}_+\\
 \bar{\theta}_{+}\\
 \bar{\varphi}_-\\
 \bar{\theta}_-
\end{bmatrix}=
  \frac{1}{2}\begin{bmatrix}1 & 1& 1& 1 \\
 1 & -1 & 1 & -1 \\
 1 &  1 &-1 & -1 \\
 1 & -1 &-1 &  1 
\end{bmatrix}
\begin{bmatrix}\tilde{\phi}_{L\uparrow}\\
 \tilde{\phi}_{R\uparrow}\\
 \tilde{\phi}_{L\downarrow}\\
 \tilde{\phi}_{R\downarrow}
\end{bmatrix}.
\end{equation}
In this new basis, the Hamiltonian
(\ref{soft_action}) splits into two uncoupled Luttinger liquids, described by the
Hamiltonian $H_+$ and $H_-$, where
\begin{equation}
 H_i=\frac{1}{2}\int dx v_i\left((\partial_x\bar{\varphi}_i)^2K_i+\frac{(\partial_x\bar{\theta}_i)^2}{K_i}\right),
\end{equation}
with $v_{\pm}=\frac{10}{9}\sqrt{(v_F+a_1\pm\beta_1)^2-(b_1\pm\beta_2)^2}$ and
\begin{equation}\label{Luttinger_strong}
K_\pm= K\left(\frac{1\pm x}{1\pm y}\right)^\frac{1}{2}.
\end{equation}
Here $K$ is the Luttinger parameter for vanishing interspin interactions (\ref{Luttinger_nointer}),
$x=\frac{\beta_1+\beta_2}{v_F+a_1+b_1}$ and $y=\frac{\beta_1-\beta_2}{v_F+a_1-b_1}$.

The fields $\bar{\varphi}_i$ and $\bar{\theta}_i$ are conjugates, and satisfy the commutation relations
\begin{equation}
 [\bar{\varphi}_a(x),\partial_y\bar{\theta}_b(y)]=-i\frac{3}{2}\delta_{ab}\delta(x-y).
\end{equation}

It follows from here that the projected operator (\ref{proj_op_10}) has scaling dimension
$\Delta_{6,{\rm proj}}^d=\frac{2K_-}{3}.$ (For 
$\mathcal{O}_{5,{\rm proj}}^d$ we have $\Delta_{5,{\rm proj}}^d=4\Delta_{6,{\rm proj}}^d$.) Within the FTI phase, this scaling dimension is much smaller than 2 making it 
highly relevant in the RG sense. The projected operator $\mathcal{O}_{6,{\rm proj}}^d$ will open a gap $m_{\rm low}$ in the low energy theory, as it backscatters 
quasiparticles between the two edges.

We can estimate the value of the gap $m_{\rm low}$ compared to the gap of the FTI precursor by applying a logic analogous to the one leading to Eq. \eqref{eq:FTIgap}.
We find
\begin{eqnarray}
 m_{\rm low}(\ell_1)&\sim&m_{\rm FTI}(\ell_1)\sqrt{2\pi K_-}(\tilde{g}_{6}^d(\ell_1))^\frac{1}{2-\Delta^d_{6,{\rm proj}}},
\end{eqnarray}
where we used that the high energy cutoff $v/a$ (with $v$ of the order of $v_\pm$) is to be interpreted as $m_{\rm FTI}(\ell_1)$, the FTI gap at scale $\ell_1$ relative to the weak 
coupling RG starting point.  

\begin{figure}[ht!]
	\centering
		\includegraphics[width=\linewidth]{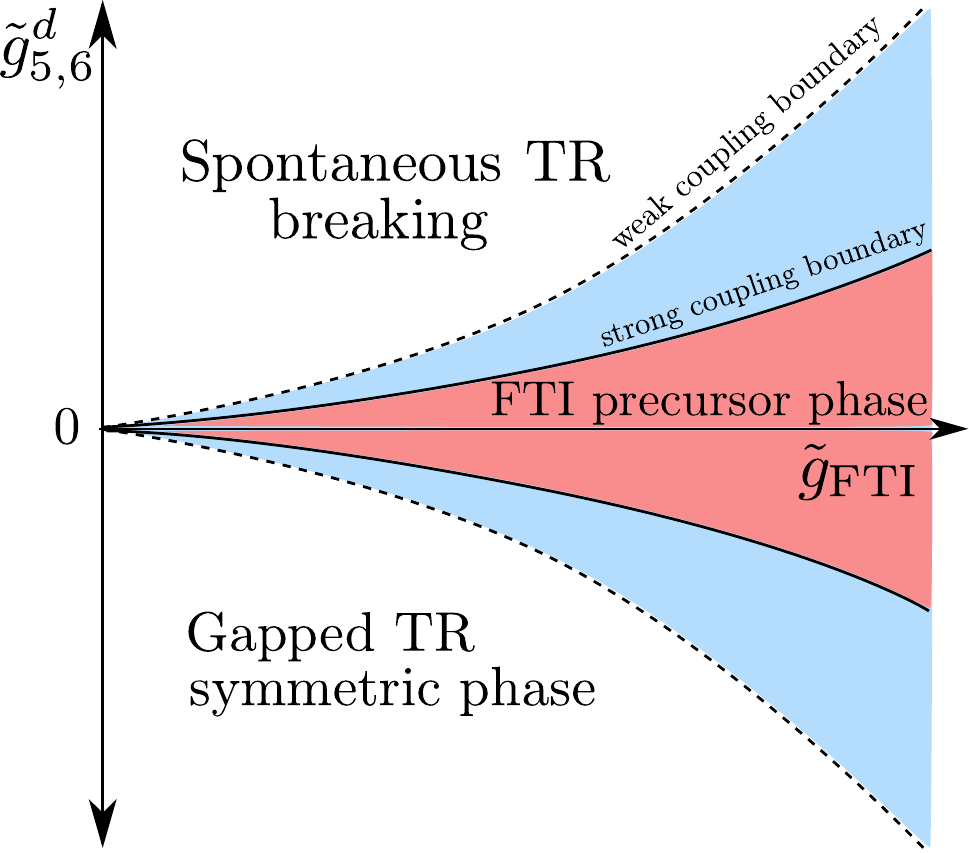} 
	\caption{(Color online) Sketch of the phase diagram
	for the SO ladder with interspin interactions, including the strong coupling analysis.
	In the strong coupling phase the most relevant operators that survive are $\mathcal{O}_{5,{\rm proj}}^d$ and $\mathcal{O}_{6,{\rm proj}}^d$, which induce quasiparticle backscattering in each effective 
	spin layer in the FTI phase. These backscattering terms open a gap in the spectrum of the low energy modes. To consider a FTI precursor, the ratio 
	$\frac{m_{\rm low}}{m_{\rm FTI}}$ between the backscattering induced gap to the FTI induced gap has to be small. For 
	$\frac{m_{\rm low}}{m_{\rm FTI}}\sim 0.1$, $K_{-}\sim 1$ and $\Delta_{6,{\rm proj}}^d\sim 1.8$, the different phase boundaries for weak and strong coupling are represented here by the light blue and red regions.}\label{fig:Coupled_spin_2}
\end{figure}

The coupling strength $\tilde{g}_{6}^d$ at scale $\ell_1$ is in turn related to the bare coupling $\tilde{g}_{6}^{d*}$ via the weak coupling RG flow
\begin{equation}
 \tilde{g}_{6}^d(\ell_1)\sim\tilde{g}_{6}^{d*}(\tilde{g}_{\rm FTI}^*)^\frac{2-\Delta_{6}^d}{\Delta_{\rm FTI}-2}.
\end{equation}
Using these two previous relations, we find the ratio between the gap of the FTI precursor to the gap induced by $\mathcal{O}_{6,{\rm proj}}^d$ to be
\begin{equation}\label{eq:gapratio}
 \frac{m_{\rm low}}{m_{\rm FTI}}\sim\sqrt{2\pi K_-}(\tilde{g}_{6}^{d*})^{\frac{1}{2-\Delta^d_{6,\rm proj}}}(\tilde{g}_{\rm FTI}^{*})^\frac{2-\Delta^d_{6}}{(\Delta_{\rm FTI}-2)(2-\Delta^d_{6,{\rm proj}})}.
\end{equation}
As long as the ratio $\frac{m_{\rm low}}{m_{\rm FTI}}\ll 1$, it is sensible to talk about an FTI precursor state. A qualitative diagram
including both the weak coupling 
and the strong coupling analysis is given in Fig. \ref{fig:Coupled_spin_2}.
 
\section{Perturbation away from the inversion symmetric point}\label{sec:alpha}

\begin{figure}[ht!]
	\centering
		\includegraphics[width=\linewidth]{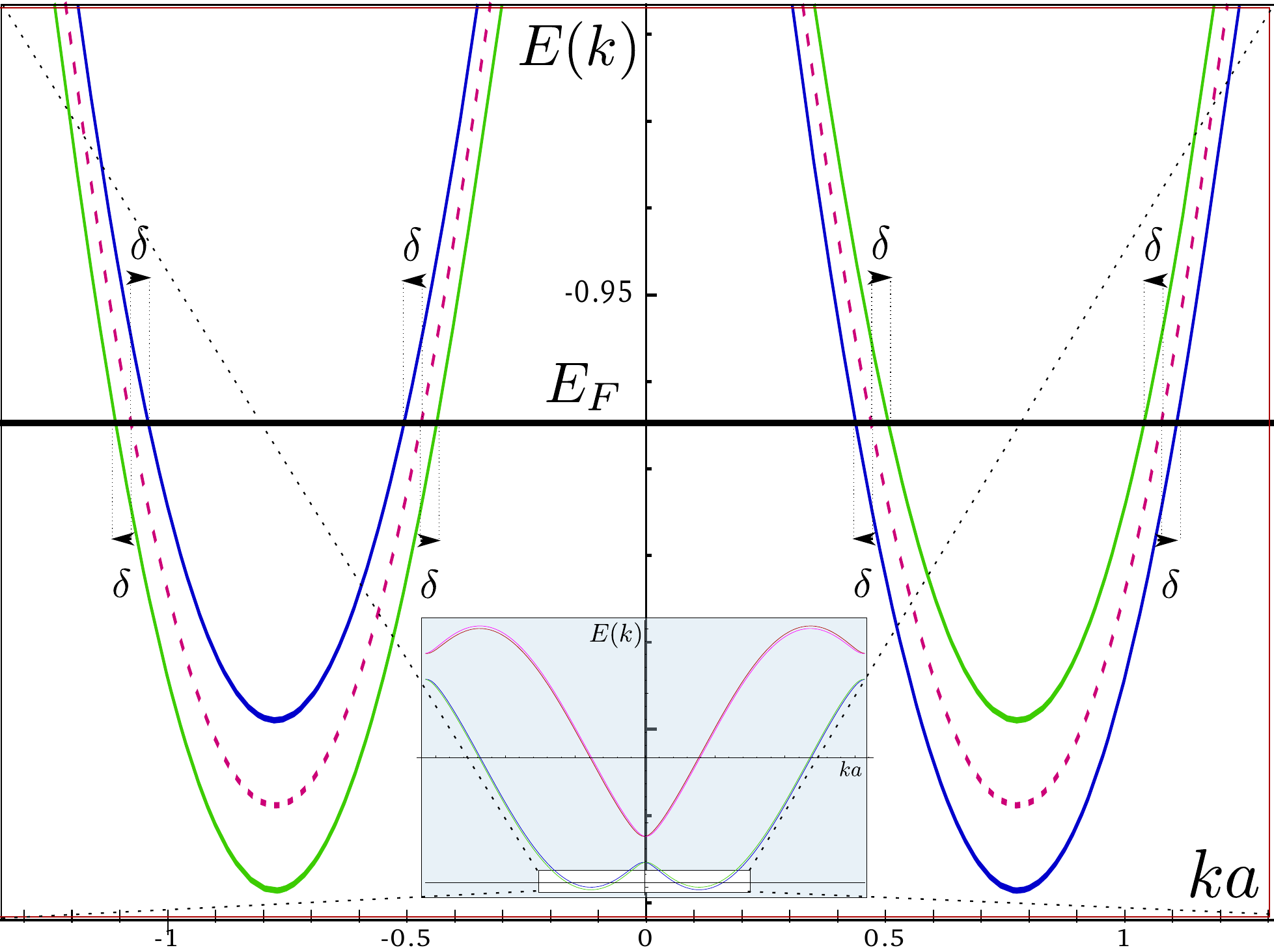}
	\caption{(Color online) Change in
	the single particle spectrum between $\alpha_{\rm so}=0$ (dashed curve) and $\alpha_{\rm so}/t=0.01 $ (solid curves) around the Fermi energy $E_F$ (horizontal black line).
	Here we plot the bottom of the lower bands of the single particle spectrum as a function of momentum $ka$. The displacement of the Fermi momenta between the
	two cases is denoted by $\delta$. Inset: Full band spectrum.}\label{fig:single_part_alpha}
\end{figure}

So far we have ignored the effect of SO coupling that breaks $S^z$ symmetry. We analyze the consequences of including such process in this section.
For small SO coupling
\begin{equation}
 \left|\frac{\alpha_{\rm so}}{t(\sin\frac{\Phi}{2})^2}\right|\ll 1
\end{equation}
the single particle band structure is only slightly modified compared to the case of vanishing $\alpha_{\rm so}$. This modification is sketched in Fig. \ref{fig:single_part_alpha}.
By a direct computation, we find that the original Fermi momenta 
$\bm{k}_F=(k_{F,\uparrow,2}^L,k_{F,\uparrow,2}^R,k_{F,\uparrow,1}^L,k_{F,\uparrow,1}^R,k_{F,\downarrow,2}^L,k_{F,\downarrow,2}^R,k_{F,\downarrow,1}^L,k_{F,\downarrow,1}^R)$ 
change from
\begin{equation}
 \bm{k}_F^0=\frac{\Phi}{3a}(-2, -1,1,2,-2,-1,1,2)
\end{equation}
for $\alpha_{\rm so}=0$ to 
\begin{equation}
 \bm{k}_F^{\alpha\neq0}= \bm{k}_F^0+\delta\left(1, -1,-1,1,-1,1,1,-1\right),
\end{equation}
with $\delta=\frac{\alpha_{\rm so}}{\sin\frac{\Phi}{6}}$, for non vanishing perpendicular SO coupling. This implies that the operator that induces the FTI precursor
still conserves momentum in the presence of small $\alpha_{\rm so}$ SO coupling for the exact same effective 1/3 per helicity. A non-zero $\alpha_{\rm so}$ nevertheless 
breaks inversion symmetry, so we cannot obtain closed expressions for the Luttinger parameters. For $\alpha_{\rm so}\neq0$, it is always possible to fix the Fermi energy to
satisfy the requirement of momentum conservation that gives rise to the FTI operator, as long as the combinations of $\tilde{\alpha}=\alpha_{\rm so}/t$, the Fermi energy 
$\tilde{E}_F=E_F/t$, the interleg tunnelling $\tilde{t}_\perp=t_\perp/t$ given by
\begin{eqnarray}
 A_1=4\tilde{E}_F\cos\Phi/2,\quad A_2=4\tilde{\alpha}\sin\Phi/2,\\
 B=4(\tilde{E}_F^2-\tilde{\alpha}^2-\tilde{t}_\perp^2)+2\cos\Phi.
\end{eqnarray}
reside inside the simplex shown in Fig. \ref{fig:alpha_surface}. A detailed derivation of this is given in the Appendix \ref{app:simplex}.
\begin{figure}[ht!]
	\centering
	\includegraphics[width=\linewidth]{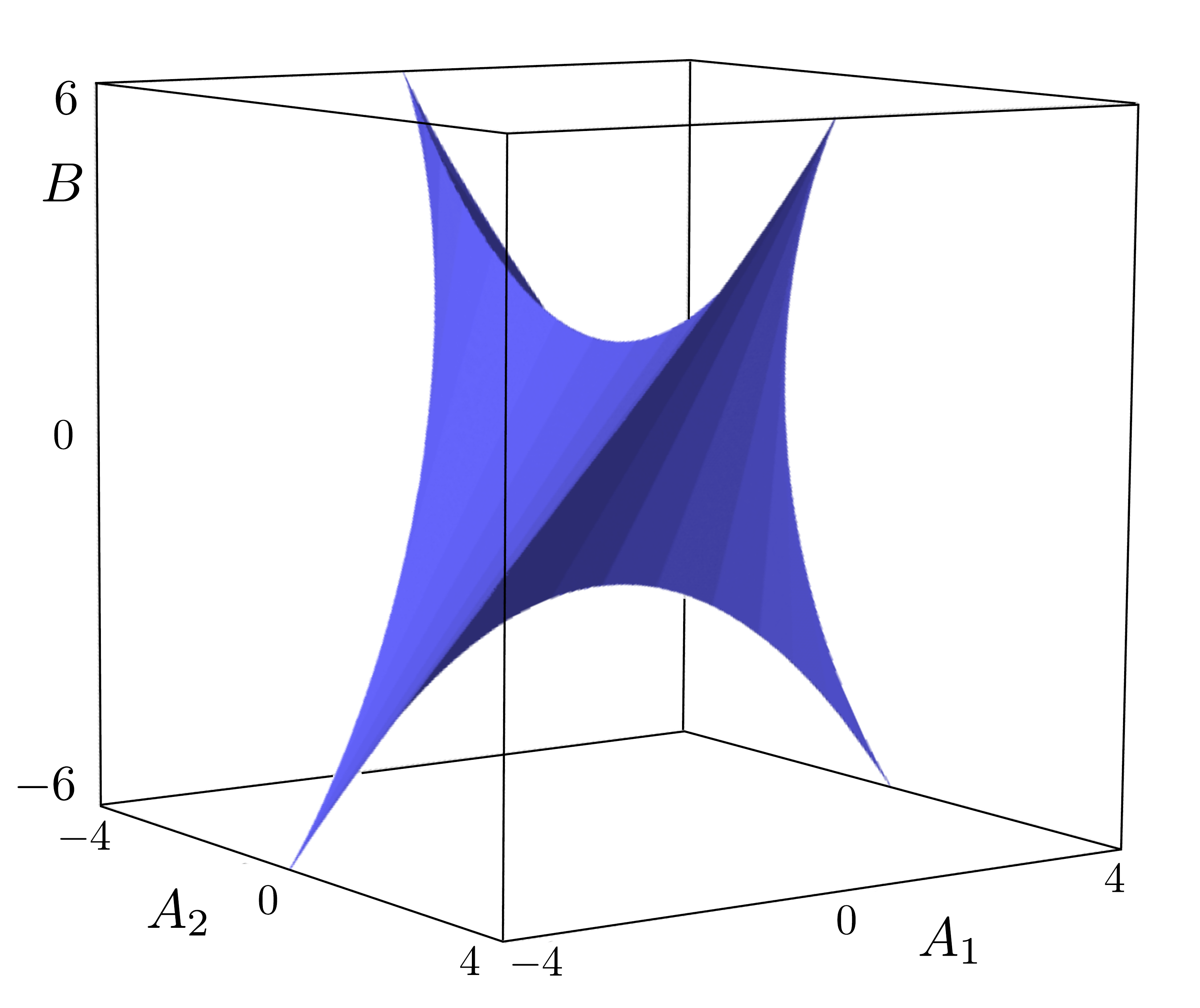}
	\caption{(Color online) Simplex-like volume in the parameter space where the FTI operator can exist. If the parameters $t,\alpha_{\rm so},t_\perp$ and $\Phi$
	reside inside the region, it is always possible to set the Fermi energy such that the FTI operator conserves momentum.}\label{fig:alpha_surface}
\end{figure}

\section{TR breaking external perturbations}\label{sec:TRbreaking}

Having established the existence of the FTI precursor phase, we can consider its stability against TR symmetry breaking perturbations. The findings of 
Ref.~\onlinecite{Beri2012} based on a phenomenological FTI edge model suggest that  the system may display a degree of robustness against weak TR symmetry breaking, and 
that moderately strong TR breaking perturbations may be used to probe the FTI phase. Here we show when such robustness may arise in terms of microscopic interactions, and 
suggest a quantised signature of the FTI precursor. The physical origin of TR breaking depends on the particular realization of the system. In solid state realizations, 
it may correspond to a Zeeman field (due to external magnetic field or arising, e.g., from coupling to a ferromagnet), while in cold atomic realizations where TR symmetry is 
synthetic (e.g., is based on conditions  on the optical coupling \cite{Goldman2010,Juzeliunas2010,Lin2011,Beri2011,Mei2012,Galitski2013,Aidelsburger2013,Grusdt2017})
it may arise from the appropriate detuning from the TR symmetric point. 

\begin{figure}[ht!]
	\centering
		\includegraphics[width=\linewidth]{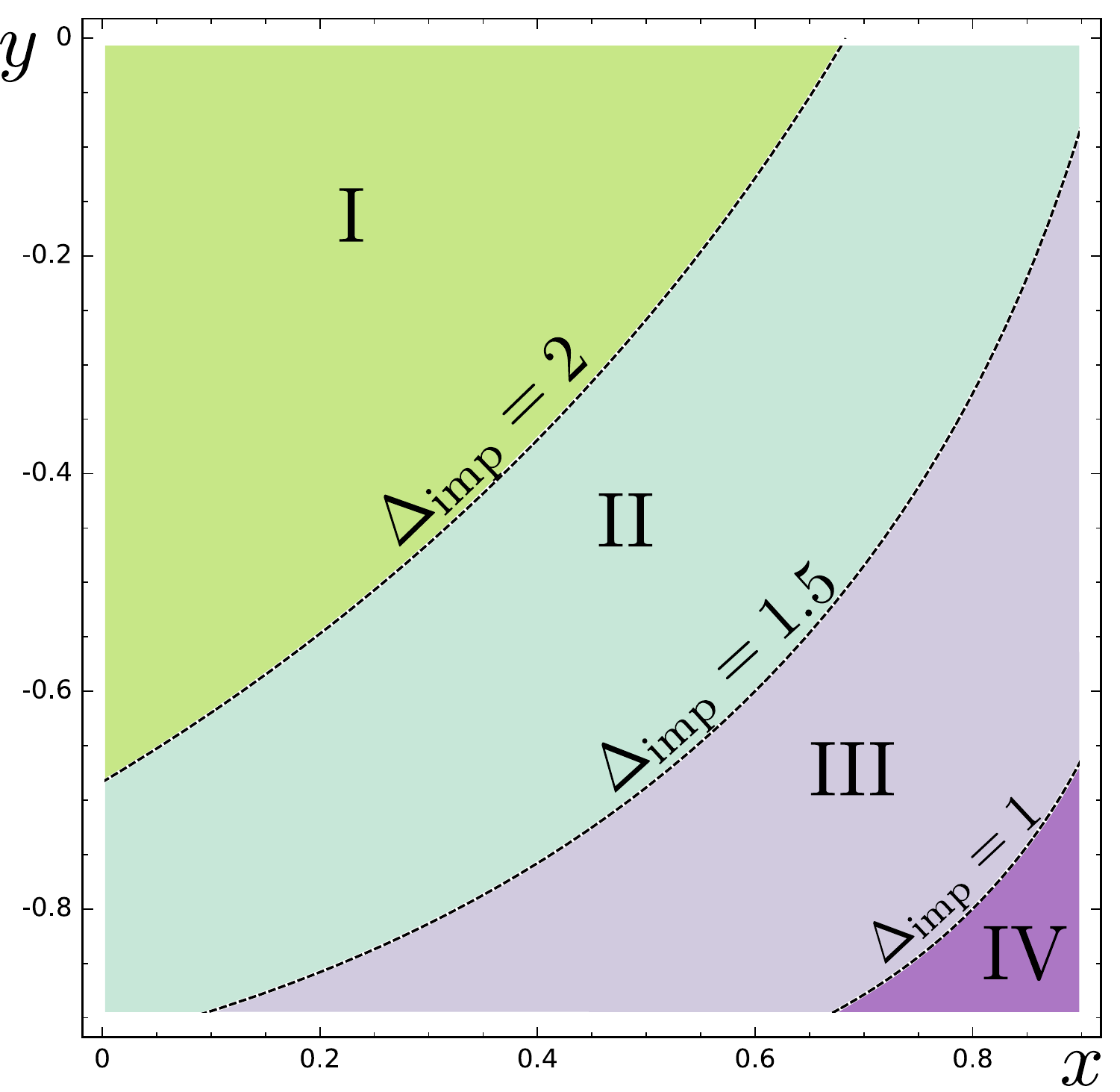}
	\caption{(Color online) Scaling dimension
	of the magnetic impurity operator $(\psi_{e,\sigma}^R)^\dagger\psi_{e,\bar{\sigma}}^L$.
	The Luttinger parameter $K$ and $x,y$ are given by Eq. (\ref{Luttinger_strong}).
	For vanishing interspin interaction $x=y=0$, the scaling dimension $\Delta_{\rm imp}\geq3$. For sufficiently strong interspin interaction of the impurity may become RG relevant.
	In region I the TR breaking term is irrelevant. In region II an extended TR breaking term is relevant. In region III, TR breaking 
	random disorder	and an extended TR breaking term are relevant in RG sense. In region IV all forms of TR breaking terms that involve
	the process of backscattering between Kramers pairs are relevant 
        This figure corresponds to the case $K=1$.}\label{fig:scalin}
\end{figure}

A simple example of TR symmetry breaking is that of an impurity that allows for the hybridization of  Kramers pairs. The perturbation that couples Kramers pairs corresponds 
to the backscattering of electrons at one ``edge" of the FTI precursor. The electron operator of chirality $\eta$ and spin $\sigma$ in the FTI precursor phase corresponds to 
\begin{equation}
 \psi_{e,\sigma}^\eta=e^{i\eta\sqrt{4\pi}\tilde{\phi}_{\eta\sigma}}.
\end{equation}
The backscattering between Kramers pairs thus corresponds to
\begin{equation}\label{eq:TRbreak_scatt}
(\psi_{e,\sigma}^R)^\dagger\psi_{e,\bar{\sigma}}^L=e^{-i\sqrt{4\pi}(\tilde{\phi}_{R\sigma}+\tilde{\phi}_{L\bar{\sigma}})}.
\end{equation}
This operator has scaling dimension 
\begin{equation}
\Delta_{\rm imp}=\Delta[(\psi_{e,\sigma}^R)^\dagger\psi_{e,\bar{\sigma}}^L]=\frac{3}{2}\left(\frac{1}{K_+}+{K_-}\right).
\end{equation}
The Luttinger parameters $K_{\pm}$ are defined in Eq.~\eqref{Luttinger_strong}. For a single 
impurity, the first order RG equation for the backscattering coupling constant is \cite{GiamarchiBook2003}
\begin{equation}
 \frac{dg_{\rm imp}}{d\ell}=(1-\Delta_{\rm imp})g_{\rm imp}.
\end{equation}
As shown in Fig. \ref{fig:scalin}, for a considerable part of the FTI phase we have $\Delta_{\rm imp}>1$ which means that the FTI precursor can be made robust against 
such a TR breaking perturbation. In particular, for vanishing interspin coupling, this operator is irrelevant for all values of interaction, as $K_\pm=K$ and 
$\frac{3}{2}(K+K^{-1})\geq 3$. Analogous robustness against perturbations seemingly at odds with a topological phase has also been noticed for strongly interacting 
integer topological insulator edge modes \cite{Santos2015b}. 
Including the interspin interaction so that $K_\pm$ are split, 
the magnetic impurity operator can become relevant. The required values of $x$ and $y$ in Eq.~\eqref{Luttinger_strong}, however correspond to  
strong interspin interactions. 

Spatially extended forms of TR breaking perturbations can also be considered. In this case, the RG equation is analogous to Eq.~\eqref{RG_eq}, and irrelevancy in the RG 
sense requires $\Delta_{\rm imp}>2$. A similar equation also holds for the case of an spatially extended region with magnetic impurities of random coupling strength; RG irrelevancy in this case requires $\Delta_{\rm imp}>3/2$.   These, more stringent, criteria can also be satisfied in a nonvanishing part of the FTI precursor phase, as shown in Fig.~\ref{fig:scalin}. 

\subsection{TR symmetry-breaking-based Thouless pump}

While the presence of a large region with $\Delta_{\rm imp}<2$ may seem as a shortcoming, its existence can be exploited to obtain quantized signatures of 
the FTI precursor, as we now discuss. Our suggestion is based on the observation of Ref.~\onlinecite{Beri2012} that when a TR breaking perturbation like
Eq.~\eqref{eq:TRbreak_scatt} gaps an FTI edge over a spatially extended region, gradually shifting the phase of the coupling by $\pi$ to rotate it from 
a starting configuration to its time-reversed conjugate, $\frac{1}{2}\nu$ charge is pumped between the ends of the rotated perturbed region \cite{Qi2008}.
Similar fractional pump mechanism has been described in ultracold atoms subject to synthetic gauge in synthetic dimensions \cite{Taddia2017}.

In the case of the FTI precursor, we find that due to the quasi-one 
dimensional nature of the system and the existence of the FTI precursor phase, the process that can gap the edge appears at third order in perturbation theory.
By controlling the phase in the coupling between the different spin projections, it is possible to induce a charge pumping of $\frac{1}{6}$, 
as in the edge of a true 2D FTI.

It must be noted that we do not assume in the following that the interspin interactions vanish, instead we assume that the FTI precursor phase exists and the FTI gap
is the largest scale, and that the perturbations considered in this section have sufficiently large amplitudes so that they control the gap of the FTI edge modes. This 
allows for small interspin interactions (in a similar sense to that in Sec.~\ref{sec:SCInterspin_int}) to be still present.

Specifically, we introduce two microscopic perturbations corresponding to same spin and interspin processes. The same-spin perturbation that we consider,
 \begin{equation}
   H_N=\sum_{i,\sigma} N_i{c}_{i\sigma}^{I\dagger}{c}_{i\sigma}^I,
 \end{equation}
preserves time reversal symmetry. This perturbation corresponds to a modulation of the density along leg $I$ and generates backscattering between modes within the same 
spin projection on that same leg. 
The interspin perturbation that we consider, 
\begin{equation}\label{TRbreaking}
 H_M=\sum_i \left(M_i (e^{i\chi}{c}_{i\uparrow}^{I,\dagger}{c}_{i\downarrow}^I+{c}_{i\uparrow}^{II,\dagger}{c}_{i\downarrow}^{II})+\text{h.c.}\right),
\end{equation}
explicitly breaks TR symmetry by effectively implementing Zeeman terms of magnitude $M_i$ and along the $\sigma_1$ direction for leg $II$ and in the $\sigma_1-\sigma_2$ 
plane in an angle set by $\chi$ for leg $I$. In what follows, we assume
that $N_i$ and $M_i$ extend over a length $L_{\rm TR}$ and that
the strength of the backscattering potential $N_i$ is larger than that of the TR breaking terms $M_i$.  

After diagonalization of the single particle Hamiltonian, projecting to the lower band, and discarding the $H_{N,M}$ induced forward scattering terms 
as they do not open a gap, we just consider the Fourier components at $ka=\frac{n\Phi}{3},(n=1,4)$ of the potentials $N(x)$,$M(x)$ in the continuum  which provide the 
momentum necessary to backscatter the low energy modes. Using the bosonized expressions for the fermionic operators at the Fermi points, the potential term becomes
\begin{equation}\label{eq:HNbackscatt}
  H_N=\sum_{\beta\beta'\sigma}\int dx |n_{\beta\beta'}^\sigma|\cos(\sqrt{4\pi}(\phi_{\sigma,\beta,R}+\phi_{\sigma,\beta',L})),
\end{equation}
where $n^\sigma_{\beta\beta'}=u^R_{\sigma\beta}u^L_{\sigma\beta'}\int e^{i(k^R_{F,\beta}-k^L_{F,\beta'}) x}\frac{N(x)}{2\pi a}dx$. The Zeeman terms become
\begin{equation}\label{eq:HMbackscatt}
 H_M=\sum_{\beta\beta'\sigma}\int dx |m^{\sigma}_{\beta\beta'}|\cos(\sqrt{4\pi}(\phi_{\sigma,\beta,R}+\phi_{\bar{\sigma},\beta',L})+\tilde{\chi}^\sigma_{\beta\beta'}),
\end{equation}
where the amplitude is, assuming for simplicity that the potential $M(x)$ is an even function  
\begin{equation}
m^{\sigma}_{\beta\beta'}=(u^R_{\sigma\beta}u^L_{\bar{\sigma}\beta'}e^{i\chi}+v^R_{\sigma\beta}v^L_{\bar{\sigma}\beta'})\int e^{i(k^R_{F,\beta}-k^L_{F,\beta'}) x}\frac{M(x)}{2\pi a}dx.  
\end{equation}
and $\tilde{\chi}^\sigma_{\beta\beta'}={\rm Arg}(m^\sigma_{\beta\beta'})$. The tensors $u$ and $v$ are defined in Eq.~\eqref{eq:uvtensor}. As seen from Eqs.~\eqref{eq:HNbackscatt} and \eqref{eq:HMbackscatt}
these backscattering perturbations can be written as
$H_a=H_a^\parallel+H_a^{\perp,2}+H_a^{\perp,4}$ where $a= N,M$.
The perturbations $H_{N,M}^\parallel$ connect Fermi points of different chirality within the same valley, and their amplitudes are controlled by the Fourier components of $N(x)$ and $M(x)$ at $ka=\frac{\Phi}{3}$, while $H_{N,M}^{\perp,n}$ connect opposite chirality states between different valleys and 
exist provided that the Fourier component of the potentials at $ka=\frac{n\Phi}{3}$ does not vanish and that tunnelling between the legs of the ladder is nonzero. Note that due to the nonvanishing tunnelling between the legs of the ladder, even focusing on a particular leg ($I$) produces terms involving the other ($II$), but with 
parametrically small strength.

In the FTI precursor phase, these perturbations are suppressed by the existence of the FTI gap $m_{\rm FTI}$, as they generate bulk excitations. This occurs in a way similar 
to the process (\ref{H_2_strong}). Performing the projection of high energy states, all the backscattering processes do not contribute at first or second order.  
At third order, the only processes that does not vanish after projecting out the high energy degrees of freedom are 
\begin{equation}\nonumber
H_{\rm MG}=\sum_\sigma g_{\rm MG}^\sigma\int dx \cos(\sqrt{4\pi}(\tilde{\phi}_{R,\sigma}+\tilde{\phi}_{L,\bar{\sigma}})+\tilde{\chi}^\sigma_{21})
\end{equation}
where $g_{\rm MG}\sim\frac{a^2|n^\sigma_{22} n^{\tilde{\sigma}}_{11} m^\sigma_{21}|}{m_{\rm FTI}^2}+O(|m|^2)$. Note that the main contribution to this effective amplitude is 
first order in the TR breaking process. In the region where this term is relevant (regions II, III and IV in Fig. \ref{fig:scalin}) and/or has sufficiently large coupling 
$g_{\rm MG}^\sigma$ compared to $g_6^d$ to control the gap of the FTI edge modes, it will also control the ($\tilde{\chi}^\sigma_{21}$ dependent) value to where of the 
field combination $\tilde{\phi}_{R,\sigma}+\tilde{\phi}_{L,\tilde{\sigma}}$ is locked.

\subsubsection*{Pumping protocol}

Once the edge has been gapped according the procedure described above, it is possible to manipulate the configuration to create a domain wall trapping $1/6$ charge (i.e., particle density integrated across the domain wall). 

\begin{figure}[ht!]
	\centering
		\includegraphics[width=0.8\linewidth]{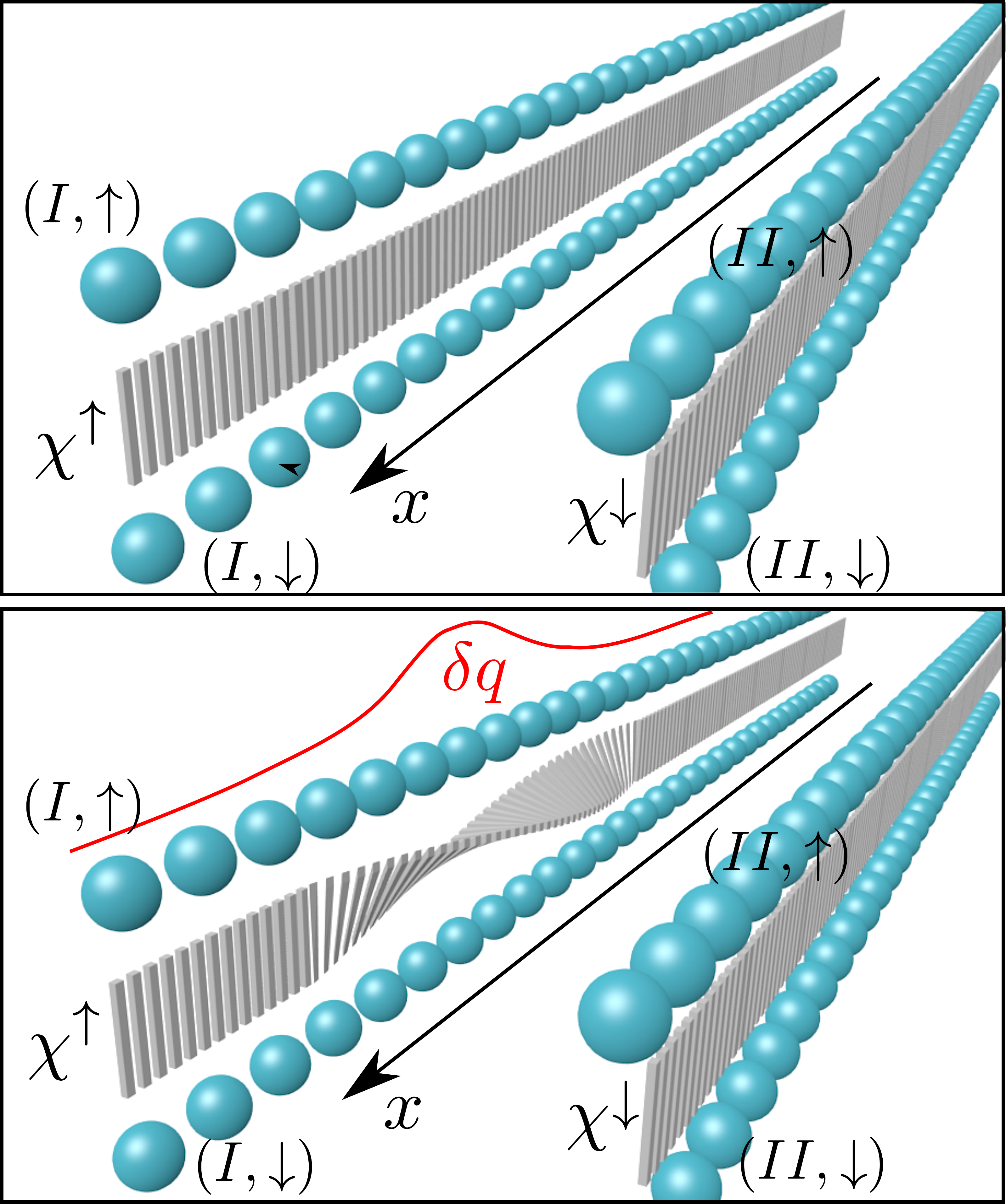}
	\caption{(Color online) A TR breaking perturbation
	can gap out a pair of edge modes. This mechanism locks the field $\tilde{\theta}_\uparrow$ to a value tracking the  
	angle $\chi$ of the rotating Zeeman term (illustrated by the ribbon of vertical bars) between the fermions to the left of the diagram. Upper box: initial gapping configuration, corresponding to a constant $\chi$
	throughout the edge. Lower box: adiabatically changing the value of $\chi$ within a sector of the gapped edge produces 
	a domain wall, represented by the twisted ribbon. A domain wall between TR conjugate configurations has $1/6$ fractional charge, corresponding to the excess accumulated particle density depicted in red.}\label{fig:3Dladder}
\end{figure}

\begin{figure}[ht!]
	\centering
		\includegraphics[width=\linewidth]{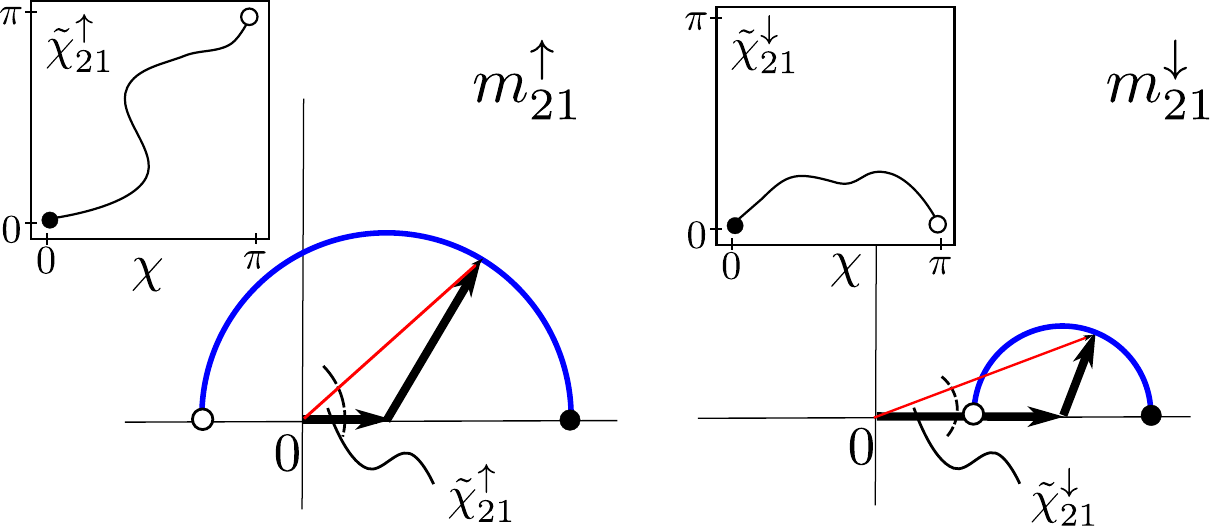}
	\caption{(Color online) An adiabatic change
	of the microscopic phase $\chi$ from $0$ to $\pi$ advances only one of the phases $\tilde{\chi}_{21}^\sigma$, and it does so by the same amount. 
	This occurs because the complex number $m_{21}^\uparrow$ encircles the origin during the adiabatic change (left diagram), while $m_{21}^\downarrow$
	does not (right diagram). In the left (right) diagram the black horizontal arrow represents $v^R_{\uparrow\beta}v^L_{\bar{\downarrow}\beta}$ 
	$v^R_{\sigma\beta}v^L_{\bar{\sigma}\beta}$ while
	the other arrow that traverses the semi-circumference is $u^R_{\sigma\beta}u^L_{\bar{\sigma}\beta}e^{i\chi}$.}\label{fig:argument}
\end{figure}

The edge mode combination $\tilde{\theta}_\sigma=\tilde{\phi}_{R,\sigma}+\tilde{\phi}_{L,\bar{\sigma}}$ is locked into a minimum that depends on the angle 
$\tilde{\chi}^{\sigma}_{21}$ throughout the region of length $L_{\rm TR}$  (that can correspond to the whole length of the ladder as well). This angle can in principle
be manipulated by rotating the leg $I$ Zeeman term in Eq.~\eqref{TRbreaking} to change the parameter $\chi$ . By adiabatically advancing the angle $\chi$ (see also Fig. \ref{fig:3Dladder}) to its TR conjugate value $\chi+\pi$  in  a segment of size $L_{\rm DW}$ within the region of length $L_{\rm TR}$, 
two domain walls are created separated by a distance $\sim L_{\rm DW}$. We note that, although $\chi$ governs both $\tilde{\chi}^{\uparrow}_{21}$ and $\tilde{\chi}^{\downarrow}_{21}$, and hence the locking value of $\tilde{\theta}_\uparrow$ and $\tilde{\theta}_\downarrow$ characterizing opposite edges, advancing $\chi$ by $\pi$ has a topologically distinct effect on
$\tilde{\chi}^{\uparrow}_{21}$ and $\tilde{\chi}^{\downarrow}_{21}$, provided the interleg tunnelling is sufficiently small.  In this case, while the complex number $m^\uparrow_{21}$ encircles the origin in the complex plane (Fig. \ref{fig:argument} left) thus advancing $\tilde{\chi}^\uparrow_{21}$ by $\pi$ as well, the complex number $m^\downarrow_{21}$ does not enclose the origin (Fig. \ref{fig:argument} right) and thus the phase of $\tilde{\theta}_\downarrow$ returns to its original value. As a result, in terms of the FTI edges, the domain walls arise only in one of these, the other one returns to being uniformly gapped along the length $L_{\rm TR}$. The appealing feature of being able to advance $\tilde{\chi}^\uparrow_{21}$ only with our protocol is a physically intuitive consequence of the rotating part of the microscopic perturbation Eq.~\eqref{TRbreaking} being concentrated in one leg, and that of the tunnelling being small.

The charge accumulated in the domain walls is conveniently obtained using the bosonization language. Recalling that in the FTI phase the charge density per edge is given by 
\begin{equation}
 \rho_{\rm edge}=-\frac{1}{3\sqrt{\pi}}(\partial_x\tilde{\phi}_{L\uparrow}+\partial_x\tilde{\phi}_{R\downarrow}),
\end{equation}
the accumulated charge across a domain wall, is given by 
\begin{equation}
 \delta q=\int_{\substack{{\rm dom.}\\ {\rm wall}}}\!\!\!\!\rho_{\rm edge}=-\left.\frac{1}{3\sqrt{\pi}}(\tilde{\phi}_{L\uparrow}+\tilde{\phi}_{R\downarrow})\right|_{\substack{{\rm dom.}\\ {\rm wall}}}=\pm\frac{1}{6},
\end{equation}
as the locking values for the field combination $\sqrt{4\pi}(\tilde{\phi}_{L\uparrow}+\tilde{\phi}_{R\downarrow})$ at the two sides of a domain wall differ by $\pm\pi$. Our protocol thus pumps charge $1/6$ between the two domain walls per half-cycle (defined such that a full cycle corresponds to $\chi\rightarrow \chi+2\pi$, i.e., returning the Zeeman fields to the original configuration).  

Although this signature of the topological phase can be seen in principle, clearly the quasi-one dimensional nature of the system conspires against the existence of a truly topological ordered state. In the next section we provide some arguments towards the stabilization of the true topological system by coupling FTI precursor states.

\section{Extension towards a 2D system}\label{sec:2Dsyst}

As we have seen, a key process competing against the emergence of the FTI precursor is the backscattering between ``opposite" FTI edges. One may hope that upon extending 
the ladder towards a 2D system,  such processes may be suppressed. By considering a multileg ladder system consisting of several 
FTI precursors coupled together, we show that this is indeed the case. While a microscopic description for such multileg ladders is beyond the scope of this work, we will, 
in the spirit of the coupled wire constructions \cite{Kane2002,Teo2014}, show that if a process between neighboring ladders can be generated that dominates over the 
intra-ladder quasiparticle backscattering and suitably gaps the neighbouring ``edge" modes, the quasiparticle backscattering that survives is exponentially suppressed 
in the number of ladders. 

\begin{figure}[b!]
	\centering
		\includegraphics[width=1\linewidth]{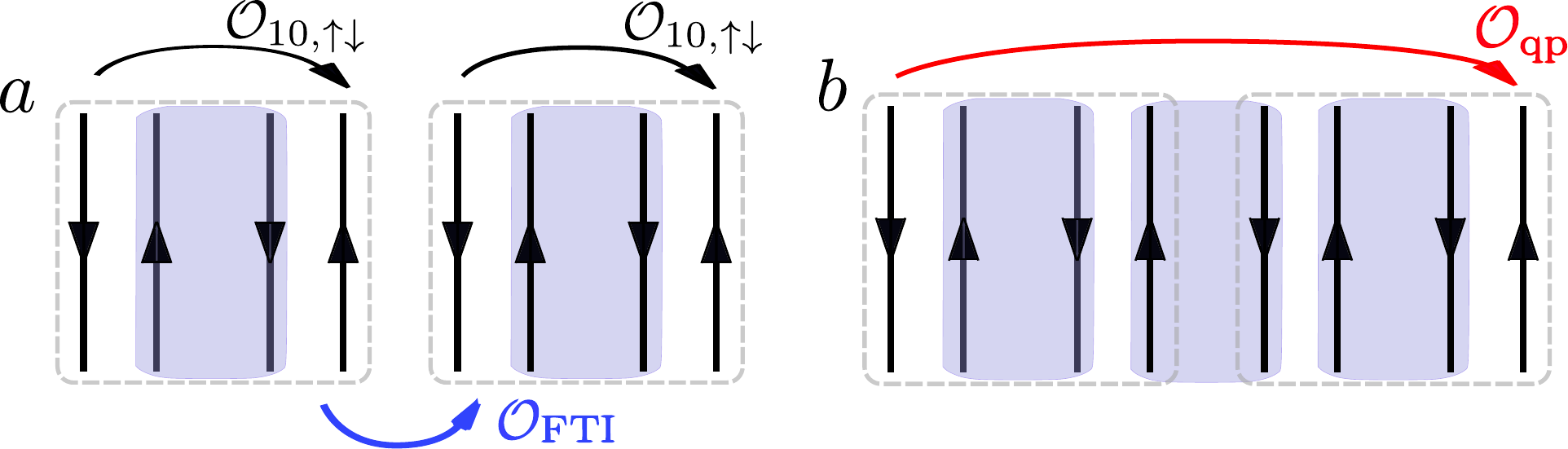}
	\caption{(Color online) Coupling scheme towards a two dimensional realization. $a$.- The coupling two FTI precursors, such that their closest edges are gapped
	by an FTI gap opening term $\mathcal{O}^\sigma_{\rm FTI}$, still contains residual terms of $\mathcal{O}_{6}^d$, $b$.- Once the gap between the 
	two FTI copies of the FTI precursor is opened, the residual terms become quasiparticle tunnelling operators between the different edges, but suppressed by an
	exponential factor in the number of copies of the FTI precursors involved in the construction.}\label{fig:two_dim}
\end{figure}

We start by considering two copies of the FTI precursors,  labelled I and II (see Fig. \ref{fig:two_dim}).  
We assume that an inter-ladder process can be generated that pins
the combination of fields 
\begin{equation}
 \theta_{{\rm link},\sigma}=\tilde{\phi}_{R\sigma,{\rm I}}+\tilde{\phi}_{L\sigma,{\rm II}}. 
\end{equation}
The conjugate field to $\theta_{{\rm link},\sigma}$ is given by 
\begin{equation}
 \varphi_{{\rm link},\sigma}=\tilde{\phi}_{R\sigma,{\rm I}}-\tilde{\phi}_{L\sigma,{\rm II}}.
\end{equation}
Deep in the gapped phase as described in the strong coupling section, a quasiparticle can tunnel from one edge of the FTI precursor to the other via the process 
$\mathcal{O}_{\rm 6, proj}^d$. Once the gap is opened between the two FTI precursor copies forming a larger correlated state, we have to project out 
the high energy degrees of freedom of the low energy theory. To do so, we first write the backscattering operators in terms of the massive degrees of freedom
\begin{eqnarray}\nonumber
 \mathcal{O}_{\rm 6, proj}^{d,{\rm I}}&=&\left(e^{i\frac{\sqrt{\pi}}{3}(\theta_{{\rm link},\uparrow}+\varphi_{{\rm link},\uparrow}+\theta_{{\rm link},\downarrow}-\varphi_{{\rm link},\downarrow})}\right.\\
  &\times&\left.{e^{i\frac{\sqrt{4\pi}}{3}(\tilde{\phi}_{L\uparrow \rm I}+\tilde{\phi}_{R\downarrow \rm I})}}+\text{h.c.}\right),
\end{eqnarray}
and
\begin{eqnarray}\nonumber
 \mathcal{O}_{\rm 6, proj}^{d,{\rm II}}&=&\left(e^{i\frac{\sqrt{\pi}}{3}(\theta_{{\rm link},\downarrow}+\varphi_{{\rm link},\downarrow}+\theta_{{\rm link},\uparrow}-\varphi_{{\rm link},\uparrow})}\right.\\
  &\times&\left.e^{i\frac{\sqrt{4\pi}}{3}(\tilde{\phi}_{L\downarrow \rm II}+\tilde{\phi}_{R\uparrow \rm II})}+\text{h.c}\right).
\end{eqnarray}

These processes now create high energy excitations involving solitons in the FTI precursor phase, which have energy of the order of the gap. By projecting out these 
high energy states, the operators above do not contribute to first order in the interspin interaction coupling $\tilde{g}_{6}^d$. In second order perturbation theory
the combination $\mathcal{O}_{\rm 6, proj}^{d,{\rm I}}\mathcal{O}_{\rm 6, proj}^{d,{\rm II}}$ creates processes that survive the high energy projection,
in particular
\begin{eqnarray}\nonumber
[\mathcal{O}^{d,{\rm I}}_{6}\mathcal{O}^{d,{\rm II}}_6]^{\rm proj}&=&\left(
e^{i\frac{\sqrt{4\pi}}{3}(\tilde{\phi}_{R\downarrow \rm I}+\tilde{\phi}_{L\downarrow \rm II}+\tilde{\phi}_{L\uparrow \rm I}+\tilde{\phi}_{R\uparrow \rm II})}+\text{h.c.}\right).
\end{eqnarray}
This projected operator corresponds to correlated quasiparticle tunnelling between the edges of the extended system. The prefactor of this operator is
$g^{(2)}_{\rm qp}\sim (\tilde{g}_{6}^d)^2/m(\ell_1)$.

Following the same procedure, we find that in the case of $N$ copies of the FTI precursors, the coupling of the backscattering operators $g_{\rm qp}^{(N)}$ scales as
\begin{equation}
 g_{\rm qp}^{(N)}=\tilde{g}_{6}^d\left(\frac{\tilde{g}_{6}^d}{m(\ell_1)}\right)^N,
\end{equation}
which decreases exponentially in the transversal size of the 2D system, as expected. This exponential decrease of the coupling with the transversal size of the system
helps to stabilize the FTI phase, as the competing order induced by $\mathcal{O}_{6}^d$ becomes negligible. In terms of the ratio between the gaps generated 
by both terms, we find that $m_{\rm low}/m\rightarrow 0$ as $N$ increases. 

\section{Summary and discussion}
In this work we have analysed the physics of SO ladders focusing on the possibility of creating a precursor of an FTI phase.
Given the quasi one-dimensional nature of the system, we could take an analytical, microscopically motivated approach, providing a complementary perspective to 
numerics on small 2D systems \cite{Neupert2011,Neupert2011b,Regnault2011,Sheng2011,Wu2012,Repellin2014,Fialko2014} 
and phenomenological constructions \cite{Klinovaja2014b,Klinovaja2014,Santos2015}.
We considered a microscopic TR invariant Hamiltonian,
describing a spinful two leg ladder with interactions and SO coupling. The SO coupling generates an effective flux $\pm\Phi$ of opposite sign for opposite spins and we 
focused on the case of a fractional effective filling $\nu = 1/3$ per spin. 

To locate the region of the parameter space where the FTI precursor may appear,   we first investigated the spin-decoupled case, i.e., when the $z$ component of the spin is 
conserved and  interspin interactions vanish. In this case the physics is equivalent to that of interacting spinless fermion ladders in a magnetic flux at $\nu = 1/3$ 
filling, or equivalently, 1D spinful fermions with spin-orbit coupling and Zeeman fields, both systems of considerable interest, in part due to the possibility of Laughlin 
precursor states \cite{Petrescu2015,Cornfeld2015,Calvanese2017,Petrescu2017},
and fractional helical liquids (of potential utility for quantum computation) \cite{Oreg2014},
respectively.

Our results include novel findings pertinent already to these cases, originating in the interrelation of forward and backscattering parameters highlighted by our 
microscopic and symmetry analysis. Using a weak coupling RG approach and considering weak interleg tunnelling $t_\perp$ while allowing also strong microscopic interactions, we found three phases: a fully gapless Luttinger 
liquid with dominant BDW, OAF or triplet superconducting QLRO; a phase induced by correlated density fluctuations in each leg, characterized by an interleg partial gap (the leg analogue of the familiar spin-gap in spinful 1D 
fermion systems) displaying RDW, CDW or singlet superconducting QLRO; and the FTI (Laughlin) precursor phase. 
For the concrete form of the interactions we considered, we could also establish that the repulsive part of the leg-SU($2$) invariant line
leads into the interior of the FTI precursor region. Along this line, due to the aforementioned interrelation, the interleg partial gap, while competing with the FTI phase, is suppressed logarithmically. With suitable interaction anisotropy, a stronger suppression of the interleg-gap can also be achieved, thus eliminating this competition. Further discussion of this spin-decoupled 
case is provided in Sec.~\ref{sec:spindecoupled_discuss}. 
The phases of the spin-decoupled SO ladder are summarized in Fig.~\ref{fig:phases_spindecoupled}. 

With the FTI part of the spin-decoupled phase diagram identified,
we performed a detailed study in this regime, first assessing the stability against weak interspin interactions. 
A weak coupling RG analysis shows that interspin interactions introduce competition against the FTI phase. This competition can result in three outcomes depending on the 
nature of the interspin interaction (attractive or repulsive), and its strength. For sufficiently small interspin interaction, the FTI phase survives. Focusing on the most RG-relevant competition, $\mathcal{O}_{5,6}^d$, we find that beyond a critical value of interspin repulsion the system shows BDW and OAF QLRO order with combined orbital and spin structure that is odd under TR symmetry, indicating the onset of TR symmetry breaking. On the other hand, for interspin attraction beyond a critical value, the system develops BDW and OAF QLRO consistent with TR 
symmetry.  
We also noted that while the effective strength of interspin interactions is influenced directly by the microscopic interaction parameters, the $\mathcal{O}_{5,6}^d$ couplings are suppressed as $O(t_\perp^2)$ for small $t_\perp$, scaling favorably compared to the $O(t_\perp)$ FTI term. This may allow the FTI precursor to survive also for stronger interspin interactions, provided the regime can be reached where $\mathcal{O}_{5,6}^d$ is the only competitor against the FTI precursor phase. 
Thus, while influenced by the presence of interspin interactions, the FTI precursor persists in a significant part of the parameter space. The FTI and adjacent regions of the phase diagram in the presence of interspin interactions are depicted in Fig.~\ref{fig:interspin_ops}.

To obtain a perspective complementary to our weak coupling RG analysis, we have also examined the FTI precursor phase at strong coupling, i.e., its nature and stability 
from a starting point with an FTI partial gap. Here we found that in the absence of interspin interactions, the FTI phase is robust against the perturbation 
of correlated density fluctuations in each leg that promotes the 
interleg-gap. This process generates high-energy anyonic excitations in the bulk of system and renormalizes the gapless FTI edge modes. 
An emergent FTI precursor low energy physics is thus compatible with this competition. 
The presence of interspin interactions generates a process that leads to correlated backscattering of quasiparticles between the gapless edges and this does compete against 
the FTI precursor at strong coupling by promoting a tendency to open a gap for the FTI precursor edge modes. The FTI precursor can be viewed as being present only when this 
edge mode gap is negligible compared to the FTI gap. The resulting strong coupling phase diagram is shown in Fig.~\ref{fig:Coupled_spin_2}.

We have also verified that the FTI precursor is robust against the inclusion of a small SO coupling that does not preserve the $z$ component of the spin. However, 
large such SO couplings eliminate the term driving the system to the FTI phase. Regarding the particle density, while we focused on precisely $\nu=1/3$ filling, 
the FTI precursor is expected to be robust against small deviations from this value up to a commensurate-incommensurate transition \cite{Petrescu2015,GiamarchiBook2003}.

Motivated by the possibility to include  TR breaking perturbations
we have also studied how a fractional Thouless pump may be created. Using a protocol \cite{Qi2008,Beri2012}
based on advancing the orientation of Zeeman-like terms to their TR conjugate configuration in an extended spatial region, we showed that  $\pm1/6$ charge is pumped between 
the corresponding domain walls as in the case of a true 2D FTI. This quantized signal of the topological nature of the FTI precursor state is remarkable in the view that 
owing to the quasi-1D nature  of the system, true topological order is absent,
as indicated by the existence of local order parameters displaying QLRO. 

For the case of nonzero interspin interactions and/or spin-$z$ nonconserving SO coupling, our results complement exact diagonalization numerics on 2D 
FTIs \cite{Repellin2014,Neupert2011}. While the $\nu=2/3$ per spin fermionic systems of Ref.~\onlinecite{Neupert2011}
are more complicated (and expected to be less stable \cite{Levin2009}) than the $\nu=1/3$ Laughlin case we considered, 
the $\nu=1/2$ bosonic Laughlin study of Ref.~\onlinecite{Repellin2014} provides a closer comparison. It finds similar conditions for stability as our results, though with 
tolerance to stronger interspin interactions. 

A closer comparison in this regard requires extrapolating our findings from the quasi-1D to the 2D regime. To this end, we have developed a coupled wire \cite{Kane2002,Teo2014} inspired procedure that provides qualitative insights using our quasi-1D ladders as elementary building blocks for a 2D system. The results from this procedure suggest that the competing processes due to interspin interactions are suppressed exponentially in the number of constituent ladders, indicating that the FTI precursor becomes increasingly robust upon moving towards 2D.  

While in our starting phase diagram (Fig.~\ref{fig:phases_spindecoupled}) we focused on a concrete, up to next-nearest-neighbor, form of interactions, our findings are expected to generalize. In particular, as reaching the required FTI Luttinger parameter regime requires only the long-wavelength Fourier component of the interactions to be strong, an interesting generalization is to allow for less strong interactions but to increase their range. By taking advantage of this freedom, we expect to be able to significantly expand the zoo of models supporting the FTI precursor phase.

Ultracold atomic systems provide a natural platform towards the experimental realization of FTI precursors due to existing schemes for imprinting large synthetic, including TR-invariant, fluxes \cite{Liu2009,Aidelsburger2011,Lin2011,Dalibard2011,Aidelsburger2013,Kennedy2013,Miyake2013,Beeler2013,Galitski2013,Atala2014,Mancini2015,Garcia2015,Stuhl2015,Zhai2015,Li2016,Song2016,Beri2011,Zhu2011,Jiang2011,Mei2012,Wei2012,Liu2012,Cocks2012,Nascimbene2013,Orth2013,Gopa2013,Jotzu2014,Wall2016}
and the control of interactions. Here, challenging aspects include reaching the quantum degenerate regime combining fluxes with strong and/or long-range interactions. Particularly challenging is to generate the requisite repulsion between (the degrees of freedom corresponding to) same spin species  while keeping interspin interactions sufficiently weak (or reaching a regime where their $t_\perp$-based control suffices).  
Dipolar Fermi gases \cite{Lu2012,Burdick2016} may offer a promising starting point as they furnish two of the key ingredients: they experience long-range interactions and they may be subjected to the magnetic field gradient scheme of Refs.~\onlinecite{Aidelsburger2013,Kennedy2013} for imprinting uniform fluxes of opposite sign for internal states with opposite magnetic moment, i.e., the form of the spin-orbit fluxes we focused on.

The FTI precursors studied here may motivate new research on SO ladders and our work will provide useful guidance for such future investigations. A particularly 
interesting next step would be to study the strongly interacting regime from a fully microscopic perspective (e.g., using the density matrix renormalization group), 
which may confirm and refine the conditions we find for stabilizing the Laughlin and FTI precursor states, and demonstrate the quantized pumping signature we predict 
in numerical simulations. 
Investigating further indicators of topology (such as string order parameters \cite{Nijs89,Nakamura2010} or entanglement features \cite{Calabrese_2004,Garcia08,Li08,Thomale10}) in the FTI precursor regime is another interesting future direction. 
Looking ahead, the line of inquiry initiated here, in conjunction with such new studies and the rapid progress in ultracold atom systems, will 
hopefully lead to a clear path towards creating and detecting FTI precursor states, and ultimately 2D FTIs, in experiments.

\acknowledgments
We thank Sam Carr, Nigel Cooper, and Ulrich Schneider for useful discussions. This research was supported by a Birmingham Research Fellowship, the Royal Society, the EPSRC grant 
EP/M02444X/1 and the ERC under the European Union's Horizon 2020 research and innovation programme under grant agreement No. 678795 TopInSy. 

\bibliographystyle{my_apsrev4-1}
\bibliography{biblio_prx}

\begin{appendix}

 \widetext
 \section{Integration of high-energy band modes}\label{app:Integrating_out}
 
 The single particle energy bands
 $E^{+}_\pm(\tilde{k})$ of Eq. \eqref{Sing_part_dispersion} are separated from the lower energy bands $E^{-}_\pm(\tilde{k})$ by an energy
 gap of the order of the interleg tunnelling strength $t_\perp$. As we work at small fillings, such that the upper bands are completely empty,
 we consider just the states related with the lower bands, discarding the contributions from higher bands. In a truly two-dimensional scenario,
 this approximation corresponds to a projection into the lowest Landau level.
 
 As we will be interested in small momentum around the Fermi points, we introduce four branches of fermion fields, each one associated with a
 particular Fermi point, and linearise the dispersion relation around these points. These branches correspond to Eq. \eqref{eq:fermion-branchdecomp}.
 These four branches can be thought as capturing the correct degrees of freedom for a small momentum and energy window around the Fermi points and Fermi 
 energy. Nevertheless, note that by modifying the UV theory, these branches can be made to be the exact description for the fermions. 
 
 Concerning the high energy modes of the lower band, it is important to clarify their role in the renormalization of the interaction 
 parameters. In particular we are interested in the FTI term
 which appears by considering second order processes in the interactions.
 Here we explore the effect of higher energy modes in a simplified model where we consider the following expansion of the fermion operator, 
 that contains the previously discussed four branches, and a high energy mode $\psi^+_{i,\sigma}$ describing the states around $k=0$
  \begin{equation}\label{fermion_expansion}
   \begin{pmatrix}c^I_{i,\uparrow} \\
 c^{II}_{i,\uparrow}
\end{pmatrix}=\sum_\eta\begin{pmatrix} \cos\alpha^\eta_2 & - \sin\alpha^\eta_1\\
 \sin\alpha^\eta_2 &  \cos\alpha^\eta_1
\end{pmatrix}\begin{pmatrix} \psi_{i,\uparrow,2,\eta}\\
\psi_{i,\uparrow,1,\eta}
\end{pmatrix}+\begin{pmatrix}\psi^+_{i,\uparrow}\\\psi^+_{i,\uparrow}\end{pmatrix},
 \end{equation}
  and similarly for the opposite spin components. 
  
  Note that although it is convenient to think of this splitting of the microscopic fermion fields in terms of branches as a linearisation
  around the Fermi points, conceptually it is possible to argue that by modifying the UV model,
  this branch decomposition becomes exact. This change in the UV does not affect the low energy description of the 
  system. The main observation is then that to first order in the interaction, the matrix elements with respect to the original fermions
  are the same as the matrix elements of the interaction with respect to the branch splitted fermion.
  
\begin{figure}[ht!]
	\centering
	\includegraphics[width=0.6\linewidth]{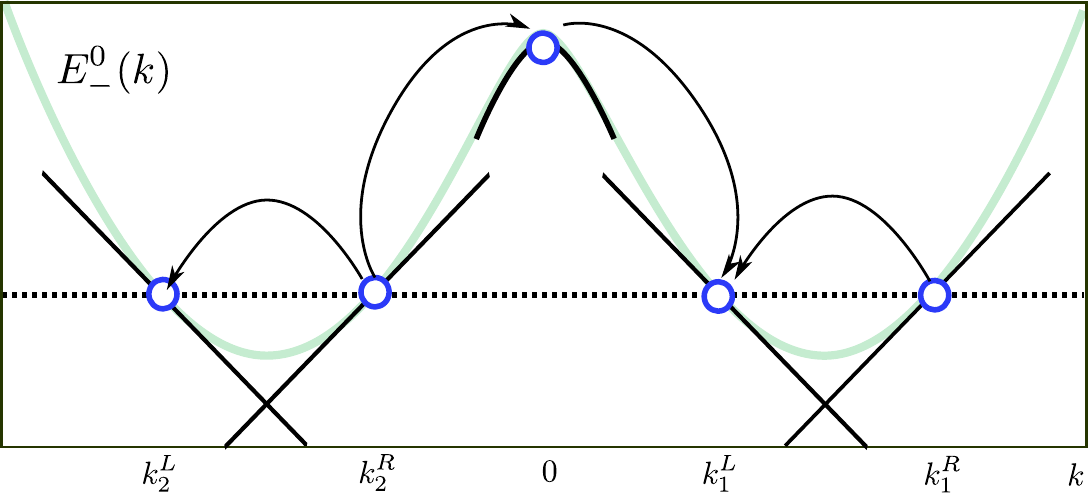}
	\caption{(Color online) The FTI process appears in second order in the interactions, after integrating out higher energy modes.
	Each process involving four fermions conserves momentum. The processes are depicted by black arrows.}\label{fig:second}
\end{figure}

 Once we have split the fermion mode as discussed above, we consider a derivative
expansion of the Hamiltonian for the band modes. For simplicity we consider the case $\alpha_{\rm so}=0$. This Hamiltonian is composed of two 
pieces, the kinetic term $H_{\rm kin}$ and the interaction $H_{\rm int}$. In the continuum we have
\begin{equation}
 H_{\rm kin}=i\sum_{\sigma,\alpha,\eta}\int dx v_F\psi^\dagger_{\sigma,\alpha,\eta}(\partial_x-ik_{F,\alpha}^\eta)\psi_{\sigma,\alpha,\eta}
 +\int dx\left(\frac{\partial_x\psi^{\dagger+}_\sigma\partial_x\psi^+_\sigma}{2m_{\rm eff}}+\mu_{\rm eff}\psi^{\dagger+}_\sigma\psi^+_\sigma\right),
\end{equation}
for the kinetic energy of the fermions, where $a^2m_{\rm eff}^{-1}=t\left(\cos\frac{\Phi}{2}-\frac{t}{t_\perp}\sin^2\frac{\Phi}{2}\right)$ and
$\mu_{\rm eff}=-t_\perp+2t\sin\frac{\Phi}{3}\sin\frac{\Phi}{6}$. This effective description is valid for momenta $k\ll \frac{t_\perp}{at}\sin\frac{\Phi}{2}$.
Note that the kinetic term of the high energy mode has the opposite sign to the usual term, to reproduce the inverted parabolic dispersion around $k=0$.
The interaction Hamiltonian, including same leg interaction on site $V^s_\parallel$, interleg interaction on the same rung $V_{\perp,0}^s$ and interleg interaction
between next nearest neighbors $V_{\perp,1}^s$,
focusing on vanishing interspin interactions reads
\begin{eqnarray}\nonumber
 H_{\rm int}=\int dx V^s_\parallel \sum_{A=I,II}c^{A\dagger}_{\sigma}(x)c^{A}_{\sigma}(x)c^{A\dagger}_{\sigma}(x+a)c^{A}_{\sigma}(x+a)
 +2V^s_{\perp,0} c^{I\dagger}_{\sigma}(x)c^{I}_{\sigma}(x)c^{II\dagger}_{\sigma}(x)c^{II}_{\sigma}(x)\\
 +V^s_{\perp,1}\left( c^{I\dagger}_{\sigma}(x+a)c^{I}_{\sigma}(x+a)c^{II\dagger}_{\sigma}(x+a)c^{II}_{\sigma}(x+a)
 + c^{I\dagger}_{\sigma}(x+a)c^{I}_{\sigma}(x+a)c^{II\dagger}_{\sigma}(x)c^{II}_{\sigma}(x)\right),
\end{eqnarray}
 where the fermion field $c^A_\sigma$ given by \eqref{fermion_expansion} receives contributions from the four Fermi points and the mode $\psi_\sigma^+$. For economy of notation we use
 $H_{\rm int}=H_{\rm int}[\psi_{\sigma,\alpha,\eta};\psi^+_\sigma]$ as the interaction Hamiltonian is a functional of the the lower band fields
 $\psi_{\sigma,\alpha,\eta}$ and the high energy mode $\psi^+_\sigma$. At tree level, the equation of motion of the high energy field $\psi^+_\sigma$ is
  \begin{figure}[ht!]
	\centering
	\includegraphics[width=0.6\linewidth]{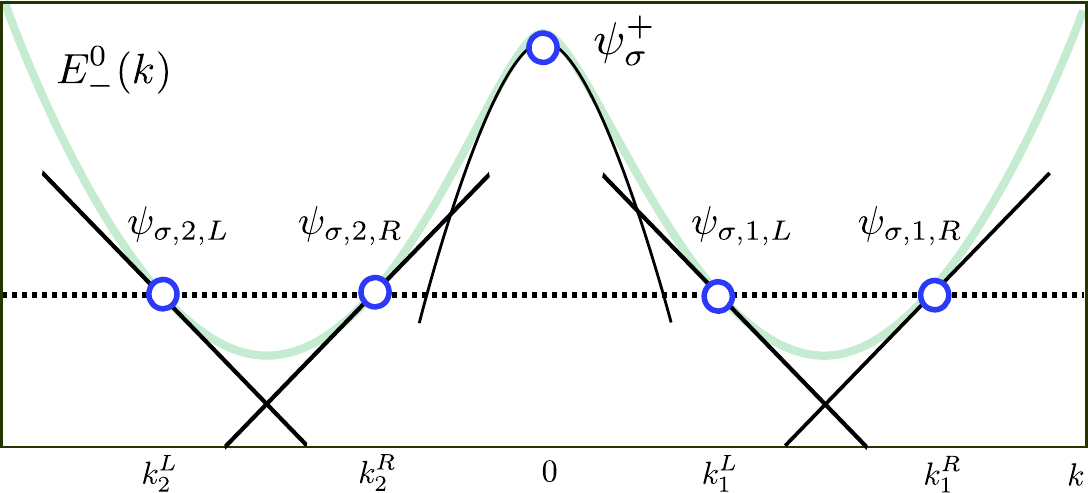}
	\caption{(Color online) Band spectrum approximation for the lower partially filled band, and the high energy state around lattice 
	momentum $k=0$. The full band spectrum is depicted with tenuous lines.
	We use a linear approximation for the fermion dispersion relations in the lower band, around the Fermi points (straight black lines). 
	For the higher energy mode, we use an inverted quadratic dispersion relation.}\label{fig:High_energy_int}
\end{figure}
\begin{equation}
 \psi^+_\sigma(x)=-\left.\frac{1}{\mu_{\rm eff}}\frac{\delta H_{\rm int}}{\delta \psi^{\dagger +}_\sigma(x)}\right|_{\psi_\sigma^+=0}+O(\mu_{\rm eff}^{-2}).
\end{equation}
where $\frac{\delta}{\delta\phi(x)}$ is the functional derivative with respect to $\phi(x)$.  
Using this equation of motion to solve for the heavy field $\psi_\sigma^+$, and inserting back into the Hamiltonian, we find to first order in the inverse mass 
$\mu_{\rm eff}$ the Hamiltonian
\begin{equation}
 H=i\sum_{\sigma,\alpha,\eta}\int dx v_F\psi^\dagger_{\sigma,\alpha,\eta}(\partial_x-ik_{F,\alpha}^\eta)\psi_{\sigma,\alpha,\eta}+H_{\rm int}[\psi_{\sigma,\alpha,\eta};0]
 -\frac{1}{\mu_{\rm eff}}\int dx\sum_\sigma\left|\frac{\delta H_{\rm int}}{\delta \psi^+_\sigma(x)}\right|_{\psi_\sigma^+=0}^2.
\end{equation}
The integration of the higher energy modes generates a renormalization of the interaction parameters at second order. It is also the responsible
for the generation of the FTI term $\mathcal{O}_{\rm FTI}^\sigma.$ This can be seen directly by expanding the interaction term using the 
decomposition (\ref{fermion_expansion}). Assuming a local interaction in the sense of Sec. \ref{subsec:lowEHam}, the expansion of the interaction
generates many different terms. Writing explicitly just a few
\begin{equation}
 H_{\rm int}=\int dx \left\{V^s_\parallel(\cos\alpha_1^L)^2\cos\alpha_1^R(\psi_{\uparrow,1,L}^\dagger\psi_{\uparrow}^+)(\psi_{\uparrow,1,L}^\dagger\psi_{\uparrow,1,R})+\right.
 V^s_\parallel\cos\alpha_2^L(\cos\alpha_2^R)^2(\psi_{\uparrow}^{+\dagger}\psi_{\uparrow,2,R})(\psi_{\uparrow,2,L}^\dagger\psi_{\uparrow,2,R})+\dots
\end{equation}
where the ellipsis indicates that many more terms are generated, including the ones proportional to $V_{\perp,0}^s$ and $V_{\perp,1}^s$
This leads to 
\begin{eqnarray}
\left| \frac{\delta H_{\rm int}}{\delta \psi^+_\sigma(x)}\right|_{\psi_\sigma^+=0}^2&=&(V^{s}_\parallel)^2(\cos\alpha_1^L\cos\alpha_2^R)^2\cos\alpha_1^R\cos\alpha_2^L(\psi_{\uparrow,1,L}^\dagger)^2\psi_{\uparrow,2,L}^\dagger(\psi_{\uparrow,2,R})^2\psi_{\uparrow,1,R}+\dots\\
&=&(V_\parallel^s)^2(\cos\alpha_1^L\cos\alpha_2^R)^2\cos\alpha_1^R\cos\alpha_2^L\mathcal{B}^{\sigma\dagger}_{\rm FTI}+\dots
\end{eqnarray}
with $\mathcal{B}^{\sigma}_{\rm FTI}$ given in (\ref{FTI_psi}) the operator that leads to the FTI phase. The whole prefactor is obtained considering all appropriate terms 
in the expansion above. It is written explicitly in (\ref{app:FTIpref}). This second order process can be visualized
in Fig.~\ref{fig:second}.

 \section{Interaction strength of different cosine terms}\label{app:pref_signs}
 
 In this section we discuss in detail the magnitude and sign of the prefactors corresponding to each cosine term considered in the main text that opens a gap. These operators
 are $\mathcal{O}_{1\sigma}^s$, $\mathcal{O}_{5}^d$, $\mathcal{O}_{6}^d$ and $\mathcal{O}^\sigma_{\rm FTI}$.
 
 Any four particle term is given generically in our model by the expression 
\begin{equation}
\mathcal{A}^{\eta_1\eta_2\eta_3\eta_4\sigma\sigma'}_{a_1a_2a_3a_4}\psi^\dagger_{\sigma a_1\eta_1}(x)\psi_{\sigma a_2\eta_2}(x)\psi^\dagger_{\sigma' a_3\eta_3}(x')
\psi_{\sigma' a_4\eta_4}(x'),
\end{equation}
 where $\mathcal{A}$ is a prefactor that depends on the interaction parameters 
 $V_\parallel^{s,d}$ and $V_{\perp,i}^{s,d}$.
 To first order on the interaction strength, it is given by the expression (\ref{eq:lowEintcoeffs}). Up to an overall positive 
 combinatorial factor, we find that the operator $\mathcal{O}_{1\sigma}^s$ appears in the Hamiltonian in the following form
 \begin{eqnarray}\nonumber
  \bar{g}_{1\sigma}^s\mathcal{O}_{1\sigma}^s=\mathcal{A}^{LRRL\sigma\sigma}_{1122}(\psi^\dagger_{\sigma,1L}\psi_{\sigma,1R}\hat{\psi}^\dagger_{\sigma,2R}\hat{\psi}_{\sigma,2L}+
  \hat{\psi}^\dagger_{\sigma,1L}\hat{\psi}_{\sigma,1R}\psi^\dagger_{\sigma,2R}\psi_{\sigma,2L})\\
 -\mathcal{A}^{LLRR\sigma\sigma}_{1122}(\hat{\psi}^\dagger_{\sigma,1L}\psi_{\sigma,1R}\psi^\dagger_{\sigma,2R}\hat{\psi}_{\sigma,2L}+
  \psi^\dagger_{\sigma,1L}\hat{\psi}_{\sigma,1R}\hat{\psi}^\dagger_{\sigma,2R}\psi_{\sigma,2L})+\text{h.c.}
 \end{eqnarray}
 where we have used the $\hat{\psi}$ to denote that the field is evaluated at a different spatial point that $\psi$. In term of the slow fields 
 $R_{\sigma,a}, L_{\sigma,a}$, to first order in the interaction parameters, using the expression (\ref{eq:lowEintcoeffs}), we find that this contribution is given by
 $\bar{g}_{1\sigma}^s(R^\dagger_{\sigma 1}L_{\sigma 1}L^\dagger_{\sigma 2}R_{\sigma 1}+\text{h.c})$ with
 \begin{eqnarray}
  \bar{g}_{1\sigma}^s&=&\frac{a}{(2\pi a)^2}\left(V_\parallel^s\sin2\alpha_2^R\sin2\alpha_2^L(\cos{\pi a\rho}-\cos\Phi)+2V^s_{\perp,0}\cos2\alpha_2^R\cos2\alpha_2^L\right.\\\nonumber
  &+&\left.{V_{\perp,1}^s}\left[\left(\cos{\pi a\rho}-\cos\Phi\right)+\left(\cos{\pi a\rho}+\cos\Phi\right)\cos2\alpha_2^R\cos2\alpha_2^L\right]\right)
 \end{eqnarray}
 where we have left explicit the dependence on the 1D density $\rho = 2N_{\rm leg}/L$ per spin. For the particular filling $\nu=1/3$ that we are interested, this becomes
  \begin{eqnarray}\label{app:g1s}
  \bar{g}_{1\sigma}^s&=&\frac{a}{(2\pi a)^2}\left(V_\parallel^s\sin2\alpha_2^R\sin2\alpha_2^L\left(\cos\frac{\Phi}{3}-\cos\Phi\right)+2V^s_{\perp,0}\cos2\alpha_2^R\cos2\alpha_2^L\right.\\\nonumber
  &+&\left.{V_{\perp,1}^s}\left[\left(\cos\frac{\Phi}{3}-\cos\Phi\right)+\left(\cos\frac{\Phi}{3}+\cos\Phi\right)\cos2\alpha_2^R\cos2\alpha_2^L\right]\right)
 \end{eqnarray}
 where the angles are $\alpha_a^\eta=\frac{1}{2}\arctan\left(\frac{-t_\perp}{t\sin(k_{Fa}^\eta)\sin\frac{\Phi}{2}}\right)$. 
 Note that for zero interleg tunnelling $t_\perp=0$, the prefactor $\bar{g}_{1\sigma}^s$ simplifies to
 \begin{equation}
  \bar{g}_{1\sigma}^s=\frac{a}{2(\pi a)^2}(V_{\perp,0}^s+V_{\perp,1}^s\cos \pi a \rho)=\frac{a}{2(\pi a)^2}\left(V_{\perp,0}^s+V_{\perp,1}^s\cos \frac{\Phi}{3}\right),
 \end{equation}
 where the second expression is valid for filling $\nu=1/3$.

 We also find for the operators $\mathcal{O}_5^d$ and $\mathcal{O}_6^d$ the following expressions for their prefactors, in first order in the interaction strength
 \begin{equation}
  \bar{g}_{5}^d=\frac{a}{2(2\pi a)^2}(V_\parallel^d+V_\perp^d)(\sin2\alpha^R_2)^2\quad \mbox{and}\quad  \bar{g}_{6}^d=\frac{a}{2(2\pi a)^2}(V_\parallel^d+V_\perp^d)(\sin2\alpha^L_2)^2.
 \end{equation}
 
 The prefactor of the FTI operator is a little more difficult to obtain. Up to an overall combinatorial factor absorbed in $\epsilon_{t_\perp}$ below, the prefactor of 
 $\mathcal{O}^\sigma_{\rm FTI}$ is given, at second order in the interaction by
 \begin{equation}\label{app:FTIpref}
 \bar{g}_{\rm FTI}=-\frac{\epsilon_{t_\perp}^2}{2\pi^3v_F}\left(V_\parallel\cos^2\left(\frac{\Phi}{3}\right)(\cos\alpha^R_2\cos^2\alpha^L_2+\sin\alpha^R_2\sin^2\alpha^L_2)
  +\frac{\sin2\alpha^L_2}{2}(\cos\alpha^R_2+\sin\alpha^R_2)\left(V_{\perp,0}^s+V_{\perp,1}^s\cos^2\left(\frac{\Phi}{3}\right)\right)\right)^2
 \end{equation}
 To obtain this result, we have projected out the high energy single particle band. For weak $t_\perp$, given that the FTI process is a single interleg tunnelling event dressed by interactions, we 
 have $\bar{g}_{\rm FTI}\propto t_\perp$. This property is generic, valid also beyond the weakly interacting regime. 

 \section{Klein factors}\label{app:Klein}
 
 The fermionic fields $\psi_{\sigma,\alpha,\eta}(x)=\frac{\kappa_{\sigma\alpha}^\eta}{\sqrt{2\pi a}}e^{i\eta\sqrt{4\pi}\phi_{\sigma,\alpha,\eta}}$, satisfy anticommutation 
 relations due to the commutation relations $[\phi_{\sigma,\alpha,\eta}(x),\phi_{\tilde{\sigma},\beta,\tilde{\eta}}(x')]=\frac{i}{4}(\sigma_{3})_{\eta\tilde{\eta}}\delta_{\sigma\tilde{\sigma}}\delta_{\alpha\beta}\text{sgn}(x-x'),$
 and the Klein factors $\kappa_{\sigma\alpha}^\eta=e^{i\pi\sum_{(\sigma',\alpha',\eta')<(\sigma,\alpha,\eta)}N_{\sigma'\alpha'}^\eta}$, where (following Refs. \onlinecite{Teo2014} and \onlinecite{Senechal1999})
 \begin{equation}
 N_{\sigma\alpha}^\eta=\frac{1}{\sqrt{\pi}}\int dx \partial_x\phi_{\sigma,\alpha,\eta} \quad\mbox{with}\quad[N_{\sigma\alpha}^\eta,\sqrt{4\pi}\phi_{\sigma',\alpha',\eta'}(x)]
 =i\delta_{\sigma\sigma'}\delta_{\alpha\alpha'}(\sigma_3)_{\eta\eta'}.
 \end{equation}
 We use the ordering 
 $\uparrow<\downarrow$, $1<2$, and $L<R$, in this order of prevalence. So for example
 \begin{equation}
  (\uparrow,1,L)<(\uparrow,1,R)<(\uparrow,2,L)<(\uparrow,2,R)<(\downarrow,1,L)<(\downarrow,1,R)<(\downarrow,2,L)<(\downarrow,2,R).
 \end{equation}
 This ordering makes the Klein factor commute and the fermions anticommute. The interaction operators $\mathcal{O}_i$ after bosonization, acquire a string of Klein operators,
 with eigenvalues $\pm1$.
 So for example, the interaction $\mathcal{O}_{1\sigma}^s$ is 
 \begin{equation}
  \mathcal{O}_{1\sigma}^s=\psi^\dagger_{\sigma,1,R}\psi_{\sigma,1,L}\psi^\dagger_{\sigma,2,L}\psi_{\sigma,2,R}=-e^{i\pi(N_{\sigma,1}^{L}+N_{\sigma,2}^{L})}
  e^{-i\sqrt{4\pi}(\phi_{\sigma,1,R}+\phi_{\sigma,1,L}-\phi_{\sigma,2,L}-\phi_{\sigma,2,R})}.
 \end{equation}
 The factor $\kappa_{\sigma,1}^L\kappa_{\sigma,2}^L=e^{i\pi(N_{\sigma,1}^{L}+N_{\sigma,2}^{L})}$ commutes with the exponential operators, as they do not change
 the number of total left movers. To determine the correct sign of a prefactor in an operator we compare the bosonic expressions with the original fermion expressions.
 Note that the scaling dimensions of the operators are insensitive to the Klein factors. The other operators considered explicitly in the text in their bosonized form are,
 \begin{eqnarray}
  \mathcal{O}_{\rm FTI}^\sigma&=&e^{i(\sqrt{4\pi}\theta_{g\sigma}+\pi(N_{\sigma,1}^L+N_{\sigma,1}^R+ N_{\sigma,2}^L))}+\text{h.c.},\\
  \mathcal{O}_{9,\uparrow\downarrow}=\mathcal{O}_5^d&= &e^{i\frac{\sqrt{4\pi}}{3}(\theta_{g\uparrow}-\theta_{g\downarrow}+2(\tilde{\phi}_{L\uparrow}+\tilde{\phi}_{R\uparrow}-\tilde{\phi}_{L\downarrow}-\tilde{\phi}_{R\downarrow}))+i\pi(N_{\uparrow,1}^L+N_{\uparrow,1}^R+N_{\uparrow,2}^L-(N_{\downarrow,1}^L+N_{\downarrow,1}^R+N_{\downarrow,2}^L))}+\text{h.c},\\
  \mathcal{O}_{10,\uparrow\downarrow}=\mathcal{O}_6^d&=&e^{i\frac{\sqrt{4\pi}}{3}(2(\theta_{g\uparrow}-\theta_{g\downarrow})+\tilde{\phi}_{L\uparrow}+\tilde{\phi}_{R\uparrow}-\tilde{\phi}_{L\downarrow}-\tilde{\phi}_{R\downarrow})+i\pi(N_{\uparrow,1}^R-N_{\downarrow,1}^R)}+\text{h.c}.
 \end{eqnarray}
 When discussing the FTI phase at strong coupling in Sec.~\ref{sec:strong_coupling},
 the projection of the massive degrees of freedom is carried out without eliminating the Klein factors. Therefore,  
 \begin{eqnarray}
  \mathcal{O}_{\rm FTI}^\sigma&=&e^{i(\sqrt{4\pi}\theta_{g\sigma}+\pi(N_{\sigma,1}^L+N_{\sigma,1}^R+ N_{\sigma,2}^L))}+\text{h.c.},\\
  \mathcal{O}_{9}^{\rm proj}&= &e^{i\frac{2\sqrt{4\pi}}{3}(\tilde{\phi}_{L\uparrow}+\tilde{\phi}_{R\uparrow}-\tilde{\phi}_{L\downarrow}-\tilde{\phi}_{R\downarrow})+i\pi(N_{\uparrow,1}^L+N_{\uparrow,1}^R+N_{\uparrow,2}^L-(N_{\downarrow,1}^L+N_{\downarrow,1}^R+N_{\downarrow,2}^L))}+\text{h.c},\\
  \mathcal{O}_{10}^{\rm proj}&=&e^{i\frac{\sqrt{4\pi}}{3}(\tilde{\phi}_{L\uparrow}+\tilde{\phi}_{R\uparrow}-\tilde{\phi}_{L\downarrow}-\tilde{\phi}_{R\downarrow})+i\pi(N_{\uparrow,1}^R-N_{\downarrow,1}^R)}+\text{h.c}.
 \end{eqnarray}
 
 \subsection{Commutation relations and Klein factors}
 
 It is convenient to absorb the Klein factors in the definition of the boson fields. Doing this changes the commutation relations of the fields by a central element, i.e.
 an element that commutes with all the fields of the theory. This central element accounts for the difference between the functions ${\rm sgn}(x)$ and $2\Theta(x)$. In 
 particular the fields that describe the bulk of the FTI state,
 \begin{equation}
  \theta_{g\sigma}=\phi_{\sigma,1,L}+2\phi_{\sigma,1,R} + 2\phi_{\sigma,2,L} + \phi_{\sigma,2,R},\quad
  \varphi_{g,\sigma}=\phi_{\sigma,1,L}+2\phi_{\sigma,1,R} - 2\phi_{\sigma,2,L} - \phi_{\sigma,2,R},
 \end{equation}
 have a commutation relation $[\theta_{g\sigma}(x),\varphi_{g\sigma'}(y)]=i\frac{3}{2}{\rm sgn}\delta_{\sigma\sigma'}(x-y)$.
 Keeping explicitly the Klein factors, the fields become
 \begin{equation}
  \theta_{g\sigma}\rightarrow\theta_{g\sigma}+\frac{\sqrt{\pi}}{2}\left[N_{\sigma,1}^L+N_{\sigma,1}^R+N_{\sigma,2}^L\right],\quad
  \varphi_{g\sigma}\rightarrow\varphi_{g\sigma}-\frac{\sqrt{\pi}}{2}\left[N_{\sigma,1}^L+3N_{\sigma,1}^R+N_{\sigma,2}^L\right],
 \end{equation}
 with commutation relation $[\theta_{g\sigma}(x),\varphi_{g\sigma'}(y)]=3i\delta_{\sigma\sigma'}\left(\frac{1}{2}{\rm sgn}\delta_{\sigma\sigma'}(x-y)+\frac{1}{2}\right)=3i\delta_{\sigma\sigma'}\Theta(x-y)$. 
 This difference is important in determining the effect of the kink operators in the field configurations of the bosons. We choose the Klein factors of the field
 $\varphi_{g\sigma}$ such that the kink operator that appears in section \ref{sec:strong_coupling} induces changes between different minima of the field $\theta_{g\sigma}$.
 The difference disappears in the commutation relation of the densities, which involve derivatives of the boson fields.

\section{Luttinger parameters}\label{app:Defs}

The computation of the forward scattering parameters
in Eq.(\ref{Gaussian}-\ref{int_matrix_same_spin}) involves a point splitting procedure. Different regularizations provide slightly different results.  Here we follow a 
procedure similar to that in Ref.~\onlinecite{GiamarchiBook2003} whereby we keep the summation over $m$ in Eq.~\eqref{interactions} discrete and take the corresponding 
field arguments to be $x$ and $x+ma$ and subsequently perform a derivative expansion of our slow fields. The parameters of the forward scattering matrix defined in the text 
[Eq.(\ref{Gaussian}-\ref{int_matrix_same_spin})] are related to the microscopic parameters through [using $\bar{V}=aV$ where $V$ are the {microscopic couplings} in Eq.~\eqref{interactions}]
\begin{eqnarray}\label{coefficients}
&f_{mn}  &= \frac{\bar{V}_{\parallel}^s}{2}\left(1-\cos\frac{\Phi}{3}\right)(c^+_{mn}+c^-_{mn})+{\bar{V}^s_{\perp,0}} s_{mn}^-
+\frac{\bar{V}_{\perp,1}^s}{2}\left(s_{mn}^+\left(1-\cos\frac{\Phi}{3}\right)+s_{mn}^-\left(1+\cos\frac{\Phi}{3}\right)\right)+O(V^2),\\\nonumber
&g_{mn} &= \frac{\bar{V}_{\parallel}^s}{2}\left(1-\cos\frac{(m+n)\Phi}{3}\right)(s^+_{mn}+s^-_{mn})+{\bar{V}^s_{\perp,0}} c_{mn}^-
+\frac{\bar{V}_{\perp,1}^s}{2}\left(c_{mn}^++c^-_{mn}+\cos\frac{(m+n)\Phi}{3}(c_{mn}^--c^+_{mn})\right)+O(V^2),\\\nonumber
&h_{mn} &= \frac{\bar{V}_{\parallel}^{d}}{2}\left(s^+_{mn}+s^-_{mn}\right)+\frac{\bar{V}_{\perp}^{d}}{2}\left(c^+_{mn}+c^-_{mn}\right)+O(V^2),\quad\mbox{and}\quad
\tilde{h}_{mn} = \frac{\bar{V}_{\parallel}^{d}}{2}\left(c^+_{mn}+c^-_{mn}\right)+\frac{\bar{V}_{\perp}^{d}}{2}\left(s^+_{mn}+s^-_{mn}\right)+O(V^2).
\end{eqnarray}
where we have introduced the notation $c_{mn}^\pm=\cos^{2}(\alpha_{m}^0\pm \alpha_{n}^0)$, $s_{mn}^\pm=\sin^{2}(\alpha_{m}^0\pm\alpha_{n}^0)$ and $\alpha^0_m \equiv \alpha^0_{\frac{m\Phi}{3}}$,
with $\alpha_k^0$ defined below Eq.~\eqref{alpha_k}.
The expressions above have been verified to display the correct behaviour in the limit of $t_\perp=0$, and also for the case $V_{\parallel,\perp}^d=0$ of a spinless ladder.
As we have discussed in Appendix \ref{app:Integrating_out}, including 
higher energy process renormalizes the interaction parameters. The effect of these renormalization corresponds to the $O(V^2)$
terms appearing in the definition of the forward scattering parameters $f_{mn},g_{mn},h_{mn}$ and $\tilde{h}_{mn}$.

\subsection{Small interleg tunnelling $t_\perp/t\ll 1$}

To gain further insight into the general phase diagram, and to simplify the relations between the different Luttinger
liquid parameters, we focus on the reflection symmetric case. Reflection symmetry $\mathcal{R}$ acts  on the low energy fermion branches $\psi_{\sigma,\alpha,\eta}$ as
$\mathcal{R}\psi_{\sigma,\alpha,\eta}(x)\mathcal{R}^{-1}= \psi_{\sigma,\alpha,\bar{\eta}}(-x)$.
The phenomenological parameters of a reflection symmetric system satisfy $w_{11}=w_{22}=w$, with $w=f,g,h,\tilde{h}$. This is an exact symmetry for $t_\perp=0$, 
which receives corrections in the forward scattering matrix by terms of order $(t_\perp/t)^2$. This implies that we can consider this symmetry to be present when working
up to and including terms of order $t_\perp/t$.
It is convenient to define the basis of charge and neutral modes per spin projection
\begin{equation}\label{basis_app}
\left(\begin{array}{c}
       \Upsilon^c_\sigma\\
       \Theta^c_\sigma  \\
       \Upsilon_\sigma\\
       \Theta_\sigma 
      \end{array}\right)=
\left(
\begin{array}{cccc}
 1 & 1 & 1 & 1 \\
 -1 & 1 & -1 & 1 \\
 -1 & 1 & 1 & -1 \\
 1 & 1 & -1 & -1 \\
\end{array}
\right)\left(\begin{array}{c}
       \phi_{\sigma,1,L}\\
       \phi_{\sigma,1,R}\\
       \phi_{\sigma,2,L}\\
       \phi_{\sigma,2,R} 
      \end{array}\right).
\end{equation}
For a reflection symmetric system (assuming that TR and inversion are symmetries as well), the scaling dimension matrix $\Lambda$ simplifies considerably.
In the basis (\ref{basis_app}) is given by
$\Lambda=\frac{1}{8\pi}\sum_{r=\pm} P_r\otimes\text{diag}[K_{\rho,r},K_{\rho,r}^{-1},K_{\beta,r}^{-1},K_{\beta,r}]$.
The Luttinger parameters are given correspondingly by
\begin{equation}
K_{\rho,\pm}=\sqrt{\frac{1+\frac{f-g-f_{12}+g_{12}\pm(h-\tilde{h}-h_{12}+\tilde{h}_{12})}{4\pi v_F}}{1+\frac{f+g+f_{12}+g_{12} \pm (h+h_{12}+\tilde{h}+\tilde{h}_{12})}{4\pi v_F}}}\quad \text{and}\quad
K_{\beta,\pm}=\sqrt{\frac{1+\frac{f+g-f_{12}-g_{12}\pm (h+\tilde{h}-h_{12}-\tilde{h}_{12})}{4\pi v_F}}{1+\frac{f-g+f_{12}-g_{12}\pm(h+h_{12}-\tilde{h}-\tilde{h}_{12})) }{4\pi v_F}}},
\end{equation}
in terms of the phenomenological parameters.

The scaling dimensions of the four fermion operators considered in the main text are
\begin{eqnarray}\nonumber
 \Delta_1^s&=&K_{\beta,+}+K_{\beta,-},\quad \Delta_1^d=K_{\rho,-}+K_{\beta,-},\\
 \Delta_2^d&=&K_{\rho,-}+K_{\beta,+}, \quad \Delta_3^d=K_{\beta,-}+K_{\beta,-}^{-1},\\\nonumber
 \Delta_4^d&=&K_{\beta,-}^{-1}+K_{\beta,+},\quad \Delta_5^d=\Delta_6^d=K_{\rho,-}+K_{\beta,-}^{-1}.
\end{eqnarray}
Additionally the scaling dimension of the FTI operator is $\Delta_{\rm FTI}=\frac{9}{4}(K_{\rho,+}+K_{\rho,-})+\frac{1}{4}(K_{\beta,-}^{-1}+K_{\beta,+}^{-1})$.
Note that given these relations, we can write
\begin{equation}\label{relation_FTI}
 \Delta_{\rm FTI}=\frac{9}{4}K_{\rho,+}+\frac{1}{4}K_{\beta,+}^{-1}+2K_{\rho,-}+\frac{1}{4}\Delta^d_5,
\end{equation}
which serves to show that there are no regions where the only relevant operator is the FTI term. To see this, let's assume that both
$\Delta^d_1$ and $\Delta^d_5$ are larger than 2, so the corresponding operators are irrelevant. This implies that
$K_{\rho,-}>2-{\rm min}(K_{\beta,-},K_{\beta,-}^{-1})\geq 1$. Using Eq. (\ref{relation_FTI}), we see that the FTI term will also be irrelevant.

Working to first order in $t_\perp/t$ and using the microscopic interactions that we consider, the previous expressions simplify further. We find
\begin{equation}
 K_{\rho,\pm}=\sqrt{\frac{1}{1+\frac{(\tilde{V}^s_\parallel+V^s_{\perp,0}+V^{s}_{\perp,1} \pm (V^d_\parallel+V^d_\perp)) a}{2\pi v_F}}}\quad \text{and}\quad
K_{\beta,\pm}=\sqrt{\frac{1}{1-\frac{(V^s_{\perp,0}+V^{s}_{\perp,1} \pm V^d_\perp-(\tilde{V}_\parallel \pm V^d_\perp)) a}{2\pi v_F}}},
\end{equation}
with $\tilde{V}^s_\parallel=V^s_\parallel(1-\cos\Phi/3)$.

Small interleg tunnelling $t_\perp/t\ll 1$ modifies the scaling dimension matrix, which to first order in $(t_\perp/t)^2$ becomes
$\Lambda=(1+\delta U)(\Lambda_0+\delta \Lambda)(1-\delta U)$, with the matrices $\delta U$ and $\delta\Lambda$ first order in $(t_\perp/t)^2$. Specifically we find
$\delta\Lambda_0=\frac{1}{16\pi}\sum_{r=\pm}P_r\otimes{\rm diag}[\frac{1}{v_{\rho,r}}(\delta\lambda_{2,r}-{\delta\lambda_{1,r}}K_{\rho,r}^2),
\frac{1}{v_{\rho,r}}({\delta\lambda_{1,r}}-{\delta\lambda_{2,r}}K_{\rho,r}^{-2}),
\frac{1}{v_{\beta,r}}(\delta\lambda_{4,r}-{\delta\lambda_{3,r}}K_{\beta,r}^{-2}),\frac{1}{v_{\beta,r}}({\delta\lambda_{3,r}}-{\delta\lambda_{4,r}}K_{\beta,r}^2)]$
and
\begin{equation}
 \delta U=\sum_{r=\pm}P_r\otimes\left(
\begin{array}{cccc}
  &  & -\delta_{+,r} &  \\
  &  &  & -\delta_{-,r} \\
 \delta_{+,r} &  &  &  \\
  & \delta_{-,r} &  &  \\
\end{array}
\right),\quad \text{with}\quad \delta_{a,r}=\frac{\epsilon_{a,r} \gamma_r^a+\epsilon_{-a,r}}{(v_{\rho,r}+v_{\beta,r})(K_{\rho,r}^a-K_{\beta,r}^{-a})}+\frac{\epsilon_{a,r}}{v_{\rho,r} K_{\rho,r}^{-a}-v_{\beta,r} K_{\beta,r}^a},
\end{equation}
and $\gamma_r=\frac{v_{\beta,r} K_{\beta,r}^{-1}-v_{\rho,r} K_{\rho,r}}{v_{\rho,r} K_{\rho,r}^{-1}-v_{\beta,r} K_{\beta,r}}$.
In terms of the original microscopic parameters $\lambda_{a,r}(t_\perp=0)\equiv \lambda_{a,r}^0$, the velocities are $v_{\rho,r}=\sqrt{\lambda_{1,r}^0\lambda_{2,r}^0}$
and $v_{\beta,r}=\sqrt{\lambda_{3,r}^0\lambda_{4,r}^0}$. The differences $\delta \lambda_{a,r}=\lambda_{a,r}(t_\perp)-\lambda^0_{a,r}$  and the parameters
$\epsilon_{\pm,r}=\frac{\delta f_{11}-\delta f_{22}\pm(\delta g_{11}-\delta g_{22})}{8\pi}+r\frac{(\delta h_{11}-\delta h_{22}\pm(\delta\tilde{h}_{11}-\delta\tilde{h}_{22}))}{8\pi}$ are
first order in $(t_\perp/t)^2$.

\subsection{Exact expressions for Luttinger parameters}

Using the forward scattering matrix $\mathcal{M}$ defined in the text (Eqs. (\ref{Gaussian}-\ref{int_matrix_same_spin})), we can obtain the Luttinger parameters for our system
as follows. The forward scattering matrix $\mathcal{M}$ can be written as
\begin{equation}
\mathcal{M}=\sum_{r=\pm}P_{r}\otimes\left(v_{F}\openone_{4}+\frac{1}{4\pi}(\mathbb{V}+r\mathbb{W})\right),
\end{equation}
where $P_{\pm}=\frac{1}{2}(\openone_{2}\pm \sigma_1)$ is the projector onto the eigenvalue $\pm1$ of the $\sigma_{1}$
Pauli matrix. This decomposition splits the matrix into two orthogonal
subspaces that can be diagonalized independently
\begin{eqnarray*}
\mathcal{M} & = & P_{+}\otimes U_{+}^{\dagger}D_{+}U_{+}+P_{-}\otimes U_{-}^{\dagger}D_{-}U_{-},
 \end{eqnarray*}
with the matrices
$D_\pm={\rm diag}(\lambda_{1,\pm},\lambda_{2,\pm},\lambda_{3,\pm},\lambda_{4,\pm})$. The eigenvalues
are explicitly
\begin{eqnarray}\nonumber\label{eigenvalues}
\lambda_{1,\pm}&=&v_{F}+\frac{1}{4\pi}\left[f_{22}+g_{22}-(f_{12}+g_{12})\tan\left(\frac{\beta_{1}^{\pm}}{2}\right) \pm (h_{22}+\tilde{h}_{22}-(\tilde{h}_{12}+h_{12})\tan\left(\frac{\beta_{1}^{\pm}}{2}\right))\right],\\
\lambda_{2,\pm}&=&v_{F}+\frac{1}{4\pi}\left[f_{11}+g_{11}+(f_{12}+g_{12})\tan\left(\frac{\beta_{1}^{\pm}}{2}\right) \pm (h_{11}+\tilde{h}_{11}+(\tilde{h}_{12}+h_{12})\tan\left(\frac{\beta_{1}^{\pm}}{2}\right))\right],\\\nonumber
\lambda_{3,\pm}&=&v_{F}+\frac{1}{4\pi}\left[f_{11}-g_{11}+(g_{12}-f_{12})\tan\left(\frac{\beta_{2}^{\pm}}{2}\right) \pm (h_{11}-\tilde{h}_{11}+(\tilde{h}_{12}-h_{12})\tan\left(\frac{\beta_{2}^{\pm}}{2}\right))\right],\\\nonumber
\lambda_{4,\pm}&=&v_{F}+\frac{1}{4\pi}\left[f_{22}-g_{22}-(g_{12}-f_{12})\tan\left(\frac{\beta_{2}^{\pm}}{2}\right) \pm (h_{22}-\tilde{h}_{22}-(\tilde{h}_{12}-h_{12})\tan\left(\frac{\beta_{2}^{\pm}}{2}\right))\right],
\end{eqnarray}
with 
\begin{equation*}
\tan\beta_{1}^{\pm}=\frac{2(f_{12}+g_{12}\pm(\tilde{h}_{12}+h_{12}))}{f_{11}-f_{22}+g_{11}-g_{22}\pm(h_{11}-h_{22}+\tilde{h}_{11}-\tilde{h}_{22})},\quad
\tan\beta_{2}^{\pm}=\frac{2(g_{12}-f_{12}\pm(\tilde{h}_{12}-h_{12}))}{f_{11}-f_{22}+g_{22}-g_{11}\pm(h_{11}-h_{22}+\tilde{h}_{22}-\tilde{h}_{11})}.
\end{equation*}
The unitary matrix $U_{\pm}$ is in turn 
\begin{eqnarray*}
U_{\pm}^{\dagger} =\frac{1}{\sqrt{2}}\begin{bmatrix}R\left(\frac{\beta_{1}^{\pm}}{2}\right) & \sigma_{1}R\left(\frac{\beta_{2}^{\pm}}{2}\right)\\
                                                   \sigma_{1}R\left(\frac{\beta_{1}^{\pm}}{2}\right) & -R\left(\frac{\beta_{2}^{\pm}}{2}\right)
\end{bmatrix},
\end{eqnarray*}
where $R(\theta)=e^{i\sigma_{2}\theta}$ and $\sigma_2$ is the second Pauli matrix.

Using the forward scattering matrix $\mathcal{M}$ and the $\mathcal{K}$ matrix $\mathcal{K}=-\openone_{4\times 4}\otimes\sigma_3$ 
in the basis $\bm{\phi}^T=(\phi_{\uparrow,1,L},\phi_{\uparrow,1,R},\phi_{\uparrow,2,L},\phi_{\uparrow,2,R},\phi_{\downarrow,1,L},\phi_{\downarrow,1,R},\phi_{\downarrow,2,L},\phi_{\downarrow,2,R})$,
the scaling dimension matrix $\Lambda$ is given by
\begin{eqnarray}\label{lambdaapp}
\Lambda & = & \frac{1}{8\pi}\sum_{r=\pm}P_{r}\otimes\tilde{U}_{r}^{\dagger}\text{diag}\begin{bmatrix}K^r_{12} & K^r_{21} & K^r_{34} & K^r_{43}\end{bmatrix}\tilde{U}_{r}.
\end{eqnarray}
The Luttinger parameters satisfy
$K_{12}^rK_{21}^r=\left(K_{34}^rK_{43}^r\right)^{-1}=\sqrt{\frac{\lambda_{3,r}\lambda_{4,r}}{\lambda_{1,r}\lambda_{2,r}}}$ 
in terms of the eigenvalues (\ref{eigenvalues}).

The Luttinger parameters are given by
\begin{eqnarray}
K^r_{ab}&=&\sqrt{\frac{\lambda_{b,r}}{\lambda_{a,r}}}\frac{\sqrt{\Gamma_{a,r}}+\sqrt{\Gamma_{b,r}}}{2}+\frac{\sqrt{\Gamma_{a,r}}-\sqrt{\Gamma_{b,r}}}{2}\left[\sqrt{\frac{\lambda_{b,r}}{\lambda_{a,r}}}\cos\phi^r_{ab}+\sin\phi^r_{ab}\tan\frac{\theta^r_{ab}}{2}\right].
\end{eqnarray}
The functions $\Gamma_{a,r}$ are in turn
\begin{eqnarray}
\Gamma_{1,r}	&=&	\frac{1}{2}\left[\frac{\lambda_{3,r}+\lambda_{4,r}+(\lambda_{4,r}-\lambda_{3,r})\cos(\beta_{1}^{r}-\beta_{2}^r)}{\lambda_{2,r}}+\frac{\lambda_{4,r}-\lambda_{3,r}}{\sqrt{\lambda_{1,r}\lambda_{2,r}}}\tan\frac{\phi_{12}^r}{2}\sin(\beta_{2}^r-\beta_{1}^r)\right],\\
\Gamma_{2,r}	&=&	\frac{1}{2}\left[\frac{\lambda_{3,r}+\lambda_{4,r}-(\lambda_{4,r}-\lambda_{3,r})\cos(\beta_{1}^{r}-\beta_{2}^r)}{\lambda_{1,r}}-\frac{\lambda_{4,r}-\lambda_{3,r}}{\sqrt{\lambda_{1,r}\lambda_{2,r}}}\tan\frac{\phi_{12}^r}{2}\sin(\beta_{2}^r-\beta_{1}^r)\right],\\
\Gamma_{3,r}	&=&	\frac{1}{2}\left[\frac{\lambda_{1,r}+\lambda_{2,r}+(\lambda_{2,r}-\lambda_{1,r})\cos(\beta_{1}^{r}-\beta_{2}^r)}{\lambda_{4,r}}+\frac{\lambda_{2,r}-\lambda_{1,r}}{\sqrt{\lambda_{4,r}\lambda_{3,r}}}\tan\frac{\phi_{34}^r}{2}\sin(\beta_{1}^r-\beta_{2}^r)\right],\\
\Gamma_{4,r}	&=&	\frac{1}{2}\left[\frac{\lambda_{1,r}+\lambda_{2,r}-(\lambda_{2,r}-\lambda_{1,r})\cos(\beta_{1}^{r}-\beta_{2}^r)}{\lambda_{3,r}}-\frac{\lambda_{2,r}-\lambda_{1,r}}{\sqrt{\lambda_{4,r}\lambda_{3,r}}}\tan\frac{\phi_{34}^r}{2}\sin(\beta_{1}^r-\beta_{2}^r)\right].
\end{eqnarray}
The angles $\theta^r_{ab}$ and $\phi^r_{ab}$ are defined by
\begin{eqnarray}
\tan\phi_{12}^r &=& \tan\phi_{21}^r  =	\frac{2\sin(\beta_{2}^{r}-\beta_{1}^{r})}{\frac{(\lambda_{4,r}+\lambda_{3,r})(\lambda_{1,r}-\lambda_{2,r})}{(\lambda_{4,r}-\lambda_{3,r})(\lambda_{1,r}+\lambda_{2,r})}+\cos(\beta_{2}^{r}-\beta_{1}^{r})}\left(\sqrt{\frac{\lambda_{1,r}}{\lambda_{2,r}}}+\sqrt{\frac{\lambda_{2,r}}{\lambda_{1,r}}}\right)^{-1},\\
\tan\phi_{34}^r &=& \tan\phi_{43}^r  =	\frac{2\sin(\beta_{1}^{r}-\beta_{2}^{r})}{\frac{(\lambda_{2,r}+\lambda_{1,r})(\lambda_{3,r}-\lambda_{4,r})}{(\lambda_{2,r}-\lambda_{1,r})(\lambda_{3,r}+\lambda_{4,r})}+\cos(\beta_{1}^{r}-\beta_{2}^{r})}\left(\sqrt{\frac{\lambda_{3,r}}{\lambda_{4,r}}}+\sqrt{\frac{\lambda_{4,r}}{\lambda_{3,r}}}\right)^{-1},\\
\tan\theta^r_{ab}	&=&	        \frac{2\sin\phi_{ab}^r}{\frac{\left(\sqrt{\Gamma_{a,r}}+\sqrt{\Gamma_{b,r}}\right)(\lambda_{a,r}-\lambda_{b,r})}{\left(\sqrt{\Gamma_{a,r}}-\sqrt{\Gamma_{b,r}}\right)(\lambda_{a,r}+\lambda_{b,r})}+\cos\phi_{ab}^r}\left(\sqrt{\frac{\lambda_{a,r}}{\lambda_{b,r}}}+\sqrt{\frac{\lambda_{b,r}}{\lambda_{a,r}}}\right)^{-1}.
\end{eqnarray}

The unitary $\tilde{U}^\dagger_r$ that appears in (\ref{lambdaapp}) is
\begin{equation}
\tilde{U}^\dagger_r= \frac{1}{\sqrt{2}}\begin{bmatrix}R\left(\frac{\zeta_{1}^r}{2}\right) & \sigma_{1}R\left(\frac{\zeta_{2}^r}{2}\right)\\
\sigma_{1}R\left(\frac{\zeta_{1}^r}{2}\right) & -R\left(\frac{\zeta_{2}^r}{2}\right)\end{bmatrix},
\end{equation}
where the angles $\zeta_1^r\equiv\beta_1^r+\theta_{12}^r$, $\zeta_2^r\equiv\beta_2^r+\theta_{34}^r$ with $\theta_{ab}^r$ are defined above.

It is illuminating to study the set of discrete transformations of the forward scattering matrix $\mathcal{M}$ that permute the labels of the different Luttinger parameters $K_{ab}^r$. 
To do so we first arrange the interaction parameters that appear in the forward scattering matrix as
$\boldsymbol{f}=(f_{11},f_{12},f_{22})^T$ and similarly for $\boldsymbol{g},\boldsymbol{h},\boldsymbol{\tilde{h}}$. Defining the matrices (without writing the vanishing entries)
\begin{equation}
 U_A=\left(\begin{array}{ccc}
      &&1\\
      &-1& \\
      1&&
     \end{array}\right)\quad\mbox{and}\quad U_B=\left(\begin{array}{ccc}
                                         1&  & \\
                                          &-1& \\
                                          &  & 1
                                         \end{array}\right),
\end{equation}
the discrete transformations that permute the indices are 
\begin{eqnarray}
 K_{12}^r(\boldsymbol{f},\boldsymbol{g},\boldsymbol{h},\boldsymbol{\tilde{h}})&=&K_{21}^r(U_A\boldsymbol{f},U_A\boldsymbol{g},U_A\boldsymbol{h},U_A\boldsymbol{\tilde{h}}),\\
 K_{34}^r(\boldsymbol{f},\boldsymbol{g},\boldsymbol{h},\boldsymbol{\tilde{h}})&=&K_{43}^r(U_A\boldsymbol{f},U_A\boldsymbol{g},U_A\boldsymbol{h},U_A\boldsymbol{\tilde{h}}),\\
 K_{12}^r(\boldsymbol{f},\boldsymbol{g},\boldsymbol{h},\boldsymbol{\tilde{h}})&=&K_{43}^r(U_B\boldsymbol{f},-U_B\boldsymbol{g},U_B\boldsymbol{h},-U_B\boldsymbol{\tilde{h}}),\\
 K_{21}^r(\boldsymbol{f},\boldsymbol{g},\boldsymbol{h},\boldsymbol{\tilde{h}})&=&K_{34}^r(U_B\boldsymbol{f},-U_B\boldsymbol{g},U_B\boldsymbol{h},-U_B\boldsymbol{\tilde{h}}),
\end{eqnarray}
together with the transformation $K_{ab}^+(\boldsymbol{f},\boldsymbol{g},\boldsymbol{h},\boldsymbol{\tilde{h}})=K_{ab}^-(\boldsymbol{f},\boldsymbol{g},-\boldsymbol{h},-\boldsymbol{\tilde{h}})$.
We call these transformations $A,B$ and $C$ respectively. They can be easily visualized using the following diagram

\begin{center}
\begin{tikzpicture}
  \matrix (m) [matrix of math nodes,row sep=3em,column sep=3em,minimum width=2em] 
   {  1 & 2 &   & +  \\
      4 & 3 &   & -  \\};
  \path[-stealth]
    (m-1-1) edge [blue,<->] node [above] {$A$} (m-1-2)
            edge [<->] node [left] {$B$}  (m-2-1)
    (m-2-1) edge [blue,<->] node [above] {$A$} (m-2-2)
    (m-1-2) edge [<->] node [left] {$B$} node [right] {$\quad\cong K_4,$} (m-2-2)
    (m-1-4) edge [red,<->] node [left] {$C$} node [black,right] {$\quad\cong \mathbb{Z}_2,$} (m-2-4);
\end{tikzpicture}
\end{center}
where we also indicate that the action of the $A$ and $B$ transformations, together with the identity transformation form a discrete group of order four, isomorphic to the Klein group $K_4$,
while the identity and the transformation $C$ are isomorphic to the $\mathbb{Z}_2$ group.

The scaling dimensions studied in the main text satisfy the relations
$\Delta^s_1 =\Delta_1^d + \Delta^d_4 - \frac{1}{2}(\Delta_5^d+\Delta^d_6),$ $\Delta_3^d-\Delta_1^d=\Delta^d_4-\Delta^d_2$ and $\Delta_3^d\geq 2$ for all interaction
parameters considered. 
In the AC-symmetric region $\boldsymbol{f}=U_A\boldsymbol{f},\boldsymbol{g}=U_A\boldsymbol{g}$ and $\boldsymbol{h}=\boldsymbol{\tilde{h}}=0$, we find that 
$\Delta_5^d=\Delta_6^d$ and $\Delta^d_1=\Delta^d_2=\Delta^d_3=\Delta^d_4\geq2$. Using these relations and the results for the scaling dimensions, we find that
\begin{equation}
 \Delta_{\rm FTI}=\frac{5}{2}\Delta^d_6\quad \mbox{and}\quad \Delta^s_1=2\Delta^d_3-\Delta^d_6\rightarrow\Delta_{\rm FTI}\geq10-\frac{5}{2}\Delta_1^s, 
\end{equation}
which in turn indicates that in the AC-symmetric region:
\begin{itemize}
 \item Either $\Delta_{\rm FTI}\leq 2$ or $\Delta_1^s\leq 2$,i.e., the FTI phase does not compete with the phase created by $\mathcal{O}^s_{1,\sigma}$.
 \item If $\Delta_{\rm FTI}\leq 2$ then $\Delta_5^d=\Delta^d_6\leq 2$, which indicates that the FTI phase does compete with the phase generated by the operators $\mathcal{O}_{5,6}^d$,
 \item $\Delta^d_6,\Delta^d_5\leq\Delta^d_1,\Delta^d_2,\Delta^d_3$ and $\Delta^d_4 $, signalling that the more relevant operators are indeed  $\mathcal{O}_{5,6}^d$.
\end{itemize}

Although these considerations are strictly valid in the AC-symmetric region of parameters, evaluating the scaling dimensions for small values of the parameters, we find that this
correspond to the generic situation. To see where this considerations break down, we go away from the highly symmetric AC region.

Along the C-symmetric region $\boldsymbol{h}=\boldsymbol{\tilde{h}}=0$, we have the relations $\Delta_1^d=\Delta_2^d$ and $\Delta_3^d=\Delta_4^d$.
In this region, the scaling dimension of the FTI operator satisfies
\begin{equation}\label{eq_scaling_dims}
 \Delta_{\rm FTI}=\frac{5}{2}(\Delta^d_1+\Delta^d_3)-\frac{5}{2}\Delta_1^s+\frac{3}{4}(\Delta^d_6-\Delta^d_5)+2(\Delta^d_1-\Delta^d_3).
\end{equation}
In order to have a region of parameter where both $\Delta_{\rm FTI}\leq 2$ and $\Delta_1^s\leq 2$, we find, using the relation above, that
\begin{equation}
 \frac{2}{3}+(\Delta_1^d-\Delta_3^d)-\frac{1}{6}(\Delta^d_5-\Delta_6^d)\leq 0,
\end{equation}
which corresponds to strong interactions and interleg tunnelling. In this region the analysis in the main text does not apply.

\section{Order parameters}\label{app:Order_params}
 
The order parameters $ O_{\mu,\sigma,x}=\sum_{\beta,\beta'}\left(c^{\dagger \beta}_{x,\sigma} (\tau_\mu)_{\beta\beta'}c^{\beta'}_{x,\sigma}\right),$ defined in
(\ref{order_params}) 
become after bosonization
\begin{equation}
O_{\mu,\sigma}(x)=O_{\mu,\sigma}^0(x)+\sum_\alpha \left(e^{i\Delta k_\alpha x}O_{\mu,\sigma}^\alpha(x)+\text{h.c.}\right),
\end{equation}
where $\alpha\equiv b,b^\prime,\eta,\eta^\prime$ in the differences $\Delta k_\alpha=k_{F,b}^\eta-k_{F,b^\prime}^{\eta^\prime}$ and
$O_{\mu,\sigma}^\alpha(x)$ are slowly varying operators. Focusing on the the oscillating parts that couple different chiral components as these act 
as local order parameters to distinguish  the  different phases  of  the system, we have 
\begin{eqnarray}
  O_{0\sigma}(x)&\rightarrow&O_{\sigma,{\rm CDW^+}}(x)= e^{-i\frac{\Phi}{3a}x}\left(e^{i\pi N_{\sigma,1}^L}e^{i\sqrt{\pi}(\Upsilon^c_{\sigma}+\Theta_\sigma)}+e^{i\pi N_{\sigma,2}^L}e^{i\sqrt{\pi}(\Upsilon^c_{\sigma}-\Theta_\sigma)}\right)+\text{h.c.},\\
  O_{3\sigma}(x)&\rightarrow&O_{\sigma,{\rm RDW}}(x)=   e^{-i\frac{\Phi}{3a}x}\left(e^{i\pi N_{\sigma,1}^L}e^{i\sqrt{\pi}(\Upsilon^c_{\sigma}+\Theta_\sigma)}-e^{i\pi N_{\sigma,2}^L}e^{i\sqrt{\pi}(\Upsilon^c_{\sigma}-\Theta_\sigma)}\right)+\text{h.c.},\\
  O_{1\sigma}(x)&\rightarrow&O_{\sigma,{\rm BDW}}(x)=   e^{i\frac{\Phi}{3a}x}\left(e^{-i\sqrt{\pi}(\Upsilon_{\sigma}^{c}+\Upsilon_{\sigma})}e^{i\frac{\Phi}{a} x}+e^{i\pi(N_{\sigma,2}^{L}+N_{\sigma,1}^{L})}e^{-i\sqrt{\pi}(\Upsilon_{\sigma}^{c}-\Upsilon_{\sigma})}e^{-i\frac{\Phi}{a} x}\right)e^{i\pi N_{\sigma,1}^{R}}+\text{h.c.},\\
  O_{2\sigma}(x)&\rightarrow&O_{\sigma,{\rm OAF}}(x)=  ie^{i\frac{\Phi}{3a}x}\left(e^{-i\sqrt{\pi}(\Upsilon_{\sigma}^{c}+\Upsilon_{\sigma})}e^{i\frac{\Phi}{a} x}-e^{i\pi(N_{\sigma,2}^{L}+N_{\sigma,1}^{L})}e^{-i\sqrt{\pi}(\Upsilon_{\sigma}^{c}-\Upsilon_{\sigma})}e^{-i\frac{\Phi}{a} x}\right)e^{i\pi N_{\sigma,1}^{R}}+\text{h.c.},
 \end{eqnarray}
 with $\Upsilon^c_{\sigma}$ and $\Theta_\sigma$ defined in (\ref{basis_app}). For $\Theta_\sigma=\sqrt{\pi}n$, the only order parameter with QLRO is $O_{0\sigma}(x)$ if 
 $N^L_{\sigma,1}=N^L_{\sigma,2}$. In the Luttinger liquid phase 
 with $c=4$, all correlators decay with power law and different non-universal exponents. The smallest exponents in this phase corresponds to the BDW and OAF order parameters.
 
 The orbital singlet and triplet superconducting order parameters are
 \begin{eqnarray}
 S_{0\sigma}(x)&\rightarrow&S^{\rm singlet}_{\sigma,\text{leg}}(x)       =-ie^{i\pi N_{\sigma,1}^{R}}\left(e^{i\sqrt{\pi}(\Theta_{\sigma}^{c}+\Theta_{\sigma})}+e^{i\pi(N_{\sigma,2}^{L}+N_{\sigma,1}^{L})}e^{i\sqrt{\pi}(\Theta_{\sigma}^{c}-\Theta_{\sigma})}\right),\\
 S_{3\sigma}(x)&\rightarrow&S^{\rm triplet}_{\sigma;{\rm z},\text{leg}}(x)=-e^{i\pi N_{\sigma,1}^{R}}\left(e^{i\sqrt{\pi}(\Theta_{\sigma}^{c}+\Theta_{\sigma})}-e^{i\pi(N_{\sigma,2}^{L}+N_{\sigma,1}^{L})}e^{i\sqrt{\pi}(\Theta_{\sigma}^{c}-\Theta_{\sigma})}\right),\\
 S_{2\sigma}(x)&\rightarrow&S^{\rm triplet}_{\sigma;{\rm y},\text{leg}}(x)=e^{i\pi N_{1L}}e^{i\sqrt{\pi}(\Theta_{\sigma}^{c}+\Upsilon_{\sigma})}e^{i\frac{\Phi}{a}x}+e^{i\pi N_{2L}}e^{i\sqrt{\pi}(\Theta_{\sigma}^{c}-\Upsilon_{\sigma})}e^{-i\frac{\Phi}{a}x},\\
 S_{1\sigma}(x)&\rightarrow&S^{\rm triplet}_{\sigma;{\rm x},\text{leg}}(x)=e^{i\pi N_{1L}}e^{i\sqrt{\pi}(\Theta_{\sigma}^{c}+\Upsilon_{\sigma})}e^{i\frac{\Phi}{a}x}-e^{i\pi N_{2L}}e^{i\sqrt{\pi}(\Theta_{\sigma}^{c}-\Upsilon_{\sigma})}e^{-i\frac{\Phi}{a}x}.
 \end{eqnarray}
 
 In the FTI precursor phase, the natural basis for the fields are given by (\ref{U_strong}), which relate to the fields above as
 \begin{eqnarray}\label{basis_app2}
 \begin{pmatrix}\Upsilon_{\sigma}^c\\
\Theta_{\sigma}^c\\
\Upsilon_{\sigma}\\
\Theta_{\sigma}
\end{pmatrix}
\equiv\begin{pmatrix}0 & \frac{1}{3} & \frac{1}{3} & \frac{1}{3}\\
-1 & 0 & 1 & -1\\
0 & 1 & -1 & -1\\
\frac{1}{3} & 0 & \frac{1}{3} & -\frac{1}{3}
\end{pmatrix}\begin{pmatrix}\varphi_{g\sigma}\\
\theta_{g\sigma}\\
\tilde{\phi}_{L\sigma}\\
\tilde{\phi}_{R\sigma}
\end{pmatrix}. 
\end{eqnarray}
The pinning of the $\theta_{g\sigma}$ field in the FTI phase makes the local order parameters $O_{1\sigma}$ (BDW) and $O_{2\sigma}$ (OAF) develop QLRO.
 
\section{Scaling dimensions}\label{app:Scaling}
 
In this appendix we derive Eq. (\ref{Lambda}), by which the matrix $\Lambda$ is defined. This matrix controls the scaling dimensions of the theory. Starting with the 
quadratic theory 
\begin{equation}\label{quad_act}
 S_{\rm quad}=\int dtdx \left(\partial_t\bm{\phi}^T\mathcal{K}\partial_x\bm{\phi}-\partial_x\bm{\phi}^T\mathcal{M}\partial_x\bm{\phi}\right),
\end{equation}
we are interested in the scaling dimension $\Delta_\eta$ of the operator $\mathcal{O}_{\bm{\eta}}=e^{i\bm{\eta}^T\cdot\bm{\phi}}$. In a renormalization {\it a la Wilson}, 
where a high momentum shell $\lambda'<k<\lambda$ is integrated out, the scaling dimension of the operator is found by \cite{GiamarchiBook2003}
 \begin{equation}\label{scaling_dim_def}
  \langle\mathcal{O}_\eta\rangle_{[\lambda',\lambda]}=\langle e^{i\eta^T\cdot\bm{\phi}}\rangle_{[\lambda',\lambda]}= 
  e^{-\frac{1}{2}\sum_{jl}\eta_j\langle\bm{\phi}_j\bm{\phi}_l\rangle_{[\lambda',\lambda]}\eta_l}=\left(\frac{\lambda'}{\lambda}\right)^{\Delta_{\eta}},
 \end{equation}
where the expectation value $\langle \cdot \rangle_{[\lambda',\lambda]}$ is taken over the high momentum shell $\lambda'<||\bm{q}||<\lambda$, with $\bm{q}=(k,\omega/v)$. 
We are then interested in the correlation function $\langle\bm{\phi}_j(r)\bm{\phi}_l(r)\rangle_{[\lambda',\lambda]}$, which can be computed directly from the quadratic 
action (\ref{quad_act}) by going to momentum and frequency representation (in imaginary time) and obtaining the Green's function. In particular 
\begin{equation}\label{integral2}
 \langle\bm{\phi}_j(r)\bm{\phi}_l(r)\rangle_{[\lambda',\lambda]}=\frac{1}{2}\int_{[\lambda',\lambda]} \frac{d\omega dk}{(2\pi)^2} \left[\left(ik\omega\mathcal{K}+k^2\mathcal{M}\right)^{-1}\right]_{jl}.
\end{equation}
Using that the forward scattering matrix $\mathcal{M}$ is positive definite, the matrix $ik\omega\mathcal{K}+k^2\mathcal{M}$ can be cast in the form
\begin{equation}
 ik\omega\mathcal{K}+k^2\mathcal{M}=\mathcal{M}^{\frac{1}{2}}(ik\omega\mathcal{M}^{-\frac{1}{2}}\mathcal{K}\mathcal{M}^{-\frac{1}{2}}+k^2)\mathcal{M}^{\frac{1}{2}}.
\end{equation}
The matrix $\mathcal{M}^{-\frac{1}{2}}\mathcal{K}\mathcal{M}^{-\frac{1}{2}}$ is real symmetric, so it can be diagonalized by an orthogonal transformation $O$. Defining
$O\mathcal{M}^{-\frac{1}{2}}\mathcal{K}\mathcal{M}^{-\frac{1}{2}} O^T\equiv\mathbb{V}^{-1}$, with $\mathbb{V}=\text{diag}({v_i})$ a diagonal matrix, we have 
(using $OO^T=O^TO=\openone$)
\begin{equation}
 ik\omega\mathcal{K}+k^2\mathcal{M}=\mathcal{M}^{\frac{1}{2}}O^T(ik\omega\mathbb{V}^{-1}+k^2)O\mathcal{M}^{\frac{1}{2}},
\end{equation}
which implies
\begin{equation}
 [(ik\omega\mathcal{K}+k^2\mathcal{M})^{-1}]_{jl}=\sum_m[\mathcal{M}^{-\frac{1}{2}}O^T]_{jm}\left(\frac{1}{ik\omega/v_m+k^2}\right)[O\mathcal{M}^{-\frac{1}{2}}]_{ml},
\end{equation}
This representation allows us to perform the integral (\ref{integral2}), which becomes
\begin{equation}
 \langle\bm{\phi}_j(r)\bm{\phi}_l(r)\rangle_{[\lambda',\lambda]}=\sum_m[\mathcal{M}^{-\frac{1}{2}}O^T]_{jm}\frac{|v_m|}{4\pi}\int_{\lambda'}^\lambda \frac{dq}{q}[O\mathcal{M}^{-\frac{1}{2}}]_{ml}
\end{equation}
note that the factor $|v_m|$ appears as the Jacobian of the transformation to polar coordinates. Rearranging and using 
$\text{diag}(v_m)=O\mathcal{M}^{\frac{1}{2}}\mathcal{K}^{-1}\mathcal{M}^{\frac{1}{2}} O^T$ we have
\begin{equation}\label{corr_fnt}
 \langle\bm{\phi}_j(r)\bm{\phi}_l(r)\rangle_{[\lambda',\lambda]}=\frac{1}{4\pi}\ln\left(\frac{\lambda}{\lambda'}\right)[\mathcal{M}^{-\frac{1}{2}}|\mathcal{M}^{\frac{1}{2}}\mathcal{K}^{-1}\mathcal{M}^{\frac{1}{2}}|\mathcal{M}^{-\frac{1}{2}}]_{jl}.
\end{equation}
Finally, comparing (\ref{corr_fnt}) and (\ref{scaling_dim_def}), we conclude that 
$\Delta_\eta=\frac{1}{8\pi}\bm{\eta}^T\mathcal{M}^{-\frac{1}{2}}|\mathcal{M}^{\frac{1}{2}}\mathcal{K}^{-1}\mathcal{M}^{\frac{1}{2}}|\mathcal{M}^{-\frac{1}{2}}\bm{\eta}\equiv\bm{\eta}^T\Lambda \bm{\eta}$, as expected.

\section{Luttinger parameters for strongly interacting ladders with weak interleg tunneling}
\label{app:KrhoKbeta}

In this Appendix, we consider fermion ladders dominated by strongly repulsive leg-SU($2$) invariant interactions. We focus on vanishing interleg tunneling and interspin interactions, $t_\perp,V^d_{\parallel,m},V^d_{\perp,m}\rightarrow 0$. Our goal is to support our discussion in Sec.~\ref{sec:FTIprecursor1} by showing that the system forms a $c=4$ Luttinger liquid with $K_\rho$ and $K_\beta$ in the region where only $\mathcal{O}_\text{FTI}$ is relevant. In particular, we show that the charge Luttinger parameter reaches $K_\rho\leq 0.25$ while the leg analogue of the spin Luttinger parameter satisfies $K_\beta\rightarrow 1$. 

\begin{figure}[ht!]
 \includegraphics[width=0.5\linewidth]{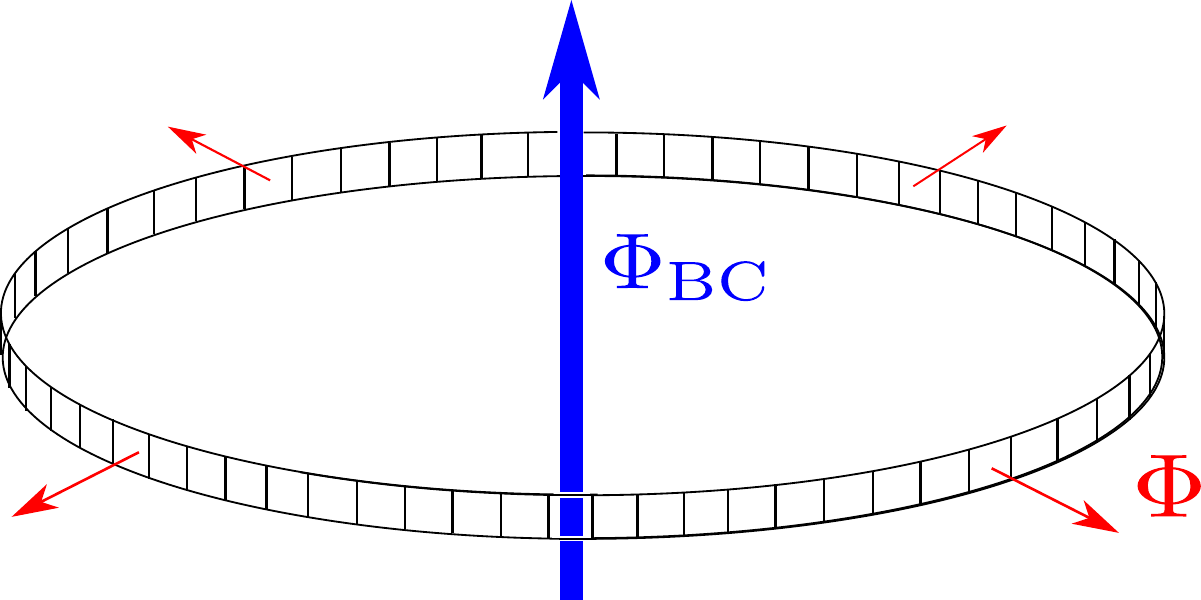}
\caption{(Color online). Flux $\Phi_{\rm BC}$ threading the ladder along a ring.  The change in the ground state energy with respect to this flux determines (through $\frac{\partial^2E_\text{GS}}{\partial \Phi_{\rm BC}^2}$) the charge stiffness. The flux $\Phi$
though the plaquettes of the ladder is responsible for the FTI precursor; however it is inoperative in the $t_\perp\rightarrow 0$ limit considered in this Appendix.}\label{fig:Fluxes}
\end{figure}

We calculate $K_\rho$ following the methods of Ref. \onlinecite{G-Santos1993}; we compute the ground state energy $E_{\rm GS}$, the inverse compressibility $L^{-1}\frac{\partial^2 E_{\rm GS}}{\partial\rho^2}$ and the phase stiffness 
$L\left.\frac{\partial^2 E_{\rm GS}}{\partial\Phi_{\rm BC}}\right|_{\Phi_\text{BC}=0}$ for a ladder along a ring of circumference $L$ threaded by a flux $\Phi_{\rm BC}$ (Fig.~\ref{fig:Fluxes}) and with one-dimensional fermion density $\rho$.  $K_\rho$ is calculated using the 
relation \cite{GiamarchiBook2003}
\begin{equation}\label{Luttinger_param}
 K_\rho = \frac{\pi}{2}\sqrt{\frac{L\frac{\partial^2 E_{\rm GS}}{\partial\Phi_{\rm BC}}}{L^{-1}\frac{\partial^2 E_{\rm GS}}{\partial\rho^2}}}.
\end{equation}
The effect of $\Phi_{\rm BC}$ is equivalent to twisting the boundary conditions: the vector potential associated to $\Phi_{\rm BC}$ can be eliminated in favor of imposing the boundary condition $\psi(L)=e^{i\Phi_\text{BC}}\psi(0)$ on the fermions in the system. Here, we choose to work with periodic boundary conditions and implement  $\Phi_{\rm BC}$ through a translation-invariant vector potential along the links, i.e., by adding a phase $a\Phi_{\rm BC}/L$ to each link of a leg \cite{GiamarchiBook2003} ($a$ is the lattice spacing along the legs).   

The interactions, Eq.~\eqref{interactions} in the main text, are parametrized by  
$V^s_{\parallel,m},V^s_{\perp,m}$ which describe the interaction between fermions $m$ sites apart of the same spin projections in the same ($\parallel$) and
different ($\perp$) legs of the ladder. In the following discussion we take
\begin{equation}
 V^s_{\perp, 0} =U \quad \mbox{and}\quad
V^s_{\perp, m}=V^s_{\parallel,m} =\begin{cases}
          V \quad &\mbox{for}\,\,  1\leq m \leq \ell_0,\\
         0 \quad &\mbox{for} \, \, m > \ell_0.
        \end{cases}
\end{equation}
For $\ell_0=0$ this model reduces to the Hubbard model, while for $\ell_0=1$ to the extended Hubbard (or $U$-$V$) model \cite{GiamarchiBook2003}.
Interactions dominate when $V,U\gg t$. For infinite $U,V$ the repulsion becomes a 
constraint: particles separated by no more than $\ell_0$ sites pay an infinite amount of energy while particles separated by 
more than $\ell_0$ sites experience no interactions. This allows one to map the system onto a fictitious one with density-dependent length, where the constraint is taken into account by the Pauli principle leaving one with a simple free-fermion problem.

\begin{figure}[ht!]
 \includegraphics[width=\linewidth]{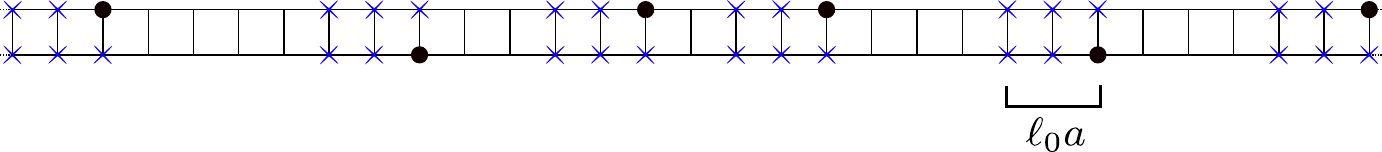}
\caption{(Color online). Occupation of the two-leg ladder as a function of particle number. Particles are represented by black circles. Blue crosses represent forbidden sites due to the infinite strong repulsion of range $\ell_0$. $L_n$ is the available space to place the $n$th particle.  In this diagram, $\ell_0=2$ }\label{fig:ladder_occupation}
\end{figure}

Focusing on a single spin species, we consider a system with $N$ particles. The infinite interactions translate to each of the particles effectively blocking out $\ell_0$ rungs, for definiteness to their left, in addition to the rung they sit on (Fig.~\ref{fig:ladder_occupation}).
Thus the number of rungs available for the $N$ particles is $L/a-\ell_0N$ translating into an effective length 
\begin{equation}
L^\prime=L-Na\ell_0=L(1-\rho a \ell_0),
\end{equation}
where $\rho=N/L$ is the 1D density of the spin species. Note that $\rho<\frac{1}{a(\ell_0+1)}$ otherwise there are not enough rungs to place the particles without violating the constraint. In this 
reduced system, the particles behave as free fermions, with the Hamiltonian given by Eq.~\eqref{eq:H0par} in the main text. The momenta are quantized as 
$k_m =\frac{2\pi m}{L'}$ (with $m$ integer). In the regime where we are interested, these fermions occupy two pockets in the lower band, in a dispersion relation analogous to Fig. \ref{fig:Dispersion_alphazero}. (The upper and lower bands, however, now touch due to $t_\perp\rightarrow 0$.) The ground state energy is
\begin{equation}
 {E_{GS}}(\Phi_{\rm BC}) = -2t\sum_{m=m_{\rm min}}^{m_{\rm max}}\cos\left(k_m a+\frac{a\Phi_{\rm BC}}{L}-\frac{\Phi}{2}\right). 
\end{equation}
Here $m_{\rm min}$ and $m_{\rm max}$ are the minimum and maximum momentum label in the right pocket. We used that, for $\Phi_\text{BC}=0$, the system is reflection symmetric to sum over just positive momenta, that for $t_\perp\rightarrow 0$ and $\Phi_\text{BC}=0$ the lower-band single-particle energies for positive momenta follow the cosine dispersion $-t\cos(ka-\Phi/2)$, and that for $\Phi_\text{BC}\neq 0$ but $\Phi_\text{BC}/L\ll 2\pi/L^\prime$ the presence of $\Phi_\text{BC}$ does not change the set of occupied momentum labels $\{m\}$.
Using now $(m_{\rm max}-m_{\rm min})/L=\rho/2$ and summing the geometric series we find
\begin{equation}
 E_{GS}(\Phi_{\rm BC})=-\frac{2tL}{a\pi}(1-\rho a\ell_0)\cos\left[\frac{\pi\rho a}{2(1-\rho a\ell_0)}+k_{\rm min}a-\frac{\Phi}{2}+\frac{a\Phi_{\rm BC}}{L}\right]\sin\left[\frac{\pi\rho a}{2(1-\rho a\ell_0)}\right].
\end{equation}
To be able to compute the inverse compressibility, we have to determine the dependence of $k_{\rm min}=2\pi m_\text{min}/L^\prime$ on the density of the system. Using that the number of occupied states to the left of the minimum $\Phi/2$ of the cosine dispersion is $\rho L/4$ we find $k_{\rm min}a = \Phi/2-\frac{\pi\rho a}{2(1-\rho a\ell_0)}$, which leads to
\begin{equation}
 E_{GS}(\rho,\Phi_{\rm BS})=-\frac{2tL}{a\pi}(1-\rho a \ell_0)\cos\left[\frac{a \Phi_{\rm BC}}{L}\right]\sin\left[\frac{\pi a\rho}{2(1-\rho a\ell_0)}\right].
\end{equation}
The gapless nature of the effective free-fermion problem indicates that the charge sector of the original system forms a Luttinger liquid. Using Eq. (\ref{Luttinger_param}), we find that the charge Luttinger parameter is
\begin{equation}\label{eq:Krhoexpr}
 K_\rho \xrightarrow[U,V\rightarrow\infty]{} \frac{1}{2}(1-\rho a\ell_0)^2.
\end{equation}
A useful check of Eq.~\eqref{eq:Krhoexpr} is the Hubbard model ($\ell_0=0$): here it is known \cite{GiamarchiBook2003} that at infinite repulsion the Luttinger parameter reaches $K_\rho =\frac{1}{2}$ which is precisely what we find. For the $U$-$V$ model ($\ell_0=1$), Eq.~\eqref{eq:Krhoexpr} shows that the charge Luttinger parameter reaches $K_\rho= 0.125$ at the limiting density $\rho=\frac{1}{a(\ell_0+1)}$. For the FTI physics we are interested in, we need two partially filled pockets as in Fig.~\ref{fig:Dispersion_alphazero}. This places a stronger constraint on the density $\rho a=\nu\frac{\Phi}{\pi}$; taking this also into account we find that the Luttinger parameter can still reach $K_\rho\approx 0.15$.

For large, but not infinite, interactions, the corrections to $K_\rho$ are of order $t/U$ or $t/V$, corresponding to small corrections to the infinite interaction result in the limit $U,V\gg t$. In the same regime, the interleg degrees of freedom also form a Luttinger liquid in a significant portion of the $U$-$V$ parameter space, including for $U$-$V$ interaction ratios where the inteleg gap is absent for weak interactions at the same ratios \cite{Sano94,Nakamura99,Nakamura00,Tsuchiizu04,Tsuchiizu04,Sandvik04,Ejima07}.
Owing to SU($2$) invariance in leg space, this implies $K_\beta\rightarrow 1$ at low energies \cite{GiamarchiBook2003} for this interleg Luttinger liquid. We therefore find that in the strongly repulsive leg-SU($2$) invariant regime, the $t_\perp\rightarrow 0$ system forms a $c=4$ Luttinger liquid with $K_{\rho,\beta}$ such that only $\mathcal{O}_\text{FTI}$ is relevant. This is the key requirement for an FTI precursor to emerge upon turning on weak $t_\perp$, as discussed in Sec.~\ref{sec:FTIprecursor1}.

\section{Renormalization group equations}\label{app:RGeqs}

In this appendix we derive the RG equations for small interleg tunnelling ($t_\perp/t\ll 1$) and zero interspin interaction.
In the basis (\ref{basis_app}), the action reads
\begin{eqnarray}
S & = &\frac{1}{2}\int dxdt\left(\partial_{t}\Theta_{\sigma}^{c}\partial_{x}\Upsilon_{\sigma}^{c}+\partial_{t}\Theta_{\sigma}\partial_{x}\Upsilon_{\sigma}\right)
- \frac{1}{4}\int dxdt\left(\left[v_{\rho}K_{\rho}(\partial_{x}\Theta_{\sigma}^{c})^{2}+\frac{v_{\rho}}{K_{\rho}}(\partial_{x}\Upsilon_{\sigma}^{c})^{2}\right]+\left[v_{\beta}K_{\beta}(\partial_{x}\Upsilon_{\sigma})^{2}+\frac{v_{\beta}}{K_{\beta}}(\partial_{x}\Theta_{\sigma})^{2}\right]\right)\nonumber \\
 & - & \frac{1}{2}\int dxdt\left(g_{\Theta}\partial_{x}\Theta_{\sigma}^{c}\partial_{x}\Theta_{\sigma}+g_{\Upsilon}\partial_{x}\Upsilon_{\sigma}^{c}\partial_{x}\Upsilon_{\sigma}+\frac{g_{1\sigma}^{s}}{(\pi a)^{2}}\cos(\sqrt{4\pi}\Theta_{\sigma})\right).
\end{eqnarray}
In terms the forward scattering parameters $f_{mn},g_{mn},h_{mn}$ and $\tilde{h}_{mn}$, the parameters above are
\begin{eqnarray}\nonumber
v_{\rho}K_{\rho}^{-1}  =  v_{F}+\frac{f_{11}+f_{22}+(g_{11}+g_{22})+2(f_{12}+g_{12})}{8\pi},&\quad&
v_{\rho}K_{\rho}  =  v_{F}+\frac{f_{11}+f_{22}-(g_{11}+g_{22})-2(f_{12}-g_{12})}{8\pi},\\\nonumber
v_{\beta}K_{\beta}^{-1}  =  v_{F}+\frac{f_{11}+f_{22}-(g_{11}+g_{22})+2(f_{12}-g_{12})}{8\pi},&\quad&
v_{\beta}K_{\beta}  =  v_{F}+\frac{f_{11}+f_{22}+(g_{11}+g_{22})-2(f_{12}+g_{12})}{8\pi},\\
g_{\Theta}  =  \frac{f_{11}-f_{22}-g_{11}+g_{22}}{2\pi},&\quad&
g_{\Upsilon}  =  \frac{f_{11}-f_{22}+g_{11}-g_{22}}{2\pi}.
\end{eqnarray}

Integrating out the fields $\Theta^c_{\sigma}$ and $\Upsilon_\sigma$ in the path integral, we are left with the action
\begin{align}\nonumber
S=\frac{1}{4}\int dxdt\left(\frac{1}{\tilde{K}_{\rho}}\left[\frac{(\partial_{t}\Upsilon_{\sigma}^{c})^{2}}{\tilde{v}_{\rho}}-\tilde{v}_{\rho}(\partial_{x}\Upsilon_{\sigma}^{c})^{2}\right]+\frac{1}{\tilde{K}_{\beta}}\left[\frac{(\partial_{t}\Theta_{\sigma})^{2}}{\tilde{v}_{\beta}}-\tilde{v}_{\beta}(\partial_{x}\Theta_{\sigma})^{2}\right]\right)\\
-\int dxdt\frac{g_{1\sigma}^{s}}{2(\pi a)^{2}}\cos(\sqrt{4\pi}\Theta_{\sigma})-\frac{1}{2}g\int dxdt\left(\partial_{t}\Theta_{\sigma}\partial_{x}\Upsilon_{\sigma}^{c}\right),
\end{align}
where we have defined $\tilde{v}_{\rho}\tilde{K}_{\rho}=v_{\rho}K_{\rho}$ , $\frac{\tilde{v}_{\rho}}{\tilde{K}_{\rho}}=\frac{v_{\rho}}{K_{\rho}}-\frac{g_{\Upsilon}^{2}}{v_{\beta}K_{\beta}}$,
 $\tilde{v}_{\beta}\tilde{K}_{\beta}=v_{\beta}K_{\beta}$ , $\frac{\tilde{v}_{\beta}}{\tilde{K}_{\beta}}=\frac{v_{\beta}}{K_{\beta}}-\frac{g_{\Theta}^{2}}{v_{\rho}K_{\rho}}$, and  $g=\left(\frac{g_{\Theta}}{v_{\rho}K_{\rho}}+\frac{g_{\Upsilon}}{v_{\beta}K_{\beta}}\right).$
Using the action above, we find the RG equations 
\begin{eqnarray}
 \frac{dg_{1\sigma}^{s}}{d\ell}=\left(2-2\tilde{K}_{\beta}\right)g_{1\sigma}^{s},\quad
 \frac{d}{d\ell}\left(\frac{1}{\tilde{K}_{\beta}}\right)=\left(\frac{g_{1\sigma}^{s}}{4\pi v_{F}}\right)^{2}-\frac{g^2}{2\pi}\tilde{K}_{\rho},\quad 
 \frac{d}{d\ell}\left(\frac{\tilde{v}_\beta}{\tilde{v}_\rho K_\rho}\right)=-\frac{\tilde{K}_\beta g^2}{2\pi}.
\end{eqnarray}
The operator product expansion of the cosine term $\cos(\sqrt{4\pi}\Theta_\sigma)$ with $\partial_x\Upsilon_\sigma^c \partial_t\Theta_\sigma$ generates the operator
$\sin(\sqrt{4\pi}\Theta_\sigma)\partial_x\Upsilon^c_\sigma$, initially not present in the Hamiltonian. We call the coupling constant of this new field $g_{\rm new}$.

Discarding terms that are either quartic in tunnelling $t_{\perp}^{4}$
or cubic in interactions $V^{3},$ the RG equations are
\begin{equation}
\frac{d}{d\ell}\left(\frac{2f_{12}-g_{11}-g_{22}}{4\pi v_{F}}\right)=\left(\frac{g_{1\sigma}^{s}}{4\pi v_{F}}\right)^{2},\quad\frac{dg_{1\sigma}^{s}}{d\ell}=\left(\frac{2f_{12}-g_{11}-g_{22}}{4\pi v_{F}}\right)g_{1\sigma}^{s},\quad \frac{dg_{\rm new}}{d\ell}=(1-2K_{\beta})g_{\rm new}-2\frac{g}{\sqrt{\pi}}\frac{g_{1\sigma}^{s}}{4\pi v_F}
\end{equation}
where $g_{\rm new}(\ell=0)=0$. Defining $\frac{2f_{12}-g_{11}-g_{22}}{4\pi v_{F}}=-y_\parallel$, $\frac{g_{1\sigma}^{s}}{4\pi v_{F}}=y$ and 
$g_{\rm new}=\frac{2g}{\sqrt{\pi}}y_{\rm new}$ these equations can be casted in the form
\begin{equation}
\frac{dy_{\parallel}}{d\ell}=-y^{2},\quad
\frac{dy}{d\ell}=-y_{\parallel}y,   \quad 
\mbox{and} \quad \frac{dy_{\rm new}}{d\ell}=-(y_\parallel+1)y_{\rm new}-y.
\end{equation}
The last equation can be solved easily. It gives $y_{\rm new}=(e^{-\ell}-1)y(\ell)$. Although this indicates that $y_{\rm new}$ flows to strong coupling as $y$ does, 
the coupling $g_{\rm new}=\frac{2g}{\sqrt{\pi}}y_{\rm new}$ stays small (as $g\ll1$ does not flow at this order of approximation). The physics is then dominated by
the RG flow of $y$ and $y_{\parallel}$.
The first two equations are the standard Kosterlitz-Thouless \cite{Kosterlitz1973} renormalization group equations. The separatrix that divides the regions where the cosine operator becomes 
relevant or irrelevant corresponds to $|y|=y_{\parallel}$. In terms of the $f,g$ parameters is given by $g_{11}+g_{22}-2f_{12}=g_{1\sigma}^{s}.$ This equation dictates 
the separatrix between the Luttinger liquid phase and the phase with dominant RDW QLRO, in the top panel of Fig. \ref{fig:phases_spindecoupled}.

\section{Interaction matrices for strong coupling analysis}\label{app:strong_coupling}
In this appendix we write the the interaction matrices of the different sectors needed for the strong coupling phase. The quadratic Hamiltonian is given by
Eq.~\eqref{H_0_strong} for decoupled spins and Eq.~\eqref{H_inter_spin_strong} for the full coupled case, where the forward interaction matrices are given by  
\begin{equation}
\tilde{\mathcal{M}}=\begin{bmatrix}\mathcal{M}^{(2)}_{hh}& \mathcal{M}^{(2)}_{hs}\\
(\mathcal{M}^{(2)}_{hs})^T & \mathcal{M}^{(2)}_{ss}
\end{bmatrix}\quad\mbox{and}\quad
{\mathcal{M}}=\begin{bmatrix}{\mathcal{M}}^{(4)}_{hh}& {\mathcal{M}}^{(4)}_{hs}\\
({\mathcal{M}}^{(4)}_{hs})^T & {\mathcal{M}}^{(4)}_{ss}
\end{bmatrix},
\end{equation}
respectively, where  $\mathcal{M}^{(m)}_{ab}$ is a $m\times m$ matrix.
The different blocks are given explicitly in terms of the microscopic parameters. For the decoupled spin sector they are
\[
\mathcal{M}^{(2)}_{hh}=\begin{bmatrix}\gamma_{3}-\eta_{3}\\
 & \gamma_{3}+\eta_{3}
\end{bmatrix}
=v\begin{bmatrix}K_{\rm eff}\\
 & K_{\rm eff}^{-1}
\end{bmatrix},\,\,
\mathcal{M}^{(2)}_{hs}=-\begin{bmatrix}\gamma_{2}-\eta_{2} & \eta_{2}-\gamma_{2}\\
\gamma_{2}+\eta_{2} & \gamma_{2}+\eta_{2}
\end{bmatrix}\quad\text{and}\quad \mathcal{M}^{(2)}_{ss}=\begin{bmatrix}\gamma_{1} & \eta_{1}\\
\eta_{1} & \gamma_{1}
\end{bmatrix}
\]

with
\begin{eqnarray*}
\gamma_{1}=\frac{1}{9}[10v_{F}+\frac{1}{2\pi}(f_{11}+4(f_{22}-f_{12})], & \quad & \eta_{1}=\frac{1}{18\pi}[g_{11}+4(g_{22}-g_{12})],\\
\gamma_{2}=\frac{1}{18}[8v_{F}+\frac{1}{2\pi}(2(f_{22}+f_{11})-5f_{12})], & \quad & \eta_{2}=\frac{1}{36\pi}[2(g_{22}+g_{11})-5g_{12}],\\
\gamma_{3}=\frac{1}{18}[10v_{F}+\frac{1}{2\pi}(f_{22}+4(f_{11}-f_{12}))], & \quad & \eta_{3}=\frac{1}{36\pi}[g_{22}+4(g_{11}-g_{12})].
\end{eqnarray*}

Including interspin interactions, the matrices are given by $\mathcal{M}^{(4)}_{ab}=\openone_2\otimes\mathcal{M}^{(2)}_{ab}+\sigma_1\otimes\bar{\mathcal{M}}^{(2)}_{ab}$,
where $\bar{\mathcal{M}}^{(2)}_{ab}$ is obtained from $\mathcal{M}^{(2)}_{ab}$ by replacing the parameters $\gamma_i,\eta_i$ by the parameters $\bar{\gamma}_i,\bar{{\eta}}_i$,
given by
\begin{eqnarray}
\bar{\gamma}_{1}=\frac{1}{18\pi}[h_{11}+4(h_{22}-h_{12})],   & \quad & \bar{\eta}_{1}=\frac{1}{18\pi}[\tilde{h}_{11}+4(\tilde{h}_{22}-\tilde{h}_{12})],\\
\bar{\gamma}_{2}=\frac{1}{36\pi}[2(h_{22}+h_{11})-5h_{12})], & \quad & \bar{\eta}_{2}=\frac{1}{36\pi}[2(\tilde{h}_{22}+\tilde{h}_{11})-5\tilde{h}_{12}],\\
\bar{\gamma}_{3}=\frac{1}{36\pi}[h_{22}+4(h_{11}-h_{12})],    & \quad & \bar{\eta}_{3}=\frac{1}{36\pi}[4\tilde{h}_{12}-\tilde{h}_{22}-4\tilde{h}_{11}].
\end{eqnarray}

\section{Region of microscopic parameters where the FTI operator can appear, for nonvanishing SO coupling $\alpha_{\rm so}$}\label{app:simplex}

Considering the single particle band structure, the FTI operator that drives the precursor of the topological phase can exist only in the presence of four Fermi points.
The FTI operator will conserve momentum if the Fermi points satisfy $2(k_{F,2}^L-k_{F,1}^R)+k_{F,2}^R-k_{F,1}^L=0$. For some values of the parameters of the single 
particle band spectrum it is impossible to fix the Fermi energy to have four Fermi points. In this appendix we explore the region of parameters where four
Fermi points are possible. A subset of this region corresponds to the parameters where the Fermi points satisfy the momentum conservation condition.

\begin{figure}[ht!]
	\centering
	\includegraphics[width=\linewidth]{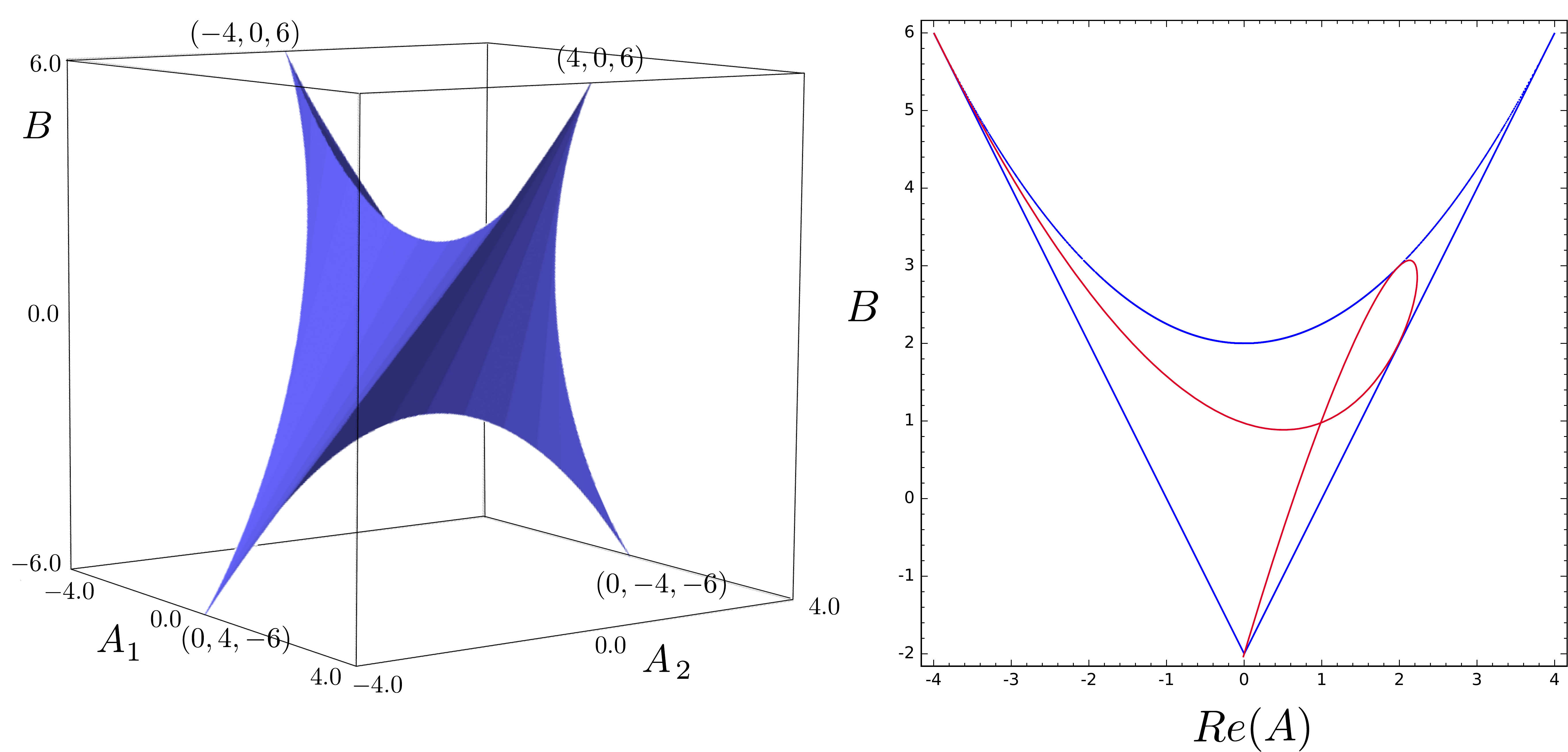}
	\caption{(Color online) Left: Simplex-like volume in the parameter space where four Fermi points exist. Right: For $\alpha_{\rm so}=0$, 
	The colored region represents the values of parameters where four Fermi points exist. The curve inside the shaded region 
	corresponds to the parameters where the four Fermi points satisfy the momentum conservation relation of the FTI operator.}\label{fig:Simplex}
\end{figure}

The single particle band structure is given by the relation $p_{\pm}(\tilde{k})=0$, where $p_\pm(\tilde{k})$ is 
\begin{equation}
 p_\pm(\tilde{k})=\left(E+t\cos\left(\tilde{k}-\frac{\Phi}{2}\right)\pm\alpha\right)\left(E+t\cos\left(\tilde{k}+\frac{\Phi}{2}\right)\mp\alpha\right)-t_\perp^2,
\end{equation}
with $t,t_\perp$ the intra and inter leg tunnelling amplitudes, $\Phi$ the SO generated flux and $\alpha_{\rm so}$ the SO coupling parameter that breaks spin $S_z$ 
conservation. We can think of $p_\pm(\tilde{k})$ as a complex polynomial in the complex plane as a function of the variable $z=e^{i\tilde{k}}$. Given than $p_-(\tilde{k})$ 
can be obtained from $p_+(\tilde{k})$ by changing the sign of $\alpha$, we consider just 
\begin{equation}
 q(z)=\frac{4z^2}{t^2}p_+(\tilde{k}(z))=z^4+Az^3+Bz^2+\bar{A}z+1,
\end{equation}
where $A=\frac{4}{t}(E\cos\left(\frac{\Phi}{2}\right)-i\alpha\sin\left(\frac{\Phi}{2}\right))$ and $B=\frac{4}{t}\left(E^2-\alpha_{\rm so}^2-t_\perp^2\right)+\frac{2}{t}\cos{\Phi}$,
$\bar{A}$ is the complex conjugate of $A$. The complex polynomial $q(z)$ satisfies $q(z)=z^4\overline{q(\bar{z}^{-1})}$ so if $z_0$ is a root of $q(z)$ then $1/\overline{z_0}$
is also a root. The existence of four Fermi points corresponds to $q(z)$ having all its roots in the unit circle. We denote this region of parameters as $W$. 
Clearly, the point $A=B=0$ corresponds to four different roots in the unit circle, so the region $W$ contains this point. Changing the parameters $A$ and $B$ the roots 
change, but the number of different roots can only change when different roots become equal as the parameters vary. Thus, the vanishing of the polynomial's $q(z)$ 
discriminant $D(q)$ defines the boundary of the region $W$. Factorizing the polynomial $q(z)$
in its roots $ q(z)=(z-z_1)(z-z_2)(z-z_3)(z-z_4)$, and assuming that all the roots live in the unit circle, implies
\begin{eqnarray}\label{prod=1}
& & z_1z_2z_3z_4=1,\\
& & z_1+z_2+z_3+z_4=A=A_1+iA_2,\\\label{pair_sum}
& &z_1z_2+z_1z_3+z_1z_4+z_2z_3+z_2z_4+z_3z_4=B.
\end{eqnarray}
To gain some insight about the shape of the region $W$, let us consider the cases 
\begin{enumerate}
\item All the roots are equal $q(z)=(z-z_1)^4$. By condition (\ref{prod=1}), $z_1=e^{\frac{i\pi n}{2}}$, $n=0,..3.$. In terms of the parameters 
$\bm{v}=(A_1,A_2,B)$, each value of $n$ correspond to a vertex of $W$: $\bm{v}_0=(-4,0,6),\bm{v}_1=(0,-4,-6),\bm{v}_2=(4,0,6),$ and $\bm{v}_3=(0,4,-6).$
\item Three equal roots $q(z)=(z-z_1)(z-z_2)^3$. Using the conditions (\ref{prod=1}-\ref{pair_sum}), we find that this situation is satisfied in the curve
defined parametrically by $\bm{v}(\theta)=(-\cos(3\theta)-3\cos(\theta),\sin(3\theta)-3\sin(\theta),6\cos(2\theta))$. Note that $\bm{v}(\frac{\pi n}{2})=\bm{v}_n$.
\item Two pairs of equal roots $q(z)=(z-z_1)^2(z-z_2)^2$. We find the curve $B=\frac{1}{4}A_1^2+2$ connecting $\bm{v}_0$ with $\bm{v}_2$ and $B=-\frac{1}{4}A_2^2-2$
connecting $\bm{v}_1$ with $\bm{v}_3$.
\item Two equal roots and two different roots $q(z)=(z-z_1)(z-z_2)(z-z_3)^2$.
\end{enumerate}
The first case defines the vertices of $W$, which corresponds to a simplex-like region shown in Fig.\ref{fig:Simplex}. The edges of $W$ corresponds to the second and third cases. The faces of 
$W$ correspond to the fourth case. The isometry group of $W$ is a subgroup of the isometry group of the 4-simplex \cite{Petersen2005}.

\begin{figure}[ht!]
	\centering
	\includegraphics[width=8cm]{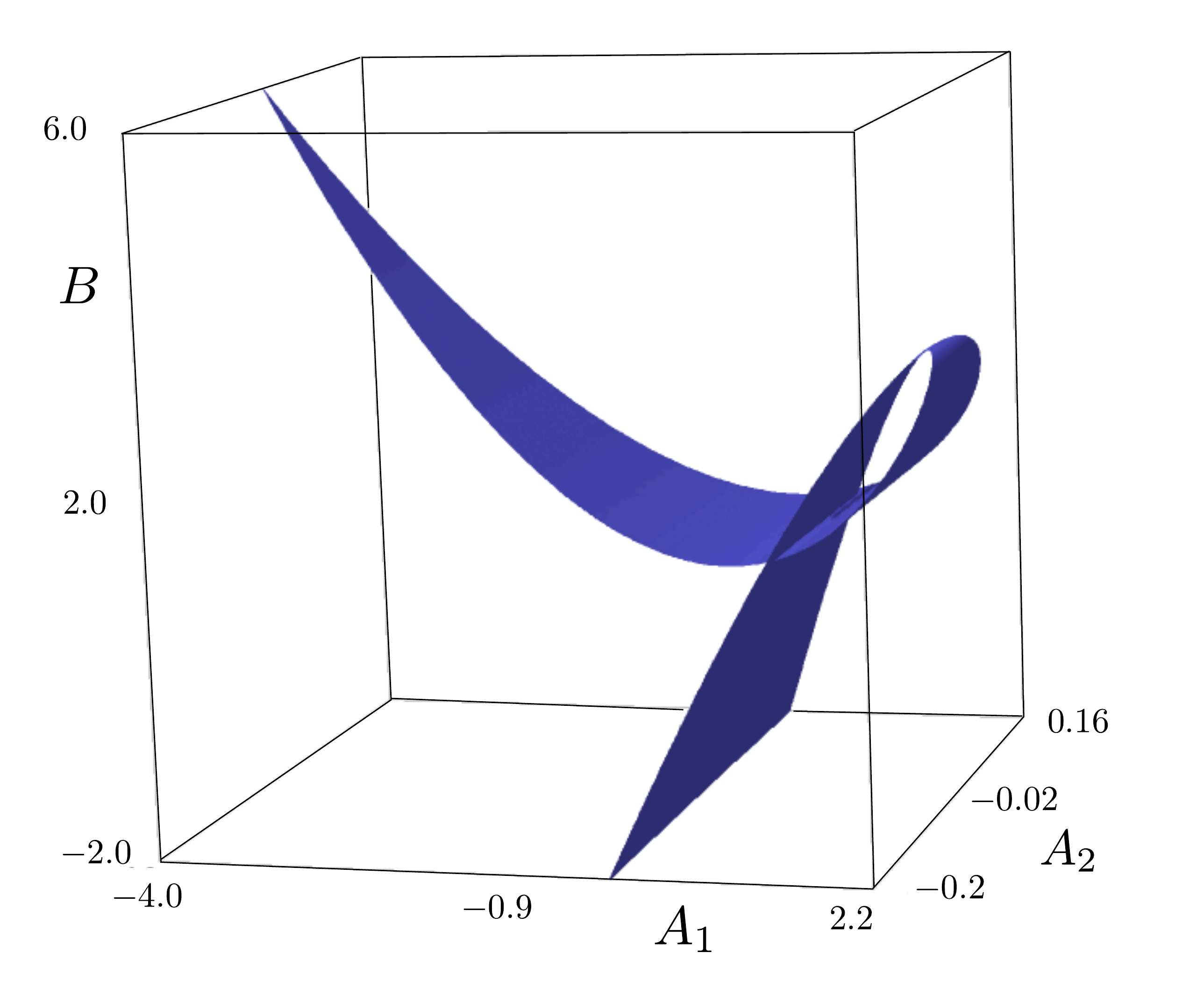}
	\caption{(Color online) For $\alpha_{\rm so}\neq0$, there is a surface of parameters where the FTI operator satisfies momentum conservation.
	Here we plot that surface, given parametrically by (\ref{surf_param}-\ref{surf_param2}) with $\theta_1=-2\theta_2+\epsilon$ with 
	$\epsilon=[-0.05,0.05]$ and $\theta_2=[0,2\pi]$.}\label{fig:region_params}
\end{figure}

The condition of momentum conservation of the FTI operator translates in this language as $z_2=z_1^2z_3^3$. The surface defined parametrically by
\begin{eqnarray}\label{surf_param}
 A_1&=&-\cos(\theta_1)-\cos(\theta_2)-\cos(2\theta_1+3\theta_2)-\cos(3\theta_1+4\theta_2),\\
 A_2&=&2\sin\left(\frac{\theta_1+\theta_2}{2}\right)\left(\cos\left(\frac{5\theta_1+7\theta_2}{2}\right)-\cos\left(\frac{\theta_1-\theta_2}{2}\right)\right),\\\label{surf_param2}
 B  &=& 4\cos(\theta_1+\theta_2)\cos(2(\theta_1+\theta_2))+2\cos(2(\theta_1+2\theta_2)).
\end{eqnarray}
with $\theta_i\in[0,2\pi)$ \textbf{and} strictly inside the simplex region $W$, determines the values where the FTI operator can exist and conserves momentum.
The simplex-like region $W$, together with a cut for vanishing and small SO coupling $\alpha_{\rm so}$ are given in Figs. \ref{fig:Simplex}-\ref{fig:region_params}.


\end{appendix}

\end{document}